\documentclass[10pt,a4paper]{article}

\usepackage{bbm}
\usepackage[final]{graphics}
\usepackage{amsmath}
\usepackage{amsfonts,amsbsy}
\usepackage{amssymb}
\usepackage{bm}
\usepackage{color}
\definecolor{linkcolor}{rgb}{0.4,0.1,0.1}
\definecolor{bibcolor}{rgb}{0.4,0.1,0.1}
\usepackage[sort&compress,merge,numbers]{natbib}
\usepackage[colorlinks,linkcolor=linkcolor,citecolor=bibcolor,pdfpagelabels,hyperindex=true]{hyperref}
\hypersetup{pdftitle={Schwinger formalism revisited}}
\hypersetup{pdfauthor={Francois Gelis, Naoto Tanji}}
\hypersetup{pdfsubject={Schwinger mechanism}}

\newcommand{\be}{\begin{equation}}
\newcommand{\ee}{\end{equation}}
\newcommand{\bea}{\begin{eqnarray}}
\newcommand{\eea}{\end{eqnarray}}

\def\empile#1\over#2{\mathrel{\mathop{\kern 0pt#1}\limits_{#2}}}
\def\bs{\boldsymbol}

\def\TODO#1{}

\def\p{{\boldsymbol p}}
\def\q{{\boldsymbol q}}

\def\k{{\boldsymbol k}}

\def\x{{\boldsymbol x}}
\def\s{{\boldsymbol s}}
\def\y{{\boldsymbol y}}

\def\z{{\boldsymbol z}}

\def\x{{\boldsymbol x}}

\def\A{{\boldsymbol A}}
\def\B{{\boldsymbol B}}
\def\E{{\boldsymbol E}}

\def\cd{\!\cdot\!}

\newcommand{\slL}{\raise.15ex\hbox{$/$}\kern-.53em\hbox{$L$}}
\newcommand{\slP}{\raise.15ex\hbox{$/$}\kern-.53em\hbox{$P$}}
\newcommand{\slD}{\raise.15ex\hbox{$/$}\kern-.67em\hbox{$D$}}
\newcommand{\slp}{\raise.1ex\hbox{$/$}\kern-.63em\hbox{$p$}}
\newcommand{\slq}{\raise.1ex\hbox{$/$}\kern-.53em\hbox{$q$}}
\newcommand{\slv}{\raise.1ex\hbox{$/$}\kern-.63em\hbox{$v$}}
\newcommand{\slR}{\raise.15ex\hbox{$/$}\kern-.53em\hbox{$R$}}
\newcommand{\slQ}{\raise.15ex\hbox{$/$}\kern-.53em\hbox{$Q$}}
\newcommand{\slK}{\raise.15ex\hbox{$/$}\kern-.53em\hbox{$K$}}
\newcommand{\slk}{\raise.15ex\hbox{$/$}\kern-.53em\hbox{$k$}}
\newcommand{\slSigma}{\raise.15ex\hbox{$/$}\kern-.53em\hbox{$\Sigma$}}
\newcommand{\slcalP}{\raise.15ex\hbox{$/$}\kern-.63em\hbox{$\cal P$}}
\newcommand{\slcalA}{\raise.15ex\hbox{$/$}\kern-.63em\hbox{$\cal A$}}
\newcommand{\slA}{\raise.15ex\hbox{$/$}\kern-.73em\hbox{$A$}}
\newcommand{\slbfA}{\raise.15ex\hbox{$/$}\kern-.73em\hbox{${\imb A}$}}
\newcommand{\slpartial}{\raise.15ex\hbox{$/$}\kern-.53em\hbox{$\partial$}}
\newcommand{\sla}{\raise.15ex\hbox{$/$}\kern-.53em\hbox{$a$}}
\newcommand{\slb}{\raise.15ex\hbox{$/$}\kern-.53em\hbox{$b$}}
\newcommand{\slc}{\raise.15ex\hbox{$/$}\kern-.53em\hbox{$c$}}
\newcommand{\slC}{\raise.15ex\hbox{$/$}\kern-.63em\hbox{$C$}}
\newcommand{\sln}{\raise.15ex\hbox{$/$}\kern-.575em\hbox{$n$}}

\newcommand{\unit}{\mbox{1}\hspace{-0.25em}\mbox{l}}

\usepackage[normalem]{ulem}  
\renewcommand\sout{\bgroup \color{blue} \ULdepth=-.5ex \ULset}

\addtocontents{toc}{\protect\setcounter{tocdepth}{1}}

\begin{document}

\title{ \vspace{1cm}\bf Schwinger mechanism revisited}
\author{Fran\c{c}ois\ Gelis,$^1$ Naoto\ Tanji$^2$}
\maketitle

\begin{itemize}
\item[{\bf 1.}] Institut de physique th\'eorique, CEA, CNRS, Universit\'e Paris Saclay\\ F-91191 Gif-sur-Yvette, France
  \item[{\bf 2.}] Institut f\"ur Theoretische Physik, Universit\"{a}t Heidelberg\\ Philosophenweg 16\,
    D-69120, Heidelberg, Germany
\end{itemize}
\vglue 20mm

\begin{abstract} 
In this article, we review recent theoretical works on the Schwinger
mechanism of particle production in external electrical
fields. Although the non-perturbative Schwinger mechanism is at the
center of this discussion, many of the approaches that we discuss can
cope with general time and space dependent fields, and therefore also
capture the perturbative contributions to particle production.
\end{abstract}

\eject
\tableofcontents

\section{Introduction}
\label{sec:intro}

\begin{figure}[htbp]
  \begin{center}
    \resizebox*{8cm}{!}{\includegraphics{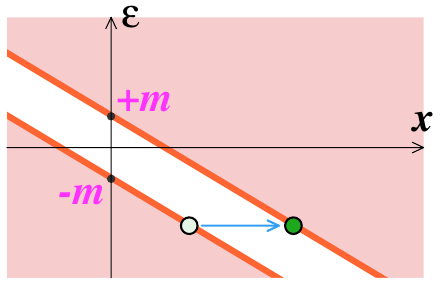}}
  \end{center}
  \caption{\label{fig:S-tunnel}Schematic picture of the tunneling
    process involved in the Schwinger mechanism. The white band is the
    gap between the anti-electron Dirac sea and the positive energy
    electron continuum, tilted by the potential $V(x)=-Ex$ in the presence
    of an external electrical field $E$.}
\end{figure}

The Schwinger mechanism \cite{Schwinger:1951nm} is a non-perturbative
phenomenon by which elec\-tron-positron pairs can be produced by a
static electrical field
\cite{Sauter:1931zz,Heisenberg:1935qt,Schwinger:1951nm}. It is the
fact that the field is time independent that makes this process
genuinely non-perturbative. Indeed, such an electromagnetic field can
be viewed as an ensemble of photons of zero frequency, which by energy
conservation are unable to produce massive particles. The
non-perturbative nature of this phenomenon is visible in the fact that
the corresponding transition amplitudes are non analytic in the
coupling: since they are proportional to $\exp(-\pi m^2/eE)$, the
coefficients of their Taylor expansion around $e^2=0$ are all zero,
and only calculations to all orders in the coupling can make this
effect manifest. The non-analytic behavior of these transition
probabilities can be understood as a tunneling phenomenon (see for
instance \cite{Itzykson:1980rh}) by which a particle from the Dirac sea is
pulled into the positive energy states: as illustrated in the figure
\ref{fig:S-tunnel}, the external gauge potential tilts the gap between
the Dirac sea and the positive energy states, allowing a hole from the
sea to tunnel through this gap and materialize as an on-shell positive
energy particle on the other side.
From this geometrical interpretation, it is clear that the length that
needs to be crossed is inversely proportional to the electrical field,
leading to a tunneling probability that is exponential in $1/E$.

From this pocket formula, we also see that the external field must be
very intense in order to lead to a significant probability of particle
production: at small coupling, this probability is of order one only
for fields that are comparable to a critical field $E_c$ inversely
proportional to the coupling constant, $E\sim E_c\equiv m^2/e$. Even
for the lightest charged particle --the electron--, these are
extremely intense fields that still surpass by several orders of
magnitude the largest fields achievable experimentally. Note that when
one studies the interaction of a charged particle with such an
external electrical field, the corresponding expansion is in powers of
$eE$. However, if $eE$ is comparable to the squared mass of the
particle, then there is no small parameter to control this expansion
and one must treat the external field to all orders. This is a
prerequisite of any formalism aiming at describing the Schwinger
mechanism.

Although the original derivation of the Schwinger mechanism was made
in the context of quantum electrodynamics, it can be generalized to
any quantum field theory coupled to some external field.  In the
context of strong interactions and quantum chromo-dynamics, where the
coupling constant $g$ is numerically much larger than in QED, the
production of pairs by the Schwinger mechanism may be achievable with
more moderate chromo-electrical fields, and this phenomenon may play a
role in the discussion of particle production in heavy ion collisions
\cite{Biro:1984cf,Kajantie:1985jh,Gatoff:1987uf,Kharzeev:2006zm}, or in the decay of ``hadronic strings''
\cite{Andersson:1983ia} in the process of hadronization. However, the
gauge fields that are generated in heavy ion collisions have two
important features: (i) since gauge fields have a direct coupling to
gluons, the leading contributions for gluon production in heavy ion
collisions are tree level contributions, that supersede the 1-loop
contributions encountered in the Schwinger mechanism; (ii) these
fields are in general space and time dependent
\cite{McLerran:1993ni,McLerran:1993ka}, and one must therefore use a
formalism that can cope with the most general type of external field,
without assuming any symmetry in its space-time dependence. Naturally,
such fields generally imply that it is no longer possible to obtain
closed analytic expressions (such analytical results have been
obtained only for very special time dependences, and for spatial
dependences that possess a high degree of symmetry). As a consequence,
it is usually necessary to resort to numerical studies, and a
recurrent concern in this review will be the practicality of various
approaches for these numerical simulations.

The Schwinger mechanism can also be discussed in the context of a much
simpler scalar field theory, coupled either to a scalar or vector
external field (as in scalar QED). Despite its lack of connection to
possible experimental realizations, this is the simplest example one
may think of and it offers a very useful playground for testing new
theoretical developments. In this review, we will often use such toy
theories to illustrate various novel approaches, because of their
didactic or phenomenological interest.

The outline of this review is as follows. In the section
\ref{sec:extsource}, we consider a scalar theory coupled to an
external source and discuss particle production at tree level and
1-loop. The goal of this section is to present general ideas about
these processes, whose range of validity is much broader than the
simple toy model used to introduce them. The section \ref{sec:corr} is
devoted to a discussion of the multi-(anti)particle correlations that
exist in the Schwinger mechanism. We first derive their general
structure and then work it out in the special case of spatially
homogeneous fields (possibly time dependent). The general discussion
of the section \ref{sec:extsource} leads to a formulation of the
Schwinger mechanism in terms of a complete basis of \emph{mode
  functions}. In the section \ref{sec:other}, we relate this
representation to other approaches: the method of Bogoliubov
transformations, the quantum kinetic equations, and the Wigner
formalism, and we discuss lattice numerical implementations of this
approach in the section \ref{sec:latt}. The section \ref{sec:WL} is
devoted to the worldline formalism, a radically different (but
equivalent) formulation of the Schwinger formalism based on
Schwinger's proper-time representation of propagators in an external
field.  Besides new methods for calculating particle production in an
external field, this approach provides a great deal of intuition on
the spatio-temporal development of the production process. In the
section \ref{sec:dynamically}, we discuss the idea of
\emph{dynamically assisted} Schwinger mechanism, where one
superimposes two fields that have vastly different timescales and
magnitudes in order to reach particle yields that are much larger than
what would have been achieved with each field separately.

\section{Quantum fields coupled to external sources}
\label{sec:extsource}

In this section, we discuss general aspects of quantum field theories
coupled to an external source \cite{Gelis:2006yv,Gelis:2006cr}. Our main goal is to
present the general aspects of such theories, focusing on their main
differences with field theories where the sole interaction are the
self-interaction of the fields. For the sake of simplicity, we
consider a scalar field theory with a quartic self-interaction, whose
Lagrangian is given by\footnote{For extra simplicity, we choose a
  potential whose absolute minimum is located at $\phi=0$, so that the
  perturbative expansion around the vacuum also corresponds to small
  field fluctuations. The mass is also important: when $m\not=0$,
  producing a particle costs some energy, which is crucial to avoid
  infrared singularities.}
\begin{equation}
  {\cal L}\equiv
  \frac{1}{2}(\partial_\mu\phi)(\partial^\mu\phi)-\frac{1}{2}m^2\phi^2-\frac{g^2}{4!}\phi^4+j\phi\; ,
  \label{eq:Lag-scal}
\end{equation}
where $j(x^0,\x)$ is an unspecified function of space-time. Note that
this source is a commuting number-valued object, rather than an
operator.  In the same way that one usually assumes that
self-interactions are adiabatically turned on and off when $x^0\to
\mp\infty$, we assume that the external source decreases fast enough
when time goes to $\pm\infty$. Moreover, in order to preserve the
unitarity of the theory, the external source must be real valued.
Since this section is devoted to a general discussion, we allow the
particles to have self-interactions. This is important for
applications to gluon production in heavy ion collisions, and it leads
to some complications. In contrast, the production of
electron-positron pairs in QED is simpler to study because the
electrons can only interact directly with the photon
field. Interactions between electrons and positrons can happen
indirectly, with the mediation of a photon, but this effect is an
extremely small correction in practice, that would arise beyond the
order considered here.

\subsection{Vacuum diagrams}
The main feature of such a theory is that it describes an \emph{open
system}: even if the system is initialized in the vacuum state that
contains no particles, the external source can --and in general will--
produce particles. This is in sharp contrast with the same theory in
the vacuum, where the in- vacuum state and the out- vacuum state are
related by a unitary transformation $\Omega(+\infty,-\infty)$~:
\begin{equation}
\big|0{}_{\rm out}\big>=\Omega(+\infty,-\infty)\,\big|0{}_{\rm in}\big>\qquad\qquad\mbox{(when $j\equiv 0$)}\; .
\end{equation}
This property ensures that when $j\equiv 0$, the vacuum state evolves
with probability one into the vacuum state, i.e. that no particle is
created. An equivalent statement is that the vacuum to vacuum
transition amplitude is a pure phase:
\begin{equation}
\big<0_{\rm out}\big|0{}_{\rm in}\big>=e^{i\,{\cal V}}\qquad \mbox{with ${\cal V}\in{\mathbbm R}$}\; .
\end{equation}
This has some important practical consequences. The perturbative
expansion for transition amplitudes generates diagrams that contain
disconnected vacuum sub-diagrams, i.e. diagrams that have no external
legs, but the above property tells us that the sum of all the vacuum
diagrams is a pure phase that, although it appears as a prefactor in
every transition amplitude, does not play any role after squaring the
amplitudes in order to obtain transition probabilities. Therefore, it
is legitimate to ignore from the start all the graphs that contain
vacuum sub-diagrams in the $j\equiv 0$ case.

\begin{figure}[htbp]
  \begin{center}
    \resizebox*{7cm}{!}{\includegraphics{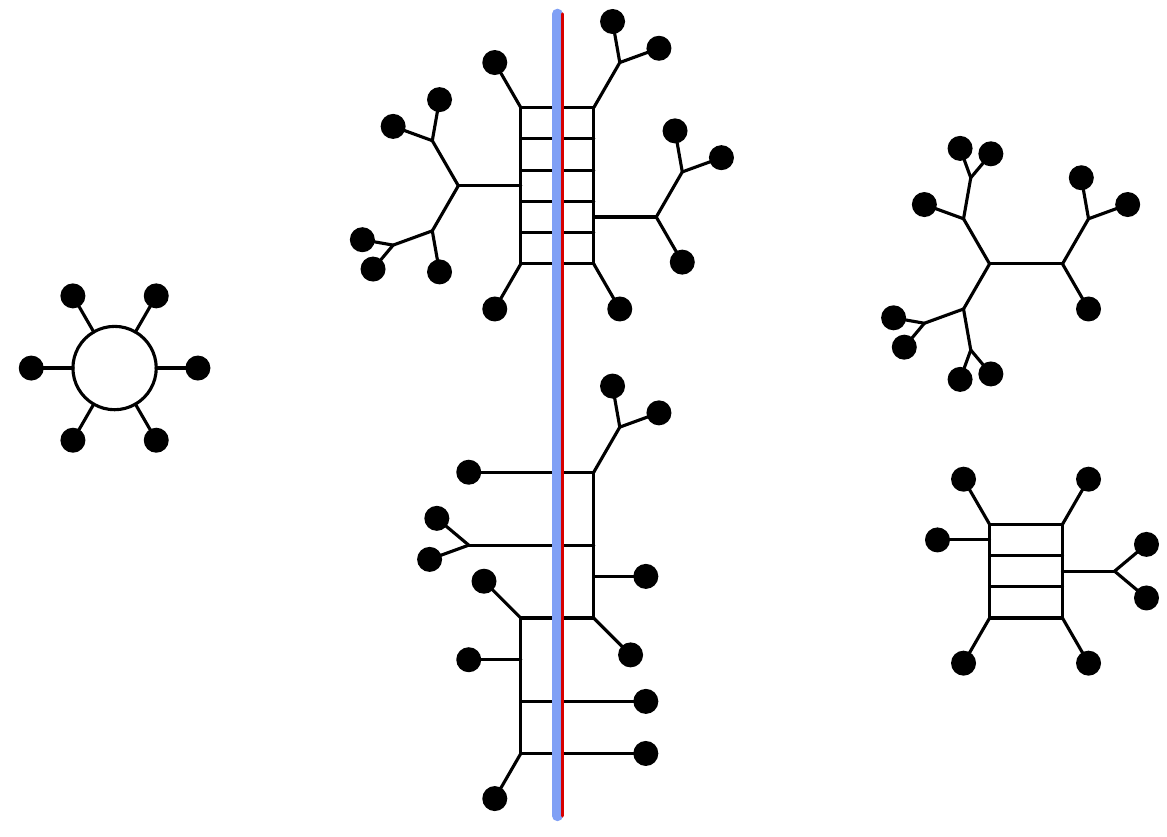}}
  \end{center}
  \caption{\label{fig:P11}Example of diagram contribution to the
    probability of producing 11 particles (here illustrated in the
    case of a scalar theory with a cubic coupling to simplify the
    diagrams). The black dots denote the insertions of the external
    source $j(x)$. The vertical line indicates the final state (left
    of this line: amplitude, right: complex conjugated amplitude).}
\end{figure}

Let us return to the case $j\not=0$ and denote $P(\alpha)$ the
transition probability from the initial vacuum state $\big|0{}_{\rm
  in}\big>$ to an arbitrary final state $\big|\alpha{}_{\rm
  out}\big>$. The unitarity of the theory implies that the sum of
these probabilities over all the possible final states is equal to
unity,
\begin{equation}
\sum_\alpha P(\alpha)=P(0)+\sum_{\alpha\not=0} P(\alpha)=1\; ,
\end{equation}
where in the first equality we have separated the vacuum and the
populated states. If particles are produced during the time evolution
of the system (this is possible only if $j\not=0$), then there is at
least one non-empty state $\alpha$ for which $P(\alpha)>0$. Since all
these probabilities are numbers in the range $[0,1]$, the previous
identity implies that $P(0)<1$, i.e. that the vacuum does not evolve
into the vacuum state with probability one anymore.  A trivial
consequence is that the vacuum to vacuum transition amplitude is no
longer a pure phase, since its square is strictly smaller than
unity. Because of this, it is no longer possible to disregard the
vacuum graphs, as illustrated in the figure \ref{fig:P11}.
In general, this complicates considerably the diagrammatic expansion
when $j\not=0$, but we will see later that inclusive quantities have a
diagrammatic expansion made only of connected diagrams. Moreover, in
the case of a strong external source (of order $1/g$), it becomes
hopelessly complicated to calculate exclusive quantities (such as the
probability $P(\alpha)$ for an individual final state) while the
inclusive ones are much easier to access.

\subsection{Power counting}
In order to discuss ways of organizing the calculation of observables
in field theories coupled to an external source, the first step is to
assess the order of magnitude of a graph in terms of its
topology. Obviously, for a graph with multiple disconnected components
such as the one in the figure \ref{fig:P11}, the order is obtained as
the product of the orders of each of its connected
sub-diagrams. Therefore, it is sufficient to consider only connected
graphs in this discussion. A connected diagram is fully characterized
by the number of sources $n_{_J}$, the number of propagators $n_{_P}$,
the number of vertices $n_{_V}$, the number of loops $n_{_L}$ and the
number of external particles $n_{_E}$ (i.e. propagators endpoints that
are not connected to a vertex or to a source). These quantities are
not all independent. A first constraint comes from the fact that each
propagator has two endpoints and each vertex receives four lines (for
the scalar theory described by the Lagrangian of eq.~(\ref{eq:Lag-scal})):
\begin{equation}
2n_{_P}=4n_{_V}+n_{_J}+n_{_E}\; .
\end{equation}
A second identity expresses the number of independent loops $n_{_L}$
in terms of the other characteristics of the graph\footnote{This can
  be proven from Euler's formula for a graph,
  $\#\mbox{nodes}-\#\mbox{edges}+\#\mbox{faces}=2-2\#\mbox{holes}$,
  and $\#\mbox{nodes}=n_{_V}+n_{_J}+n_{_E}$, $\#\mbox{edges}=n_{_P}$
  and $n_{_L}=\#\mbox{faces}+2\#\mbox{holes}-1$.}:
\begin{equation}
  n_{_L}=n_{_P}-n_{_E}-n_{_J}-n_{_V}+1\; .
\end{equation}
Thanks to these two formulas, the order of a connected graph can be
written as
\begin{equation}
  \omega(G)\equiv g^{2n_{_V}}\,j^{n_{_J}}=g^{-2+n_{_E}+2n_{_L}}\,(gj)^{n_{_J}}\; .
  \label{eq:PC}
\end{equation}
In this expression, we have combined one power of the coupling $g$
with each power of the external source $j$, because the combination
$gj$ disappears from this power counting formula in the strong source
regime where $j\sim g^{-1}$ (i.e. when particle production by the
Schwinger mechanism becomes likely).

Eq.~(\ref{eq:PC}) displays a standard dependence on the number of loops
and external legs. But it also indicates that in the strong source
regime the dependence of the order on $n_{_J}$ disappears: therefore,
infinitely many graphs contribute at each order. In conjunction with
our previous observation that vacuum graphs cannot be disregarded,
this poses a serious bookkeeping challenge for the calculation of
quantities such as the transition probability shown on the figure
\ref{fig:P11}.

\subsection{Exclusive and inclusive quantities}
Earlier, we have alluded to important differences between inclusive
and exclusive observables when it comes to calculating them. Before
going into these technical differences, let us first define them more
precisely. Loosely speaking, exclusive observables are related to a
full measurement of the final state, while inclusive observables
involve dropping a lot of information about the final state.

If we assume the initial state to be empty, all the information about
the final state is encoded in the transition amplitudes
\begin{equation}
  \big<\p_1\cdots\p_n{}_{\rm out}\big|0{}_{\rm in}\big>
  \label{eq:trans-amp}
\end{equation}
in which the final state is fully specified by giving the list of the
momenta of all the final particles (in theories with more structure
than the scalar theory under consideration in this section, one would
also need to specify other quantum numbers of the final
particles). Any observable related to this evolution can be expressed
in terms of these amplitudes. For instance, the differential
probability for producing $n$ particles can be expressed as
\begin{equation}
  \frac{dP_n}{d^3\p_1\cdots d^3\p_n}
  =
  \frac{1}{n!}\frac{1}{(2\pi)^32E_{\p_1}}\cdots \frac{1}{(2\pi)^32E_{\p_n}}\,\big|\big<\p_1\cdots\p_n{}_{\rm out}\big|0{}_{\rm in}\big>\big|^2\; ,
\end{equation}
where $E_\p\equiv\sqrt{\p+m^2}$ is the on-shell energy of a particle
of momentum $\p$. This quantity is called exclusive because only a
single final state can contribute to it, at the exclusion of all
others.

The archetype of inclusive observables are the particle spectra,
obtained from the above probability distributions by integrating out
the phase-space of all particles but a few. The simplest of them is
the single particle spectrum, defined as
\begin{equation}
  \frac{dN_1}{d^3\p}
  \equiv
  \sum_{n=0}^\infty(n+1)\int d^3\p_1\cdots d^3\p_n\; \frac{dP_{n+1}}{d^3\p d^3\p_1\cdots d^3\p_n}\; .
\end{equation}
This quantity gives the number of particles (thanks to the factor
$n+1$ included under the sum) produced in a given momentum range, but
the information about the distribution of individual final states has
been lost. As we shall see, this quantity is much easier to calculate than the
exclusive observables. Experimentally, it is also much more accessible
since it only requires to make an histogram of the momenta of the
final state particles. The single particle spectrum has obvious
generalizations: the 2-particle spectrum, the 3-particle spectrum,
etc... that provide information about the correlations between the
final state particles.

The ``vacuum survival probability'' often plays a special role in
discussions of the Schwin\-ger mechanism. This quantity is nothing but
\begin{equation}
P_0\equiv \big|\big<0{}_{\rm out}\big|0{}_{\rm in}\big>\big|^2\; .
\end{equation}
A standard result in field theory is that the vacuum transition
amplitude $\big<0{}_{\rm out}\big|0{}_{\rm in}\big>$ can be written as
the exponential of the sum of the connected vacuum diagrams,
\setbox2=\hbox to
  7cm{\resizebox*{7cm}{!}{\includegraphics{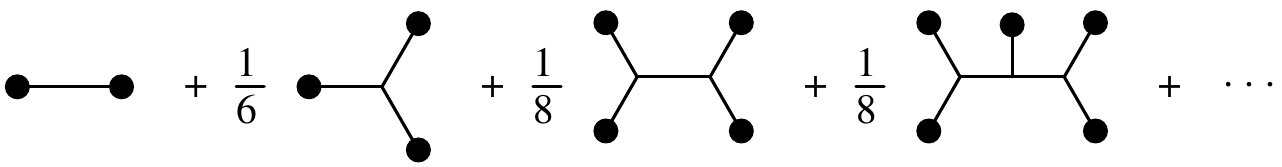}}}
\begin{equation}
  \big<0{}_{\rm out}\big|0{}_{\rm in}\big>=e^{i\,{\cal V}}\; ,\qquad i\,{\cal V}\equiv\quad\;\raise -3.5mm\box2
\label{eq:Vconn}
\end{equation}
(Only a few tree level contributions to ${\cal V}$ have been shown in the
illustration.)  After squaring this amplitude, we get
$P_0=\exp(-2\,{\rm Im}\,{\cal V})$. In the presence of an external source
that remains constant over a long interval of time and is sufficiently
spatially homogeneous, the imaginary part $2{\rm Im}\,{\cal V}$ can usually
be written as an integral over space-time,
\begin{equation}
2{\rm Im}\,{\cal V}=\int d^4x\;(\cdots)\; ,
\end{equation}
and the integrand in this formula is often interpreted as the
``particle production rate''. We will return on this interpretation
later, as it not entirely accurate in general and can be a source of confusion.

\subsection{Cutting rules and Schwinger-Keldysh formalism}
Earlier in this section, we have seen that the imaginary part of the
connected vacuum graphs, ${\rm Im}\,{\cal V}$, plays an important role
in the expression of the transition probabilities. This imaginary part
can be obtained from an extension of the standard Feynman rules known
as Cutkosky's cutting rules \cite{Cutkosky:1960sp,'tHooft:1973pz}. Here, we
do not re-derive these rules but simply state them as a recipe, that
must be applied to every graph:
\begin{itemize}
  \item[{\bf i.}] Divide the nodes (vertices and sources) contained in
    the graph into a set of $+$ nodes and a set of $-$ nodes, in all
    the possible ways. It is customary to materialize diagrammatically
    these two sets of nodes by drawing a line (the ``cut'') that
    divides the graph in two subgraphs. Even if the original graph is
    connected, the sub-graphs on each side of this cut do not have to
    be connected.
  \item[{\bf ii.}] The $+$ vertices give a factor $-ig^2$ and the
    $-$ vertices give a factor $+ig^2$. The $+$ sources give a factor
    $+ij(x)$ and the $-$ sources give a factor $-ij(x)$.
  \item[{\bf iii.}] Two nodes of types $\epsilon=\pm$ and
    $\epsilon'=\pm$ are connected by a bare propagator
    $G^0_{\epsilon\epsilon'}$. In momentum space, these four
    propagators read
    \begin{align}
      G_{++}^0(p)&=\frac{i}{p^2-m^2+i\epsilon}\;,
      &&&
      G_{--}^0(p)&=\frac{-i}{p^2-m^2-i\epsilon}&&&\nonumber\\
      G_{+-}^0(p)&=2\pi\theta(-p^0)\delta(p^2-m^2)\;,
      &&&
      G_{-+}^0(p)&=2\pi\theta(+p^0)\delta(p^2-m^2)&&&\; .
      \label{eq:cut-rules}
    \end{align}
  \item[{\bf iv.}] Multiply the outcome of these rules by $1/2$.
\end{itemize} 
The subgraph that involves only $+$ labels is obtained by the usual
Feynman rules used to calculate transition amplitudes, while the $-$
subgraph is given by the complex conjugation of these rules. The
interpretation of these cutting rules is straightforward: the $+$
sector corresponds to an amplitude and the $-$ sector to a complex
conjugated amplitude, while the $+-$ and $-+$ propagators provide the
phase-space integration for the (on-shell, hence the delta functions)
final state particles.

In fact, these rules are the perturbative realization\footnote{In the
  scalar theory under consideration in this section, the cutting rules
  realize the optical theorem at the level of single graphs. In
  non-abelian gauge theories, where ghosts cancel the unphysical gluon
  polarizations, similar cutting rules provide a realization of the
  optical theorem for groups of graphs that form a gauge invariant
  set.} of the optical theorem, that stems from unitarity. If we write
the $S$ matrix as $S\equiv 1+iT$, then $S^\dagger S=1$ can be
rewritten as
\begin{equation}
i(T^\dagger -T)=T^\dagger T\; .
\end{equation}
By taking the expectation value of this identity in the vacuum state
and inserting a complete set of states in the right hand side, it
leads to
\begin{equation}
  {\rm Im}\,\underbrace{\vphantom{\sum_\alpha}\big<0{}_{\rm in}\big|T\big|0{}_{\rm in}\big>}_{{\cal V}}
  =
  \frac{1}{2}
  \sum_\alpha\big<0{}_{\rm in}\big|T^\dagger\big|\alpha{}_{\rm in}\big>\big<\alpha{}_{\rm in}\big| T\big|0{}_{\rm in}\big>
  =
  \frac{1}{2}\underbrace{\sum_\alpha}_{+-}
  \underbrace{\vphantom{\sum_\alpha}\big<0{}_{\rm in}\big|\alpha{}_{\rm out}\big>}_{-\mbox{\scriptsize\ sector}}
  \underbrace{\vphantom{\sum_\alpha}\big<\alpha{}_{\rm out}\big|0{}_{\rm in}\big>}_{+\mbox{\scriptsize\ sector}}\; .
\end{equation}
The factor $1/2$ in the right hand side of this identity is the origin
of the rule {\bf iv} above. The Schwinger-Keldysh
\cite{Schwinger:1960qe,Keldysh:1964ud} formalism is essentially equivalent to
Cutkosky's cutting rules, but it is usually introduced as a tool for
the perturbative calculation of squared matrix elements rather than as
a technique for calculating the imaginary part of a scattering
amplitude (however, the optical theorem states that the two are
closely related).

\subsection{General remarks on the distribution of produced particles}
All the transition amplitudes such as (\ref{eq:trans-amp}) contain in
their diagrammatic expansion the disconnected vacuum graphs, as shown
in the example of the figure \ref{fig:P11}. In other words, one can
pull out a factor $\exp(i{\cal V})$ in all these amplitudes, or a factor
$\exp(-2\,{\rm Im}\,{\cal V})$ in squared amplitudes. From our previous power
counting formula (\ref{eq:PC}), we expect ${\rm Im}\,{\cal V}$ to start at
the order $1/g^2$ in the strong source regime,
\begin{equation}
2{\rm Im}\,{\cal V}=\frac{a}{g^2}\; ,\qquad P_0=e^{-a/g^2}\; ,
\end{equation}
where $a$ is an infinite series in powers of
$g^{2n_{_L}}(gj)^{n_{_J}}$. At tree level ($n_{_L}=0$) and in the
strong source regime, $a$ is therefore of order unity (but depends
non-perturbatively on the source).

Let us now turn to the probabilities of producing particles. Besides
the factor $\exp(i{\cal V})$, the transition amplitude from the vacuum to a
state containing one particle is made of graphs that connect sources
to a single final particle. From eq.~(\ref{eq:PC}), these graphs start
at the order $g^{-1}$, and the leading contribution to $P_1$ is of the
form
\begin{equation}
P_1 = e^{-a/g^2}\;\Big[\frac{b_1}{g^2}\Big]\; ,
\end{equation}
where $b_1/g^2$ is the sum of the \emph{1-particle cuts} (i.e. cuts
that cut exactly one propagator) through connected vacuum
diagrams. Its diagrammatic expansion starts with the following terms
\setbox2=\hbox to
8cm{\resizebox*{8cm}{!}{\includegraphics{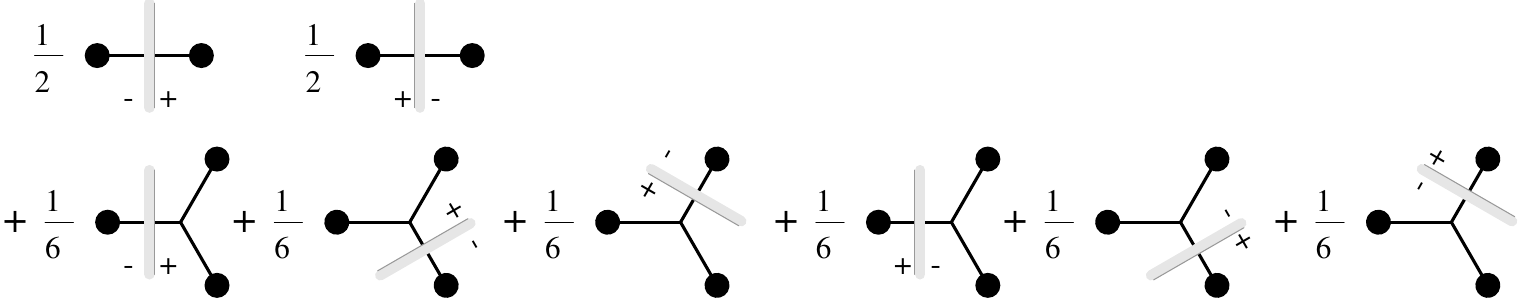}}}
\begin{eqnarray}
\frac{b_1}{g^2}&=&\;\raise -13mm\box2 \nonumber \\
&&+\cdots\; .
\label{eq:b1}
\end{eqnarray}
The probability $P_2$ of producing two particles is a bit more
complicated. It contains a term $(b_1/g^2)^2/2!$ corresponding to the
independent emission of two particles (the factor $1/2!$ is a symmetry
factor due to the fact that the two particles are indistinguishable)
and an additional term $b_2/g^2$ in which the two particles are correlated:
\begin{equation}
P_2= e^{-a/g^2} \; \left[\frac{1}{2!}\frac{b_1^2}{g^4} 
+ \frac{b_2}{g^2}\right]\; .
\end{equation}
The quantity $b_2/g^2$ can be obtained as the sum of the
\emph{2-particle cuts} of connected vacuum diagrams:
\setbox1=\hbox to
8cm{\resizebox*{8cm}{!}{\includegraphics{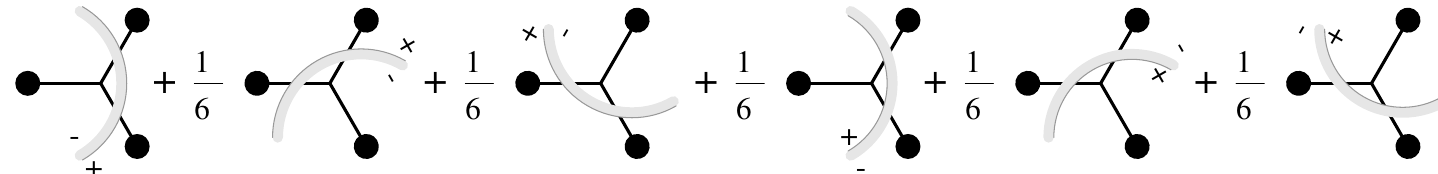}}}
\begin{eqnarray}
\frac{b_2}{g^2}&=&\;\raise -4.5mm\box1 \nonumber \\
&&+\cdots
\label{eq:b2}
\end{eqnarray}
Likewise, the probability $P_3$ of producing 3 particles reads
\begin{equation}
P_3 = e^{-a/g^2} \; \left[\frac{1}{3!}\frac{b_1^3}{g^6} 
+\frac{b_1b_2}{g^4}+ \frac{b_3}{g^2}\right]\; ,
\end{equation}
where the last term is the sum of the \emph{3-particle cuts} of connected
vacuum diagrams
\setbox1=\hbox to
5.3cm{\resizebox*{5.3cm}{!}{\includegraphics{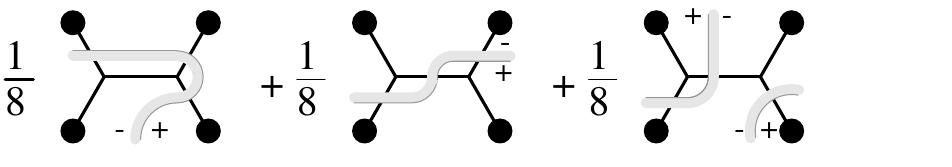}}}
\begin{eqnarray}
\frac{b_3}{g^2}&=&\;\raise -3.5mm\box1 \nonumber \\
&&+\cdots
\label{eq:b3}
\end{eqnarray}

The previous examples can be generalized easily into a formula for the
probability of producing $n$ particles,
\begin{equation}
P_n=e^{-a/g^2} \; \sum_{p=1}^n \frac{1}{p!}
\sum _{r_1+\cdots+r_p=n}
\frac{b_{r_1}\cdots b_{r_p}}{g^{2p}}\; .
\label{eq:Pn}
\end{equation}
In this formula, $p$ is the number of clusters (sets of correlated
particles) into which the $n$ particles can be divided. Note that,
except for the powers of $g^2$ that indicate the order of each term in
the strong source regime, this formula is completely generic and does
not depend on the details of the field theory under consideration: it
just expresses the combinatorics of grouping $n$ objects into $p$
clusters. Let us end this subsection by noting that unitarity requires
that $a=b_1+b_2+b_3+\cdots$ (this identity can be viewed as a
consequence of the optical theorem, since $a/g^2$ is $2\,{\rm
  Im}\,({\cal V})$ and $(b_1+b_2+b_3+...)/g^2$ is the sum of all the
cuts through ${\cal V}$).

\subsection{Generating functional}
So far, our discussion has been at a rather qualitative level. We now
turn to a more detailed discussion of what quantities can be
calculated and of what tools can do it. For bookkeeping purposes, it is
useful to introduce the following generating functional,
\begin{equation}
{\cal F}[z(\p)]\!\equiv\!\!
\sum_{n=0}^\infty\!
\frac{1}{n!}
\!\int\!\!
\frac{d^3\p_1}{(2\pi)^3 2E_{\p_1}}
\cdots 
\frac{d^3\p_n}{(2\pi)^3 2E_{\p_n}}\,
z(\p_1)\cdots z(\p_n)\,
\left|
\big<\p_1\cdots\p_n{}_{\rm out}\big|0_{\rm in}\big>
\right|^2\!\!,
\label{eq:F-def}
\end{equation}
where $z(\p)$ is a test function over the 1-particle phase-space. Any
observable which is expressible in terms of the transition amplitudes
(\ref{eq:trans-amp}) can be obtained from derivatives of
${\cal F}[z(\p)]$. In particular, the differential
probabilities and the single particle spectrum are obtained
as:
\begin{eqnarray}
  \frac{dP_n}{d^3\p_1\cdots d^3\p_n} 
&=& 
\frac{1}{n!}
\left.
\frac{\delta^n {\cal
      F}[\z(\p)]}{\delta z(\p_1)\cdots \delta z(\p_n)}\right|_{\z(\p)=0}\; ,
\nonumber\\
\frac{dN_1}{d^3\p}
&=&
\left.
\frac{\delta {\cal F}[z(\p)]}{\delta z(\p)}\right|_{z(\p)=1}\; .
\end{eqnarray}
As one can see on these two examples, observables in which the final
state is fully specified correspond to derivatives evaluated at
$z(\p)=0$ (which eliminates most of the final states from the sum in
eq.~(\ref{eq:F-def})), while inclusive observables --in which all the
final states are kept and most of their particles are
integrated out-- correspond to derivatives evaluated at
$z(\p)=1$. This is in fact a general property.  For instance, the
2-particle spectrum is
\begin{equation}
\frac{dN_2}{d^3\p d^3\q}
=
\left.
\frac{\delta^2 {\cal
      F}[\z(\p)]}{\delta z(\p)\delta z(\q)}
\right|_{z(\p)=1}\; ,
\end{equation}
and this formula has an obvious generalization to the case of the
inclusive $n$-particle spectrum.

As we shall see, inclusive observables are much simpler to calculate
than the exclusive ones. To a large extent, this simplification is due to
unitarity. The simplest consequence of unitarity is
\begin{equation}
{\cal F}[z(\p)=1]=1\; .
\label{eq:F-unitarity}
\end{equation}
At $z(\p)=0$, one would have instead obtained ${\cal F}[z(\p)=0]=P_0$,
which is a very complicated object. These considerations show that it
is much simpler to study the generating functional near the point
$z(\p)=1$ than near the point $z(\p)=0$. In terms of a diagrams, the
identity (\ref{eq:F-unitarity}) corresponds to an exact cancellation
among an infinite set of diagrams when one evaluates ${\cal F}[z(\p)]$
at $z(\p)=1$.

The reason why we discussed at length vacuum-vacuum diagrams in the
previous subsection is that they play an important role in organizing
the calculation of other quantities. The key observation here is that
the sum of the vacuum-vacuum diagrams is nothing but the generating
functional for time-ordered Green's functions. More precisely, one has
\begin{equation}
\big<0_{\rm out}\big|{\rm T}\,\phi(x_1)\cdots\phi(x_n)\big|0_{\rm in}\big>
=
\frac{\delta}{i\delta \eta(x_1)}\cdots\frac{\delta}{i\delta \eta(x_n)}
\;
\left.e^{i\,{\cal V}[j+\eta]}\right|_{\eta=0}\; ,
\label{eq:vac-as-gen-func}
\end{equation}
where the notation ${\cal V}[j+\eta]$ indicates that one should evaluate the
connected vacuum diagrams with a fictitious source $\eta$ added to the
physical source $j$. The fictitious source is set to zero after having
performed the functional differentiations.

Then, it is easy to obtain a formal but useful formula for ${\cal
  F}[z(\p)]$. Start from the
Leh\-mann--Symanzik--Zimmermann~\cite{Itzykson:1980rh} reduction formula for
the transition amplitude to a final state with $n$ particles,
\begin{eqnarray}
\big<\p_1\cdots\p_n{}_{\rm out}\big|0{}_{\rm in}\big>
&=&
i^n
\int d^4x_1 \cdots d^4x_n\;
e^{i(p_1\cdot x_1+\cdots +p_n\cdot x_n)}
\nonumber\\
&&\quad
\times
(\square_{x_1}+m^2)\cdots(\square_{x_n}+m^2)\;
\big<0_{\rm out}\big|{\rm T}\,\phi(x_1)\cdots\phi(x_n)\big|0_{\rm in}\big>\; .
\end{eqnarray}
By plugging eq.~(\ref{eq:vac-as-gen-func}) in this reduction formula
and squaring the result, we can write the squared amplitude as
\begin{equation}
\left|\big<\p_1\cdots\p_n{}_{\rm out}\big|0{}_{\rm in}\big>\right|^2
=
\left.
{\cal C}_{\p_1}\cdots {\cal C}_{\p_n}\; e^{i{\cal V}[j+\eta_+]}\,e^{-i{\cal V}^*[j+\eta_-]}
\right|_{\eta_\pm=0}\; ,
\label{eq:trans-from-vac}
\end{equation}
where the operator ${\cal C}_\p$ is defined by
\begin{equation}
{\cal C}_\p
\equiv
\int d^4x d^4y\; e^{ip\cdot(x-y)}\;
(\square_x+m^2)(\square_y+m^2)\;
\frac{\delta^2}{\delta \eta_+(x)\delta \eta_-(y)}\; .
\label{eq:CP-def-scalar}
\end{equation}
In the right hand side of eq.~(\ref{eq:trans-from-vac}), the factors
$\exp(i{\cal V})$ and $\exp(-i{\cal V}^*)$ come respectively from the
amplitude and its complex conjugate. Note that it is essential to keep
their arguments distinct--hence the separate $\eta_+$ and $\eta_-$--so
that the two derivatives in the operators ${\cal C}_\p$ act on
different factors. One should set $\eta_\pm$ to zero only after all
the derivatives have been evaluated. The final step is to substitute
eq.~(\ref{eq:trans-from-vac}) into the definition (\ref{eq:F-def}) of
${\cal F}[z(\p)]$. One obtains immediately
\begin{equation}
{\cal F}[z(\p)]
=\left.
\exp\left[\int \frac{d^3\p}{(2\pi)^3 2E_\p}\;z(\p)\,{\cal C}_\p\right]
\;
e^{i{\cal V}[j+\eta_+]}\,e^{-i{\cal V}^*[j+\eta_-]}
\right|_{\eta_\pm=0}\; .
\label{eq:F-from-vac}
\end{equation}
From this formula, one can show that the generating functional ${\cal
  F}[z(\p)]$ is the sum of all the vacuum diagrams in a \emph{modified
Schwinger-Keldysh formalism} in which all the off-diagonal ($G_{+-}^0(p)$
and $G_{-+}^0(p)$) propagators are multiplied by $z(\p)$.

\begin{figure}[htbp]
  \begin{center}
    \resizebox*{6cm}{!}{\includegraphics{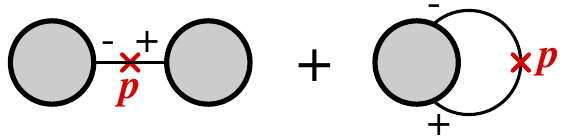}}
  \end{center}
  \caption{\label{fig:derF}Schematic diagrammatic representation of
    the two contributions to the derivative of the generating
    functions.  Left: term in $\varphi_+\varphi_-$. Right: term in
    ${\cal G}_{+-}$.}
\end{figure}

There is no simple expression for the generating functional ${\cal
  F}[z(\p)]$ itself, but it turns out that it is much easier to obtain a
formula for its first derivative $\delta{\cal F}[z(\p)]/\delta z(\p)$.
Using eq.~(\ref{eq:F-from-vac}) and the explicit form of the operator
${\cal C}_\p$, we can write this derivative as
\begin{eqnarray}
\frac{\delta\ln{\cal F}[z(\p)]}{\delta z(\p)}
&=&
\smash{\frac{1}{(2\pi)^3 2E_\p}\int} d^4x d^4y\; e^{ip\cdot(x-y)}\;
(\square_x+m^2)(\square_y+m^2)
\nonumber\\
&&\qquad\qquad\qquad\qquad\qquad\times\,\Big[
  \varphi_+(x)\varphi_-(y)
+
{\cal G}_{+-}(x,y)
\Big]
\; .
\label{eq:dF-general}
\end{eqnarray}
where $\varphi_\pm(x)$ and ${\cal G}_{+-}(x,y)$ are the connected
1-point and 2-point Green's functions in this modified
Schwinger-Keldysh formalism, respectively (they implicitly depend on
the external source $j$ and on the test function $z(\p)$), as
illustrated in the figure \ref{fig:derF}.

The order of magnitude of these objects is easily obtained from our
general results for the power counting of connected graphs~:
\begin{eqnarray}
\varphi_\pm&\sim&{\cal O}(g^{-1})\; ,
\nonumber\\
{\cal G}_{+-}&\sim&{\cal O}(1)\; .
\end{eqnarray}
Therefore, the first derivative of $\ln{\cal F}[z(\p)]$ starts at the
order $g^{-2}$. Moreover, at this order, only the first term in
$\varphi_+(x)\varphi_-(y)$ contributes. The term in ${\cal G}_{+-}$
starts contributing only at the next-to-leading order. A further
simplification at leading order is that it is sufficient to keep tree
level contributions to the 1-point functions $\varphi_\pm$. Thanks to
this tree structure, the functions $\varphi_\pm$ at Leading Order are
solutions of the classical equation of motion,
\begin{equation}
  (\square_x+m^2)\varphi_\pm(x)+U^\prime(\varphi_\pm(x))=j(x)\; ,
  \label{eq:classEOM}
\end{equation}
where $U'$ is the first derivative of the interaction potential
(e.g. $U'(\varphi)=g^2 \varphi^3/6$ in the scalar model considered in
this section).  This equation depends on the external source $j(x)$,
but not on the test function $z(\p)$. The latter comes only via the
boundary conditions
\begin{eqnarray}
f_+^{(+)}(-\infty,\p)&=&f_-^{(-)}(-\infty,\p)=0\;,
\nonumber\\
f_+^{(-)}(+\infty,\p)&=&z(\p)\;f_-^{(-)}(+\infty,\p)\; ,
\nonumber\\
f_-^{(+)}(+\infty,\p)&=&z(\p)\;f_+^{(+)}(+\infty,\p)\; .
\label{eq:BC-Fourier-1}
\end{eqnarray}
In these equations, the boundary conditions have been written in terms
of the coefficients of the Fourier decomposition of the fields
$\varphi_\pm$,
\begin{equation}
\varphi_\epsilon(y)
  \equiv
  \int\frac{d^3\p}{(2\pi)^3 2E_\p}
  \;\Big[
  f_\epsilon^{(+)}(y^0,\p)\,e^{-ip\cdot y}
  +
  f_\epsilon^{(-)}(y^0,\p)\,e^{+ip\cdot y}
  \Big]\; .
\label{eq:field-Fourier}
\end{equation}
(The Fourier coefficients are time dependent because $\varphi_\pm$ are
not free fields.)

By plugging the Fourier representation of $\varphi_\pm$ in the general
formula~(\ref{eq:dF-general}) (and setting ${\cal G}_{+-}=0$ at
this order), we get a very simple formula for the first derivative of
$\ln{\cal F}[z(\p)]$ at leading order:
\begin{equation}
\left.\frac{\delta\ln{\cal F}[z(\p)]}{\delta z(\p)}\right|_{_{\rm LO}}
=
\frac{1}{(2\pi)^3 2E_\p}
\;
f_+^{(+)}(+\infty,\p)\,
f_-^{(-)}(+\infty,\p)\; .
\label{eq:dF-LO}
\end{equation}
Note that it is in general extremely difficult to solve the non-linear
partial differential equation (\ref{eq:classEOM}) with boundary
conditions imposed both at $x^0=-\infty$ and at $x^0=+\infty$, as in
eqs.~(\ref{eq:BC-Fourier-1}).  Therefore, one should not hope to be
able to find solutions of this problem (either analytically or
numerically). As we shall see later, the only exception is when the
fields under consideration do not have self-interactions but are
solely driven by the external source. This is for instance the case
for fermion production under the influence of an external
electromagnetic field. Nevertheless, this result for the first
derivative of the generating functional ${\cal F}[z(\p)]$ is very
useful as an intermediate tool for deriving other results, as will be
shown in the rest of this section.

\subsection{Inclusive quantities at leading order (tree level)}
\subsubsection{Single particle spectrum}
Let us now show how to obtain inclusive moments (for now, at leading
order) from eq.~(\ref{eq:dF-LO}). The simplest one is the single
inclusive spectrum.  At leading order, it is simply obtained by
evaluating eq.~(\ref{eq:dF-LO}) at the special point $z(\p)=1$, since
${\cal F}[z(\p)=1]=1$. This means that one must solve the classical
equation of motion with boundary conditions (\ref{eq:BC-Fourier-1}) in
which one sets $z(\p)=1$. Setting $z(\p)=1$ in these boundary
conditions simplifies them considerably: the two fields $\varphi_+$
and $\varphi_-$ are identical,
\begin{equation}
\mbox{if\ }z(\p)\equiv 1\;,\qquad \varphi_+(x)=\varphi_-(x)\equiv\varphi(x)\; ,
\end{equation}
and obey the simple retarded boundary condition
\begin{equation}
\lim_{x^0\to -\infty}\varphi(x^0,\x)=0\; ,\qquad
\lim_{x^0\to -\infty}\partial^0\varphi(x^0,\x)=0\; .
\end{equation}
Thus, the prescription for computing the single inclusive spectrum at
leading order is the following:
\begin{itemize}
\item[{\bf i.}] Solve the classical field equation of motion with a null
  initial condition in the remote past,
\item[{\bf ii.}] At $x^0\to+\infty$, compute the
  coefficients~\footnote{Since the fields $\varphi_+$ and $\varphi_-$
    are equal, there is no need to keep a subscript $\pm$ for these
    coefficients.}  $f^{(\pm)}(+\infty,\p)$ of the Fourier
  decomposition of this classical field,
\item[{\bf iii.}] The single inclusive spectrum is then obtained as:
  \begin{equation}
    \left.\frac{dN_1}{d^3\p}\right|_{_{\rm LO}}=\frac{1}{(2\pi)^3 2E_\p}
    \;
    \left|f^{(+)}(+\infty,\p)\right|^2\; .
    \label{eq:N1-LO}
  \end{equation}
\end{itemize}
In eq.~(\ref{eq:N1-LO}), we have used the fact that the retarded classical
field $\varphi$ is purely real~\footnote{Its initial condition is
  real, and its equation of motion involves only real
  quantities.}. Since in the step {\bf i} the boundary conditions are
retarded, this problem is straightforward to solve, at least
numerically. A few important comments are in order here:
\begin{itemize}
\item If the fields are not self-interacting, then the single particle
  spectrum is simply the square of the Fourier coefficients of the
  source itself.
\item If the source has only space-like Fourier modes (this happens if
  there is frame in which it is time independent), then this is also
  the case for the solution of the classical equation of motion
  (\ref{eq:classEOM}), and the single particle spectrum is zero at
  leading order.
  \item This leading order contribution can only exist if the field
    under consideration is directly coupled to the external
    source. For instance, the direct production of electrons from an
    electromagnetic current is impossible at this order, but can
    happen at next-to-leading order (the current couples to an
    electron-positron pair, but this requires an extra coupling
    constant).
\end{itemize}

This is the form that gluon production takes in the Color Glass
Condensate framework \cite{Iancu:2000hn,Ferreiro:2001qy,Iancu:2002xk,Iancu:2003xm,Weigert:2005us,Gelis:2007kn,Gelis:2010nm,Gelis:2012ri} when applied to heavy ion
collisions. In this effective description of high energy
nucleus-nucleus collisions, the color gauge fields are coupled to two
external currents that represent the color charges carried by the fast
partons of the incoming nuclei. At high energy, it is expected that
the gluon occupation number in nuclei may reach non-perturbative
values of order $1/g^2$, which would correspond to strong sources in
the sense used in this section. The spectrum of produced gluons at
leading order in this description is obtained from the retarded
classical solutions of the Yang-Mills equations,
\begin{equation}
\big[D_\mu,F^{\mu\nu}\big]=J_1^\nu+J_2^\nu\; ,
\end{equation}
where $J_1^\nu$ and $J_2^\nu$ are the color currents of the two
projectiles. This approach has been implemented in a number of works
\cite{Kovner:1995ja,Kovchegov:1997ke,Krasnitz:1998ns,Krasnitz:1999wc,Krasnitz:2001qu,Lappi:2003bi,Lappi:2006fp,Gelis:2009wh}, and is now
included in the IP-glasma model for the matter produced immediately
after such a collision \cite{Schenke:2012wb}.

\subsubsection{Multi-particle spectra}
\label{sec:LO-mult}
The $n$-particle inclusive spectrum\footnote{Note that the
  $n$-particle spectrum defined in this way gives the expectation
  value of $N(N-1)\cdots (N-n+1)$ when integrated over the momenta
  $\p_1$ to $\p_n$~:
\begin{equation*}
\int d^3\p_1\cdots d^3\p_n\; \frac{dN_n}{d^3\p_1\cdots d^3\p_n} =
\sum_{N=n}^\infty N(N-1)\cdots(N-n+1)\;P_{_N}\; ,
\end{equation*}
where $P_{_N}$ is the total probability of producing exactly $N$
particles.} is also obtained from derivatives
of the generating functional ${\cal F}[z(p)]$ evaluated at $z(\p)=1$,
\begin{equation}
\frac{dN_n}{d^3\p_1\cdots d^3\p_n}
=
\left.
\frac{\delta^n {\cal F}[z(\p)]}{\delta z(\p_1)\cdots \delta z(\p_n)}
\right|_{z(\p)=1}\; ,
\label{eq:Nn-def-F}
\end{equation}
that can equivalently be written as~:
\begin{equation}
\frac{dN_n}{d^3\p_1\cdots d^3\p_n}
=
\prod_{i=1}^n
\left[\frac{\delta\ln{\cal F}}{\delta z(p_i)}\right]_{z(\p)=1}
+
\sum_{i<j}
\left[\frac{\delta^2\ln{\cal F}}{\delta z(p_i)\delta z(\p_j)}\right]_{z(\p)=1}
\prod_{k\not=i,j}
\left[\frac{\delta\ln{\cal F}}{\delta z(p_k)}\right]_{z(\p)=1}
+\cdots
\label{eq:Nn-general}
\end{equation}
The terms we have not written explicitly contain increasingly high
order derivatives (but less and less factors), up to a single factor
with an $n$-th derivative. However, these terms are not
needed. Indeed, we already know that at leading order $\ln{\cal F}$ is
of order $g^{-2}$ since it is a sum of connected vacuum-vacuum
diagrams. Therefore, in the right hand side of this equation, the
first term is of order $g^{-2n}$, the second term is of order
$g^{-2(n-1)}$, etc... The leading contribution is thus the first term,
and all the subsequent terms are subleading~\footnote{The second term
  will play a role in the next-to-leading order corrections.}.  We see
that, at leading order, the $n$-particle inclusive spectrum is simply
the product of $n$ single particle spectra:
\begin{equation}
\left.\frac{dN_n}{d^3\p_1\cdots d^3\p_n}\right|_{_{\rm LO}}
=
\prod_{i=1}^n
\left.\frac{dN_1}{d^3\p_i}\right|_{_{\rm LO}}\; .
\label{eq:dNn-LO}
\end{equation}
Any deviation from this factorized result is a subleading effect. Note
also that at leading order, there is no difference between the
factorial moments $\big<N(N-1)\cdots (N-n+1)\big>$ and the ordinary
moments $\big<N^n\big>$. Moreover, at this order, the multiplicity
distribution cannot be distinguished from a Poisson distribution.

In the Color Glass Condensate framework at leading order, the
correlations among the produced particles can either originate from
correlations that pre-exist in the distribution of the color sources
that produce the gauge field \cite{Gelis:2008ad,Gelis:2008sz}, or be
built up at a later stage of the evolution of the system through
collective motion of the produced particles (e.g. radial
hydrodynamical flow \cite{Voloshin:2003ud,Shuryak:2007fu}). In heavy
ion collisions, strong correlations have been observed between pairs
of hadrons
\cite{Adare:2008ae,Alver:2008aa,Abelev:2009af,Alver:2009id},
characterized by a ridge shape, very elongated in the relative
rapidity of the two particles, and peaked in their relative azimuthal
angle. By causality, the correlations in rapidity have to be created
in the very early stages of the collisions \cite{Dumitru:2008wn}, and
they can be simply understood as a consequence of the near boost
invariance of the sources of the incoming nuclei. In contrast, the
azimuthal correlations can be produced at any time, and are easily
explainable by the hydrodynamical flow that develops in the later
stages of the collision process \cite{Alver:2010gr,Alver:2010dn}.

\subsection{Exclusive quantities at leading order}
\label{subsec:excl}
Let us now consider exclusive quantities. This discussion will be very
short, and its purpose is only to illustrate the fact that the
calculation of exclusive quantities is considerably more difficult
than that of inclusive quantities.  Let us consider as an example the
calculation of the differential probability for producing exactly one
particle. It may be obtained from ${\cal F}[z(\p)]$ by the formula
\begin{equation}
\frac{dP_1}{d^3\p}
=
\left.
\frac{\delta {\cal F}[z(\p)]}{\delta z(\p)}
\right|_{z(\p)=0}
=
\underbrace{
{\cal F}[z(\p)=0]}_{\displaystyle{P_0}}
\;
\left.
\frac{\delta \ln{\cal F}[z(\p)]}{\delta z(\p)}
\right|_{z(\p)=0}\; .
\end{equation}
There are two major differences compared to the inclusive spectra
studied in the previous section~:
\begin{itemize}
\item[{\bf i.}] The derivative of $\ln{\cal F}[z(\p)]$ must be
  evaluated at the point $z(\p)=0$. At leading order, it can still be
  expressed in terms of the Fourier coefficients of a pair of
  solutions of the classical equation of motion, via
  eq.~(\ref{eq:dF-LO}). However, because we must now set $z(\p)=0$ in
  the boundary conditions (\ref{eq:BC-Fourier-1}) for these classical
  fields, they are not retarded fields anymore\footnote{It is
    precisely because in exclusive observables the final state is
    constrained that the boundary conditions for the fields cannot be
    purely retarded.}, and there is no practical way to calculate
  them.
\item[{\bf ii.}] The quantity ${\cal F}[z(\p)=0]$ appears as a
  prefactor in front of all the exclusive quantities. This prefactor
  is nothing but the probability $P_0$ for not producing anything,
  i.e. the vacuum {\sl  survival probability}. Calculating $P_0$
  directly is a very difficult task. However, if one were able to
  calculate the second factor for all the probabilities $P_1, P_2,
  \cdots$, one could then obtain $P_0$ from the unitarity condition
  $\sum_{n=0}^\infty P_n=1$.
\end{itemize}
These difficulties, observed here on the example of $dP_1/d^3\p$, are
in fact generic for all exclusive quantities.  Note however that this
is to a large extent an academic problem, since exclusive quantities
--where one specifies in minute detail the final state-- are not very
interesting for the phenomenology of processes in which the final
state has typically a very large number of particles, parametrically
of order $g^{-2}$. Indeed, in this context, the probability of
occurrence of a given fully specified final state is exponentially
suppressed, like $e^{-c/g^2}$.

\subsection{Particle production at next-to-Leading order (one loop)}
\label{sec:NLO}
In situations where the particles under consideration couple directly
to a time dependent external source, these particles can be produced
by the leading order mechanism described in the previous subsections,
and the moments of the particle distribution are expressible in terms
of the classical field generated by this source. However, when this
direct coupling does not exist or when the external source is static,
the production of particles is impossible at this order and can at
best happen at the next-to-leading order in the coupling. This is in
particular the case in the traditional setup of the Schwinger
mechanism, where an electromagnetic current couples indirectly to
fermion pairs via an electromagnetic field. Therefore, our goal in
this subsection is not to calculate the subleading corrections to
particle spectra in general, but to calculate the first nonzero
contributions they receive when the leading order contribution
vanishes.

In this subsection, we discuss the new production mechanisms that arise
at NLO, by considering the single particle spectrum at 1-loop.  This
spectrum is given by the first derivative of the generating functional
${\cal F}[z(\p)]$, for which a general formula was given in
eq.~(\ref{eq:dF-general}). At NLO, i.e. at the order $g^0$, it
involves two quantities \cite{Gelis:2008rw}:
\begin{itemize}
\item[{\bf i.}] The 1-loop corrections $\beta_\pm$ to the
  1-point functions $\varphi_\pm$,
\item[{\bf ii.}] The 2-point function ${\cal G}_{+-}$ at tree level.
\end{itemize}
However, since we are interested in the NLO corrections only in cases
where the leading order is zero, we do not need to evaluate the terms
that contain $\beta_\pm$. Indeed, $\beta_\pm$ being a 1-loop
correction to $\varphi_\pm$, it also leads to a vanishing contribution
for static sources or if the particles of interest cannot couple
directly to the external source. Therefore, only the second term is
needed. This term corresponds to the production of a pair of particles
from the external source (in the case of the single particle spectrum,
one of the two produced particles is integrated out). Moreover, since
we are not going to calculate further derivatives with respect to
$z(\p)$, it is sufficient to evaluate this quantity at the point
$z(\p)=1$ -- which simplifies considerably the calculation.  This NLO
correction to the single inclusive spectrum reads
\begin{eqnarray}
\left.\frac{dN_1}{d^3\p}\right|_{_{\rm NLO}}
=
\frac{1}{(2\pi)^3 2E_\p}
\int d^4x d^4y\;e^{ip\cdot(x-y)}\;
(\square_x+m^2)(\square_y+m^2)
\;
  {\cal G}_{+-}(x,y)
\; .\label{eq:N1-NLO-1}
\end{eqnarray}
The 2-point function ${\cal G}_{+-}(x,y)$ at tree level obeys the
following integral equation~:
\begin{equation}
{\cal G}_{\epsilon\epsilon^\prime}(x,y)
=
G_{\epsilon\epsilon^\prime}^0(x,y)
-i
\sum_{\eta=\pm}\eta\int d^4z\;
G_{\epsilon\eta}^0(x,z)\,
U^{\prime\prime}(\varphi(z))\,
{\cal G}_{\eta\epsilon^\prime}(z,y)\; ,
\label{eq:lipmann}
\end{equation}
and therefore mixes with the other three components, ${\cal G}_{-+}$,
${\cal G}_{++}$ and ${\cal G}_{--}$.  Here,
$-iU^{\prime\prime}(\varphi(z))$ is the general form for the insertion
of a background field on a propagator in a theory with a potential
$U(\varphi)$. From these equations\footnote{The simplest method is to
  write the equations of motion obeyed by ${\cal G}_{+-}$ and its
  boundary conditions, and then to check a posteriori that
  eq.~(\ref{eq:+--resummed}) satisfies both. Alternatively, it is
  possible to perform explicitly the summation implied by
  eq.~(\ref{eq:lipmann}), which avoids having to guess the structure
  of the answer.}, one can obtain the following formulas~:
\begin{eqnarray}
{\cal G}_{+-}(x,y)
&=&
\int\frac{d^3\p}{(2\pi)^3 2E_\p}\;
a_\p^*(x)a_\p(y)\; ,
\nonumber\\
{\cal G}_{-+}(x,y)
&=&
\int\frac{d^3\p}{(2\pi)^3 2E_\p}\;
a_\p(x)a_\p^*(y)\; ,
\label{eq:+--resummed}
\end{eqnarray}
where the functions $a_\p(x)$ are \emph{mode functions} defined by
\begin{eqnarray}
\big[\square_x+m^2+U^{\prime\prime}(\varphi(x))\big]\,a_\p(x)=0\; ,\quad
\lim_{x^0\to-\infty}a_\p(x)=e^{ip\cdot x}\;. 
\label{eq:small-fluct-retarded}
\end{eqnarray}
Thus, the problem of finding the Schwinger-Keldysh propagators in a
background field can be reduced to determining how plane waves propagate
on top of the classical background field (and are distorted by this
field). In the previous formulas, we have used the plane wave basis,
but other choices are possible, as long as one chooses a properly
normalized complete basis. Note also that the 4-dimensional space-time
integrals in eq.~(\ref{eq:N1-NLO-1}) can in fact be rewritten as
purely spatial integrals on a constant-$t$ surface (with $t\to
+\infty$),
\begin{eqnarray}
\left.\frac{dN_1}{d^3\p}\right|_{_{\rm NLO}}
=
\frac{1}{(2\pi)^3 2E_\p}\int\frac{d^3\k}{(2\pi)^3 2E_\k}
\;
\lim_{t\to+\infty}\left|\int d^3\x\;e^{-i\p\cdot \x}
\;(\partial_t-iE_\p)\;
  a_\k(x)\right|^2
\; .\label{eq:N1-NLO-2}
\end{eqnarray}

Before going into more specialized subjects, let us summarize this
section by diagrammatic representations  of the LO and NLO
contributions to the single particle spectrum,
\setbox2\hbox to 4cm{\resizebox*{4cm}{!}{\includegraphics{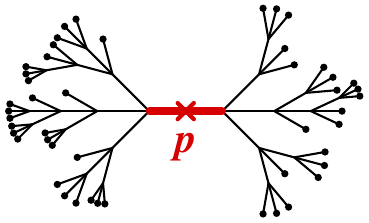}}}
\setbox3\hbox to 4cm{\resizebox*{4cm}{!}{\includegraphics{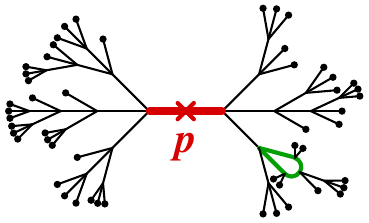}}}
\setbox4\hbox to 3.8cm{\resizebox*{3.8cm}{!}{\includegraphics{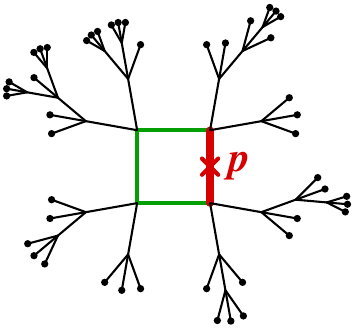}}}
\setbox5\hbox to 3.8cm{\resizebox*{3.8cm}{!}{\includegraphics{NLOb}}}
\begin{eqnarray}&&
  \frac{dN_1}{d^3\p}
  \quad=\quad
  \underbrace{\vphantom{\raise -18mm\box4}\raise -12mm\box2}_{\mbox{LO}}
  \nonumber\\&&
  \qquad\qquad+\quad
  \underbrace{
  \raise -12mm\box3
  \quad+\quad
  \raise -18mm\box5
  +\cdots
  }_{\mbox{NLO}}
  \label{eq:LO+NLO}
\end{eqnarray}
In this representation, the line represented in bold is the on-shell
propagator of momentum $\p$ corresponding to the final particle one is
measuring. In this illustration, we show only one typical graph
representative of each class. Therefore, each of the trees that appear
in these graphs should be understood as the infinite series that
comprises all such trees, and an arbitrary number of such trees can be
attached around the loop for the NLO graphs.

Let us also mention a few issues that arise in some special cases:
\begin{itemize}
\item[{\bf i.~}]{\bf Static sources~:} The LO contribution is zero unless
  the external source is time dependent, so that the classical field
  it generates has nonzero time-like Fourier components. Indeed, this
  contribution is just the square of the Fourier transform of the
  classical field generated by the external source. Likewise, the
  first of the two NLO contributions contributes only when the
  external source is time dependent.

  The second NLO contribution is also zero for a static source if one
  inserts a \emph{finite number} of trees around the loop. This means
  that this 1-loop graph does not contain any contribution analytic in
  the external field. However, it also contains a non-perturbative
  contribution that exists even for a static source. This
  non-perturbative contribution is non-analytic in the coupling
  constant. More precisely, all its Taylor coefficients at zero
  coupling vanish, which is why it cannot be seen with any finite
  order in the external field. This 1-loop diagram contains the
  contribution that one usually calls the \emph{Schwinger mechanism}.

\item[{\bf ii.~}]{\bf Time dependent sources~:} In situations where
  the external source is time dependent (especially when it is slowly
  varying compared to the natural frequency scale set by the mass of
  the particles), there may be a competition between the perturbative
  (analytic) contributions and the non-perturbative (non-analytic)
  ones, and one should avoid considering them in isolation if one
  wishes to describe the transition between static and time dependent
  fields. The general formulation that we have adopted in this
  presentation is well suited to this case. Indeed, since the external
  field is treated to all orders, it naturally ``packages'' the
  perturbative and non-perturbative contributions in the same
  formulas.

\item[{\bf iii.}~]{\bf Particles not directly coupled to the external
  source~:} The structure illustrated pictorially in
  eq.~(\ref{eq:LO+NLO}) is completely generic, despite the fact that
  we have used the example of a scalar field theory in this
  section. Of course, certain graph topologies may be impossible in
  certain theories. For instance, in QCD, if the external source is a
  color current and if we are interested in the quark spectrum, then
  the first two topologies of eq.~(\ref{eq:LO+NLO}) cannot exist
  because a single quark field does not couple directly to a color
  field. For the quark spectrum, the second 1-loop graph therefore
  constitute the lowest order contribution
  \cite{Gelis:2004jp,Gelis:2005pb,Gelis:2015eua}. In contrast, if we
  are interested in the gluon spectrum, then all the three topologies
  contribute.

\item[{\bf iv.}~]{\bf Gauge theories~:} Even if there is no
  (chromo)electromagnetic field at asymptotic times, it may happen
  that the gauge potential is a nonzero pure gauge at $t\to\pm\infty$.
  In the presence of such a pure gauge background, one should not use
  the vacuum mode functions (i.e. plane waves) in the Fourier
  decomposition of the fields in order to calculate the particle
  spectrum. Instead, one should first apply to the mode functions the
  gauge rotation that transforms the null gauge potentials into the
  pure gauge of interest.
\end{itemize}

\section{Correlations in the Schwinger mechanism}
\label{sec:corr}
\subsection{Generating functional at one loop}
In the section \ref{sec:LO-mult}, we have observed that when particles
can be produced directly at leading order from an external source,
their multiplicity distribution is a Poisson distribution at this
order. The interpretation of this absence of correlations is
that, when particles are produced at leading order by a strong (time
dependent) source, they come from 1-point functions and the graphs
that produce two or more particles can be factorized into subgraphs in
which a single particle is produced. In this case, deviations from a
Poisson distribution only arise in subleading corrections. For
instance, the second NLO graph of eq.~(\ref{eq:LO+NLO}) contains the
production of particle \emph{pairs}, which introduces correlations
among the final state particles.

In order to discuss this effect in more detail, let us consider a
simple scalar QED model in which we disregard the self interactions of
the charged scalar fields \cite{Fukushima:2009er}. Its Lagrangian
reads
\begin{equation}
{\cal L}\equiv (D_\mu \phi)(D^\mu\phi)^*-m^2\phi\phi^*\; ,
\end{equation}
where $D_\mu\equiv\partial_\mu-igA_\mu$ is the covariant
derivative. In this section, the electromagnetic potential $A_\mu$ is
assumed to be non-dynamical. In other words, it is a purely classical
field imposed by some external action, and we disregard the feedback
(such as screening effects) of the produced charged particles on the
electromagnetic field. In this model, the lowest order for the
production of charged scalars is at 1-loop, because there is no direct
coupling involving $A_\mu$'s and a single field $\phi$.

Since these particles are charged, it is interesting to keep track
separately of particles and antiparticles. For this, we generalize
the generating functional introduced in eq.(\ref{eq:F-def}) into
\begin{eqnarray}
  {\cal F}[z,\overline{z}]
  &\equiv&
\sum_{m,n=0}^\infty
\frac{1}{m!n!}
\int
\prod_{i=1}^m
\frac{d^3\p_i}{(2\pi)^3 2E_{\p_i}}\;z(\p_i)
\prod_{j=1}^n
\frac{d^3\q_j}{(2\pi)^3 2E_{\q_j}}\;\overline{z}(\q_j)
\nonumber\\
&&\qquad\qquad\qquad\qquad\qquad\times\,
\Big|
\big<
\underbrace{\p_1\cdots\p_m}_{\mbox{\scriptsize particles}}
\underbrace{\q_1\cdots\q_n}_{\mbox{\scriptsize antiparticles}}
           {}_{\rm out}\big|0_{\rm in}\big>
\Big|^2\!\!,
\label{eq:F-charged-def}
\end{eqnarray}
where $z(\p)$ and $\overline{z}(\q)$ are two independent
functions. Unitarity trivially implies ${\cal F}[1,1]=1$. By
differentiating with respect to $z$ or $\overline{z}$, we obtain the
single particle and antiparticle spectra,
\begin{equation}
  \frac{dN_1^+}{d^3\p}=
  \left.\frac{\delta{\cal F}[z,\overline{z}]}{\delta z(\p)}\right|_{z=\overline{z}=1}\;,
  \qquad
  \frac{dN_1^-}{d^3\q}=
  \left.\frac{\delta{\cal F}[z,\overline{z}]}{\delta \overline{z}(\q)}\right|_{z=\overline{z}=1}\; ,
\end{equation}
and second derivatives give the two particles and two antiparticles spectra:
\begin{equation}
  \frac{dN_1^{++}}{d^3\p_1d^3\p_2}=
  \left.\frac{\delta^2{\cal F}[z,\overline{z}]}{\delta z(\p_1)\delta z(\p_2))}\right|_{z=\overline{z}=1}\;,
  \qquad
  \frac{dN_1^{--}}{d^3\q_1d^3\q_2}=
  \left.\frac{\delta^2{\cal F}[z,\overline{z}]}{\delta \overline{z}(\q_1)\delta \overline{z}(\q_2))}\right|_{z=\overline{z}=1}\;,
\end{equation}
as well as a mixed spectrum
\begin{equation}
  \frac{dN_1^{+-}}{d^3\p d^3\q}=
  \left.\frac{\delta^2{\cal F}[z,\overline{z}]}{\delta z(\p)\delta \overline{z}(\q))}\right|_{z=\overline{z}=1}\;.
\end{equation}

At the lowest non-zero order (i.e. one loop), it is possible to obtain
a compact expression of this generating functional. This will provide
complete information about the production of charged
particles\footnote{This does not contradict what was said in the
  section \ref{subsec:excl}. In this subsection, the difficulty was
  due to the fact that we were considering a strong source that
  couples directly to the fields we want to produce, and that these
  fields had self-interactions. Here, we are considering a much
  simpler problem since the fields we want to produce have no
  self-interactions.} by an external field at this order, and make a
contact with other approaches to this problem. From the general result
that the generating functional is the sum of the vacuum diagrams in
the Schwinger-Keldysh formalism, and the fact that this sum is the
exponential of the subset of the connected vacuum diagrams, we can
first write:
\setbox2\hbox to 8cm{\resizebox*{8cm}{!}{\includegraphics{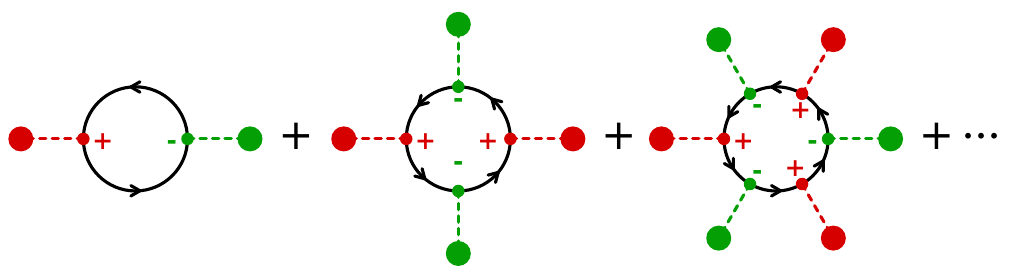}}}
\begin{equation}
  \ln{\cal F}[z,\overline{z}]=\mbox{constant}+\;\raise -10mm\box2\; ,
  \label{eq:logFdiag}
\end{equation}
where the unwritten constant is independent of $z$ and $\overline{z}$
(its value should be adjusted to satisfy unitarity, i.e. $\ln{\cal
  F}[1,1]=0$).  In eq.~(\ref{eq:logFdiag}), the objects attached to
the loop already resum infinite sequences of $+$ or $-$ vertices
respectively,
\setbox2\hbox to 7cm{\resizebox*{7cm}{!}{\includegraphics{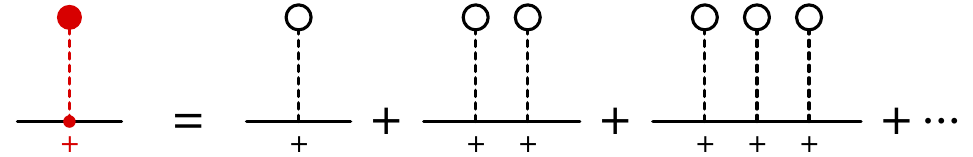}}}
\setbox3\hbox to 7cm{\resizebox*{7cm}{!}{\includegraphics{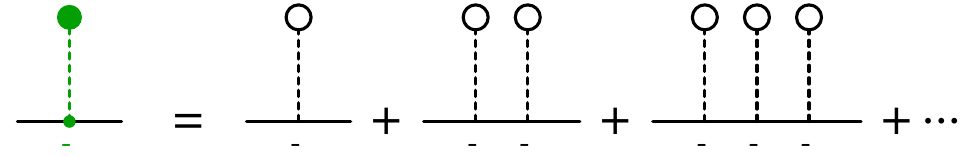}}}
\begin{eqnarray}
  {\cal T}_+&\equiv&\raise -2mm\box2\nonumber\\
  {\cal T}_-&\equiv&\raise -2mm\box3
  \label{eq:T+T-}
\end{eqnarray}
Therefore, these objects do not contain any $G_{+-}^0$ or $G_{-+}^0$
propagators, and are therefore independent of $z$ or
$\overline{z}$. The $z$ and $\overline{z}$ dependence is carried by
the propagators that appear explicitly in the loops in
eq.~(\ref{eq:logFdiag}). In the diagrammatic representation used in
eqs.~(\ref{eq:T+T-}), the dotted lines are a shorthand for the sum of
the two possible interactions that exist in scalar QED:
\setbox2\hbox to 5cm{\resizebox*{5cm}{!}{\includegraphics{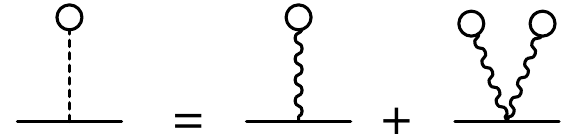}}}
\begin{equation}
\raise 0mm\box2\; .
\end{equation}
(The wavy lines terminated by a circle denote the external
electromagnetic potential.)

The sum in eq.~(\ref{eq:logFdiag}) can be written in the following
compact form,
\begin{eqnarray}
  \ln{\cal F}[z,\overline{z}]
  &=&
  \mbox{constant}+\sum_{n=1}^\infty\frac{1}{n}\,{\rm Tr}\,\left({\cal T}_+\,(zG_{+-}^0)\,{\cal T}_-\,(\overline{z}G_{-+}^0)\right)^n\nonumber\\
  &=&
  \mbox{constant}-{\rm Tr}\,\ln\Big(1-{\cal T}_+\,(zG_{+-}^0)\,{\cal T}_-\,(\overline{z}G_{-+}^0)\Big)\; ,
\end{eqnarray}
where we have made explicit that the factors $z$ and $\overline{z}$
come along with the off-diagonal propagators $G_{+-}^0$ and $G_{-+}^0$
(the functions $z$ and $\overline{z}$ carry the same momentum as the
propagator they are attached to).  The trace denotes an integration
over all the spacetime coordinates of the vertices around the
loop. The factor $1/n$ is a symmetry factor, absorbed in the second
line in the Taylor expansion of a logarithm. By using the relationship
between time-ordered and retarded propagators, as well as unitarity,
the argument of the logarithm can be rearranged and expressed in terms
of a \emph{retarded scattering amplitude} ${\cal T}_{_R}$:
\begin{equation}
  \ln{\cal F}[z,\overline{z}]
  =\mbox{constant}
  -{\rm Tr}\,\ln\Big(1+{\cal M}[z,\overline{z}]\Big)\; ,
  \label{eq:logF-ret}
\end{equation}
where ${\cal M}[z,\overline{z}]$ is a compact notation for
\begin{eqnarray}
  \Big[{\cal M}[z,\overline{z}]\Big]_{\p,\q}
  &\equiv&
  \int\frac{d^4k}{(2\pi)^4}\;
  (1-z(\k)\overline{z}(-\q))\;
           {\cal T}_{_R}(p,k)\,G_{+-}^0(k)\,{\cal T}_{_R}^\dagger(k,q)\,G_{-+}^0(q)\nonumber\\
           &=&
           2\pi\theta(q^0)\delta(q^2-m^2)
  \int\frac{d^3\k}{(2\pi)^3 2E_\k}\;
  (1-z(-\k)\overline{z}(-\q))\;
           {\cal T}_{_R}(p,-k)\,{\cal T}_{_R}^*(q,-k)\; .
           \label{eq:block-logF}
\end{eqnarray}
In these formulas, ${\cal T}_{_R}(p,k)$ is the retarded scalar
propagator in the external field, amputated of its external lines,
with an incoming momentum $k$ and outgoing momentum $p$ (in the second
line, we have changed $k\to -k$, so that $k^0>0$.).

The equations (\ref{eq:logF-ret}) and (\ref{eq:block-logF}) provide
the generating functional, from which one can in principle extract all
the information about the distribution of produced particles at one
loop. Moreover, the retarded scattering matrix it contains can be
computed by solving the wave equation in the background field under
consideration. More precisely, it can be expressed in terms of mode
functions as follows:
\begin{eqnarray}
  &&{\cal T}_{_R}(p,-k)=\lim_{x_0\to+\infty}\int d^3\x\;e^{ip\cdot x}\;
  (\partial_{x_0}-iE_\p)\;a_\k(x)
  \nonumber\\
  &&(D^2+m^2)\,a_\k(x)=0\;,\qquad \lim_{x_0\to-\infty}a_\k(x)=e^{ik\cdot x}\; .
  \label{eq:mode-eq}
\end{eqnarray}
In other words, in order to obtain ${\cal T}_{_R}(p,-k)$, one should
start in the remote past with a negative energy plane wave of momentum
$\k$, evolve over the background field until late times, and project
it on a positive energy plane wave of momentum $\p$.

\subsection{General structure of the 1-loop particle correlations}
Let us denote by $z$ one of the $z(\p)$'s or one of the
$\overline{z}(\q)$'s. From eq.~(\ref{eq:logF-ret}), we obtain
\begin{equation}
  \frac{\delta {\cal F}}{\delta z}=-{\cal F}\;{\rm Tr}\,\Big((1+{\cal M})^{-1}
  \frac{\delta{\cal M}}{\delta z}\Big)\; .
  \label{eq:F-1der}
\end{equation}
At $z(\p)=\overline{z}(\q)\equiv 1$, we have ${\cal M}=0$ and ${\cal
  F}=1$, so that this derivative simplifies into
\begin{equation}
  \left.\frac{\delta {\cal F}}{\delta z}\right|_{z=\overline{z}=1}
  =-{\rm Tr}\,\Big(
  \left.\frac{\delta{\cal M}}{\delta z}\right|_{z=\overline{z}=1}\Big)\; .
  \label{eq:mom1}
\end{equation}
This is the general structure of the single (anti)particle
spectrum. From eq.~(\ref{eq:F-1der}), we can take one more derivative
to obtain
\begin{eqnarray}
  \left.\frac{\delta^2 {\cal F}}{\delta z_1\delta z_2}\right|_{z=\overline{z}=1}
  =
  \left.\frac{\delta {\cal F}}{\delta z_1}\;
  \frac{\delta {\cal F}}{\delta z_2}\right|_{z=\overline{z}=1}
  -
  {\rm Tr}\,\Big[
  \frac{\delta^2{\cal M}}{\delta z_1\delta z_2}
  -
  \frac{\delta {\cal M}}{\delta z_1}\;
  \frac{\delta {\cal M}}{\delta z_2}
  \Big]_{z=\overline{z}=1}\; .
  \label{eq:mom2}
\end{eqnarray}
From this equation, we see quite generally that the two (anti)particle
spectra contain the product of the two corresponding single particle
spectra, plus a single trace term that contains the non-trivial
correlations. In particular, this extra term encodes possible
deviations from a Poisson distribution.

One can further differentiate with respect to $z$ or $\overline{z}$ in
order to obtain expressions for higher moments of the particle
distribution. Starting with the moment of order 3, an additional
simplification arises due to the fact that ${\cal M}$ is proportional
to $z\overline{z}$. These successive differentiations lead to
expressions in terms of traces of products of first or second
derivatives of ${\cal M}$.  Particles that appear in the same trace
are correlated, while they are not correlated if their momenta appear
in two different traces. The $n$-th moment contains a term with a
single trace, that correlates all the $n$ particles. This term provides
the genuine $n$-particle correlation, since it is not reducible into
smaller clusters.

This formulation can be turned into an algorithm for calculating the
single particle spectrum, the 2-particle spectrum, etc.. in the
presence of an external field. Indeed, after having discretized space
(and, consequently, momentum space), ${\cal M}$, $\delta{\cal
  M}/\delta z$, $\delta{\cal M}/\delta \overline{z}$ and
$\delta^2{\cal M}/\delta z \delta\overline{z}$ can be viewed as (very
large) matrices, and the evaluation of the right side of
eqs.~(\ref{eq:mom1}) and (\ref{eq:mom2}) is just a matter of linear
algebra. In order to compute these building blocks, one would have to
first obtain the retarded scattering amplitude ${\cal T}_{_R}$, which
can be done by solving the partial differential equation
(\ref{eq:mode-eq}) for each mode $\k$. Although this method can in
principle provide all the (or more realistically the first few)
moments of the distribution of produced particles, it faces a serious
computational difficulty in the general case where the background
field is a completely generic function of space-time, with no particular
symmetry: computing ${\cal T}_{_R}$ would require computations that
scale as the square of the number of lattice spacings (one power comes
from the size of the spatial domain in eq.~(\ref{eq:mode-eq}), and
another power from the number of modes $\k$). In the subsection
\ref{subsec:statistical}, we describe a method of statistical sampling
that considerably reduces this cost, at the expense of a reduced
accuracy due to statistical errors.

\subsection{Correlations in the case of a homogeneous field}
The above procedure could in principle be implemented on a lattice,
and would provide the answer for a general background field. In the
case of a spatially homogeneous background field, momentum
conservation allows us to simplify considerably the structure of the
generating functional and to completely uncover the $n$ particle
moments. Firstly, the scattering matrix becomes diagonal in momentum:
\begin{equation}
T_{_R}(p,-k)=-2iE_\p\;(2\pi)^3\delta(\p+\k)\;\beta_\p\; ,
\end{equation}
and all the information about the background field is contained in the
coefficients $\beta_\p$. This leads to
\begin{equation}
  \Big[{\cal M}[z,\overline{z}]\Big]_{\p,\q}
  =2\pi\theta(q^0)\delta(q^2-m^2)\,2E_\p\;(2\pi)^3\delta(\p-\q)\;(1-z(\p)\overline{z}(-\q))\;\big|\beta_\p\big|^2\; .
\end{equation}
Since this object is diagonal in momentum, the same momentum $\p$ runs
in the loop in the trace of eq.~(\ref{eq:logF-ret}). This implies that
only correlations of particles with the same
momentum\footnote{Obviously, this result is true only for an
  homogeneous external field.} are possible, or correlations between
particles of momentum $\p$ and antiparticles of momentum $-\p$
(because of the respective arguments of the functions $z$ and
$\overline{z}$). Using this in eq.~(\ref{eq:logF-ret}), we obtain
\begin{equation}
  \ln{\cal F}[z,\overline{z}]
  =\mbox{constant}-V\int\frac{d^3\k}{(2\pi)^3}\;\ln\Big[1-(z(\k)\overline{z}(-\k)-1)\;f_\k\Big]\; ,
\end{equation}
where we denote $f_\k\equiv |\beta_\k|^2$. The prefactor $V$ is the
overall volume of the system, that results from a $(2\pi)^3\delta(0)$
in momentum space.

By differentiating this formula with respect to $z$ and/or
$\overline{z}$, we obtain the following results for the 1 and 2
particle spectra\footnote{Our definition of the two particle spectrum
  $dN_2^{++}/d^3\p_1 d^3\p_2$ correspond to pairs of \emph{distinct
  particles}. Therefore, its integral over $\p_1$ and $\p_2$ leads to
  the expectation value $\left<{\bs N}^+({\bs N}^+-1)\right>$. This
  explains why the right hand side of the second equation contains a
  factor $f_{\p_1}$ instead of $1+f_{\p_1}$.}:
\begin{eqnarray}
  &&
  \frac{dN_1^+}{d^3\p}=n_\p\equiv \frac{V}{(2\pi)^3}\,f_\p
  \nonumber\\
  &&
  \frac{dN_2^{++}}{d^3\p_1 d^3\p_2}
  -\frac{dN_1^+}{d^3\p_1}\frac{dN_1^+}{d^3\p_2}
  =\delta(\p_1-\p_2)\;n_{\p_1}\,f_{\p_1}
  \nonumber\\
  &&
  \frac{dN_2^{+-}}{d^3\p d^3\q}
  -\frac{dN_1^+}{d^3\p}\frac{dN_1^-}{d^3\q}
  =\delta(\p+\q)\;n_{p}\,(1+f_{\p})
  \; .
\end{eqnarray}
In the limit where $f_\p\ll 1$, the right hand side of the two
particle correlation would simplify into a form consistent with a
Poisson distribution. In contrast, when the occupation number $f_\p$
is not small, deviations from a Poisson distribution arise due to
Bose-Einstein correlations.

Since there is no correlation except for particles with the same
momentum or antiparticles with opposite momentum, one can also derive
the probability distribution $P_\k(m,n)$ to produce $m$ particles of
momentum $\k$ and $n$ antiparticles of momentum $-\k$,
\begin{equation}
P_\k(m,n)=\delta_{m,n}\;\frac{1}{1+f_\k}\;\left(\frac{f_\k}{1+f_\k}\right)^n\; ,
\end{equation}
i.e. a Bose-Einstein distribution. The main difference between such a
distribution and a Poisson distribution is the existence of \emph{large
  multiplicity tails}, that are due to stimulated emission.

From these semi-explicit formulas, we can also clarify the difference
between the vacuum survival probability and the exponential of the
particle multiplicity. By evaluating the generating functional at
$z=\overline{z}=0$, one obtains the following expression for the
vacuum-to-vacuum transition probability (i.e. the vacuum survival
probability):
\begin{equation}
  P_0=\exp\Big\{ -V\int\frac{d^3\k}{(2\pi)^3}\;\ln(1+f_\k)\Big\}\; ,
  \label{eq:P0}
\end{equation}
while the total particle multiplicity is
\begin{equation}
\left<{\bs N}^+\right>=V\int  \frac{d^3\k}{(2\pi)^3}\; f_\k\; .
\end{equation}
Only when the occupation number is small in all modes, we can expand
the logarithm in eq.~(\ref{eq:P0}) and obtain
$P_0\approx\exp(-\big<{\bs N}^+\big>)$, but this relationship is not
exact in very strong fields. This is a limitation of the methods that
give only $P_0$ (for instance by providing a way to calculate the
imaginary part of the effective action in a background field): in
general, the knowledge of $P_0$ is not sufficient to obtain the
momentum dependence of the spectrum of produced particles
(proportional to $f_\k$).

\subsection{Constant electrical field in scalar QED}
\label{sec:constE}
For further reference, let us quote a useful
approximation\footnote{The exact formula has a more complicated $k_z$
  dependence, but this formula captures the main features of the
  result if the time $x_0$ is large, and in the strong field limit
  $eE\gg m^2+k_\perp^2$.} for the occupation number and vacuum
survival probability in the case of a constant and spatially
homogeneous electrical background field $E$ in the $z$ direction. The
occupation number $f_\k$ is independent of the position $\x$ and
reads:
\begin{equation}
f_\k(x^0)\approx\theta(k_z)\;\theta(eEx^0-k_z)\;\exp\left(-\frac{\pi (m^2+k_\perp^2)}{eE}\right)\; .
\end{equation}
The $k_z$ dependence is a consequence of the acceleration of the
particles by the electrical field after they have been produced (we
assume that the electrical charge is positive, $e>0$). Therefore, we
have also
\begin{equation}
  \ln(1+f_\k)
  \approx
  \theta(k_z)\;\theta(eEx^0-k_z)
  \sum_{n=1}^\infty \frac{(-1)^{n-1}}{n}\;\exp\left(-\frac{n\pi (m^2+k_\perp^2)}{eE}\right)\; ,
\end{equation}
and after insertion in eq.~(\ref{eq:P0}) we obtain
\begin{equation}
  P_0
  \approx
  \exp\left(-\frac{Vx^0}{8\pi^3}(eE)^2\sum_{n=1}^\infty \frac{(-1)^{n-1}}{n^2}\;e^{-n\pi m^2/(eE)}\right)\; .
  \label{eq:P0-constE}
\end{equation}
Note that if we had neglected the Bose-Einstein correlations, i.e. the
higher order terms in $f_\k$ in the expansion of $\ln(1+f_\k)$, we
would have obtained instead
\begin{equation}
  P_0\empile{\approx}\over{f_\k\ll 1}
  \exp\left(-\frac{Vx^0}{8\pi^3}(eE)^2\;e^{-\pi m^2/(eE)}\right)\; .
  \label{eq:P0-nocorr}
\end{equation}
We should close this section with some words about practical
applications of these formulas. In all experimental situations where
such formulas are used, one is still very far from the critical
electrical field $E_c= m^2/e$ that would make pair production an
event that has a probability of order 1. Consequently, the above
discussion about Bose-Einstein correlations and the related issue of
reconstructing the particle spectrum from the sole knowledge of the
vacuum persistence probability $P_0$ are mostly academic. The formula
(\ref{eq:P0-nocorr}) is a very good approximation in these realistic
situations, and from $P_0$ one can read directly the number of
produced pairs.

\section{Equivalent formulations of the Schwinger mechanism}
\label{sec:other}
In the previous sections, we have exposed a general formulation of the
particle production in the presence of strong external sources/fields
based on a resummed perturbation theory approach.  There is an
equivalent derivation based on the canonical quantization of the field
operators, in which particle production is described via Bogoliubov
transformations~\cite{Narozhnyi:1970uv,Gitman:1977ne,Soffel:1982pm,Ambjorn:1982bp,%
Gavrilov:1996pz,Gavrilov:2006jb,Tanji:2008ku}. 
This section is devoted to a presentation of
this alternative approach, as well as a few other related methods that
are often used in the literatures.  In this section, we use spinor QED
in an external classical gauge field in order to illustrate these
approaches and their relationships.

\subsection{Bogoliubov transformations}
\label{subsec:Bogoliubov}
\subsubsection{General case}
In the method of canonical quantization of the field operators, one
can describe the particle production in classical background fields by
a Bogoliubov transformation.  A fermion field operator $\hat{\psi}
(x)$ coupled to a classical gauge field $A_\mu(x)$ obeys the
following Dirac equation:
\begin{equation}
\left[ i\gamma^\mu D_\mu -m \right] \hat{\psi} (x) = 0 \; ,
\label{eq:Dirac_eq1}
\end{equation}
where
\begin{equation}
D_\mu = \partial_\mu +ieA_\mu \; .  
\end{equation}
We assume that there is no electromagnetic field at asymptotic times
$t\to \pm \infty$, and take the null initial condition for the gauge
field\footnote{This is always possible with a gauge transformation.}:
\begin{equation} \label{eq:gauge_ini}
\lim_{t\to -\infty} A_\mu (x) = 0\; . 
\end{equation}
Furthermore, we impose the temporal gauge condition
\begin{equation}
A^0 = 0 \; ,
\end{equation}
which largely simplifies the description of the time-evolution of the
system.  After this gauge fixing, there is still a residual invariance
under all the spatially dependent gauge transformations.  Because the
Dirac equation is linear in the field operator, its solution can be
expanded in normal modes:
\begin{equation} \label{eq:in-expansion}
\hat{\psi} (x) = \sum_{s=\uparrow ,\downarrow} \int \frac{d^3\p}{(2\pi)^3}\;
\left[ \psi_{\p ,s}^{\text{in} +} (x)\, a_{\p ,s}^\text{in}
  +\psi_{\p ,s}^{\text{in} -} (x)\, b_{\p ,s}^{\text{in} \dagger} \right]  \; ,
\end{equation}
where $a_{\p ,s}^\text{in}$ and $b_{\p ,s}^\text{in}$ are annihilation
operators for a particle and an antiparticle of momentum $\p$ and spin
$s$, respectively. The $\psi_{\p ,s}^{\text{in} \pm} (x)$ are c-number
solutions of the Dirac equation, analogous to the mode functions
introduced in the previous sections.  The superscript $+$ and $-$
distinguish the positive energy and negative energy modes, while the
superscript `in' specifies the initial condition that the mode
functions satisfy are free ones at $t\to -\infty$~:
\begin{equation} \label{eq:mode_in}
\begin{split}
\lim_{t\to -\infty} \psi_{\p ,s}^{\text{in} +} (x) &= u(\p ,s)\; e^{-ip\cdot x} \; , \\
\lim_{t\to -\infty} \psi_{\p ,s}^{\text{in} -} (x) &= v(\p ,s)\; e^{+ip\cdot x} \; . 
\end{split}
\end{equation}
The momentum space free spinors are normalized by
\begin{equation}
\overline{u} (\p ,s) \gamma^\mu u(\p ,s^\prime )
= \overline{v} (\p ,s) \gamma^\mu v(\p ,s^\prime )
= 2p^\mu \delta_{s,s^\prime} \; . 
\end{equation}
With the inner product defined by
\begin{equation}
\left( \psi |\chi \right) \equiv \int d^3 \x\; \psi^\dagger (t,\x ) \chi (t,\x ) \; ,  
\end{equation}
the mode functions should be normalized as follows
\begin{gather}
\left( \psi_{\p ,s}^{\text{in} +} \big| \psi_{\p^\prime ,s^\prime }^{\text{in} +} \right)
 = \left( \psi_{\p ,s}^{\text{in} -} \big| \psi_{\p^\prime ,s^\prime }^{\text{in} -} \right)
 = 2E_\p\; (2\pi )^3 \delta (\p -\p^\prime )\; \delta_{s,s^\prime} \; , \\
\left( \psi_{\p ,s}^{\text{in} +} \big| \psi_{\p^\prime ,s^\prime }^{\text{in} -} \right)
 = \left( \psi_{\p ,s}^{\text{in} -} \big| \psi_{\p^\prime ,s^\prime }^{\text{in} +} \right)
 = 0 \, . 
\end{gather}
One can easily confirm that the initial condition \eqref{eq:mode_in}
satisfies these conditions, and the inner product is
conserved\footnote{It is essential that the external gauge potential
  be real for this property to be true.} by unitary time evolution.
The previous orthonormality conditions are therefore satisfied at
arbitrary times for any real background gauge field.  By using the
orthonormality conditions, we can extract the creation and annihilation
operators from the field operator as follows:
\begin{equation}
a_{\p ,s}^\text{in} = \frac{1}{2E_\p} \bigl( \psi_{\p ,s}^{\text{in} +} \big| \hat{\psi} \bigr) \; ,\qquad
b_{\p ,s}^{\text{in} \dagger} = \frac{1}{2E_\p} \bigl( \psi_{\p ,s}^{\text{in} -} \big| \hat{\psi} \bigr) \; . 
\end{equation}
From the canonical anti-commutation relation for the field operators
\begin{equation}
\left\{ \hat\psi (t,\x ) , \hat\psi^\dagger (t,\y ) \right\} = \delta (\x -\y) \; ,
\end{equation}
the following anti-commutation relation for the creation and annihilation
operators can be derived:
\begin{equation} \label{eq:CCR_in}
\left\{ a_{\p ,s}^\text{in} , a_{\p^\prime ,s^\prime }^{\text{in} \dagger} \right\} 
 = \left\{ b_{\p ,s}^\text{in} , b_{\p^\prime ,s^\prime }^{\text{in} \dagger} \right\} 
 = \frac{1}{2E_\p} (2\pi )^3\; \delta (\p -\p^\prime )\; \delta_{s,s^\prime} \; . 
\end{equation}

Instead of the \emph{in}-solutions, one could have considered the
\emph{out}-solutions, that satisfy the free boundary condition at
$t\to +\infty$.  Although we have assumed that the electromagnetic
field is vanishing at asymptotic times, the gauge field $A^i (x)$ can
generally be nonzero at $t\to +\infty$. This non-zero asymptotic gauge
field must be a pure gauge since the field strength is zero.
Therefore, the out-solutions must approach free spinors that are
gauge rotated:
\begin{equation} \label{eq:mode_out}
\begin{split}
\lim_{t\to +\infty} \psi_{\p ,s}^{\text{out} +} (x) &= U^\dagger (x)\; u(\p ,s)\; e^{-ip\cdot x} \; , \\
\lim_{t\to +\infty} \psi_{\p ,s}^{\text{out} -} (x) &= U^\dagger (x)\; v(\p ,s)\; e^{+ip\cdot x} \; , 
\end{split}
\end{equation}
where $U(\x)$ is a gauge factor defined as
\begin{equation}
U(t, \x ) \equiv \exp \left[ ie\int^\x d\z \cd \A (t, \z )\right] \; .
\end{equation}
The integration path should be contained in the constant-$t$ plane,
and its starting point can be chosen arbitrary (this residual
arbitrariness amounts to multiplying the spinors by a constant
phase). Note that when the background field is a pure gauge, the gauge
link $U(t,\x)$ depends only on the endpoints of the line integral, but
not on the shape of this path. The field operator can also be expanded 
in terms of the out-mode functions,
\begin{equation} \label{eq:out-expansion}
\hat{\psi} (x) = \sum_{s=\uparrow ,\downarrow} \int \frac{d^3\p}{(2\pi)^3}\;
\left[ \psi_{\p ,s}^{\text{out} +} (x)\, a_{\p ,s}^\text{out}
  +\psi_{\p ,s}^{\text{out} -} (x)\, b_{\p ,s}^{\text{out} \dagger} \right]  \; .
\end{equation}
Since the out-mode functions satisfy the same orthonormal condition as
the in-mode solutions, the creation and annihilation operators for the
out-particles obey the same anti-commutation relations as those for
the in-particles \eqref{eq:CCR_in}.

At this point, we have two different definitions of a ``particle''; one
is based on the in-basis and the other on the out-basis.  If a
nontrivial background gauge field exists at some point of the
evolution of the system, these two definitions are in general
different.  This difference is nothing but the consequence of the
particle production from the vacuum under the influence of the
background field.  If we assume that the initial state is the vacuum,
the spectrum of particles observed at the asymptotic time $t\to +\infty$ is
represented by the in-vacuum expectation value of the out-particle
number operator:
\begin{equation} \label{eq:spec_out0}
\frac{dN_s}{d^3 \p} = \frac{2E_\p}{(2\pi)^3} \;
\langle 0_\text{in} | a_{\p ,s}^{\text{out} \dagger} a_{\p ,s}^\text{out} |0_\text{in} \rangle \; . 
\end{equation}
The in-vacuum $|0_\text{in}\rangle$ is defined by $a_{\p ,s}^\text{in}
|0_\text{in} \rangle = b_{\p ,s}^\text{in} |0_\text{in} \rangle = 0$.
In order to calculate this spectrum, we need to find the relationship
between the creation and annihilation operators of the in-basis and of
the out-basis.  By substituting the expansion \eqref{eq:in-expansion}
into
\begin{equation}
a_{\p ,s}^\text{out} = \frac{1}{2E_\p}\; \bigl( \psi_{\p ,s}^{\text{out} +} \big| \hat{\psi} \bigr) \; , \qquad
b_{\p ,s}^{\text{out} \dagger} = \frac{1}{2E_\p}\; \bigl( \psi_{\p ,s}^{\text{out} -} \big| \hat{\psi} \bigr) \; , 
\end{equation}
we obtain the following relationship between the two bases,
\begin{align}
a_{\p ,s}^\text{out} &= \frac{1}{2E_\p} 
\sum_{s^\prime} \int \frac{d^3\p' }{(2\pi)^3}\;
\left[ \bigl( \psi_{\p ,s}^{\text{out} +} \big| \psi_{\p^\prime ,s^\prime}^{\text{in} +} \bigr)\, a_{\p^\prime ,s^\prime}^\text{in} 
+\bigl( \psi_{\p ,s}^{\text{out} +} \big| \psi_{\p^\prime ,s^\prime}^{\text{in} -} \bigr)\, b_{\p^\prime ,s^\prime}^{\text{in} \dagger} \right] \; , 
\label{eq:Bogo0a} \\
b_{\p ,s}^{\text{out} \dagger} &= \frac{1}{2E_\p} 
\sum_{s^\prime} \int \frac{d^3\p' }{(2\pi)^3}\;
\left[ \bigl( \psi_{\p ,s}^{\text{out} -} \big| \psi_{\p^\prime ,s^\prime}^{\text{in} +} \bigr)\, a_{\p^\prime ,s^\prime}^\text{in} 
+\bigl( \psi_{\p ,s}^{\text{out} -} \big| \psi_{\p^\prime ,s^\prime}^{\text{in} -} \bigr)\, b_{\p^\prime ,s^\prime}^{\text{in} \dagger} \right] \; . 
\label{eq:Bogo0b}
\end{align}
We can modify the spinors by a constant phase (which is irrelevant to
physical observables) so that the following relations are fulfilled:
\begin{gather}
\bigl( \psi_{-\p ,s}^{\text{out} -} \big| \psi_{-\p^\prime ,s^\prime}^{\text{in} -} \bigr) 
 = \bigl( \psi_{\p ,s}^{\text{out} +} \big| \psi_{\p^\prime ,s^\prime}^{\text{in} +} \bigr)^* \; , \\
\bigl( \psi_{-\p ,s}^{\text{out} -} \big| \psi_{\p^\prime ,s^\prime}^{\text{in} +} \bigr) 
 = -\bigl( \psi_{\p ,s}^{\text{out} +} \big| \psi_{-\p^\prime ,s^\prime}^{\text{in} -} \bigr)^* \; .
\end{gather}
Then, eqs.~\eqref{eq:Bogo0a} and \eqref{eq:Bogo0b} take the form of a
Bogoliubov transformation,
\begin{align}
\sqrt{2E_\p}\; a_{\p ,s}^\text{out} &= 
\sum_{s^\prime} \int \frac{d^3\p^\prime }{(2\pi)^3}\; \sqrt{2E_{\p^\prime}}\;
\left[ \alpha(\p,s;\p^\prime ,s^\prime )\, a_{\p^\prime ,s^\prime}^\text{in} 
+\beta (\p,s;\p^\prime ,s^\prime )\, b_{-\p^\prime ,-s^\prime}^{\text{in} \dagger} \right] \; , 
\label{eq:Bogo1a} \\
\sqrt{2E_\p}\; b_{-\p ,s}^{\text{out} \dagger} &= 
\sum_{s^\prime} \int \frac{d^3\p^\prime }{(2\pi)^3}\; \sqrt{2E_{\p^\prime}}\;
\left[ \alpha^* (\p,s;\p^\prime ,s^\prime )\, b_{-\p^\prime ,-s^\prime}^{\text{in} \dagger} 
-\beta^* (\p,s;\p^\prime ,s^\prime )\, a_{\p^\prime ,s^\prime}^\text{in} \right] \; , 
\label{eq:Bogo1b}
\end{align}
where the coefficients $\alpha$ and $\beta$ are defined by
\begin{gather}
\alpha (\p,s;\p^\prime ,s^\prime ) \equiv \frac{1}{2\sqrt{E_\p E_{\p^\prime}}}
 \bigl( \psi_{\p ,s}^{\text{out} +} \big| \psi_{\p^\prime ,s^\prime}^{\text{in} +} \bigr) \; , \\
\beta (\p,s;\p^\prime ,s^\prime ) \equiv \frac{1}{2\sqrt{E_\p E_{\p^\prime}}}
 \bigl( \psi_{\p ,s}^{\text{out} +} \big| \psi_{-\p^\prime ,s^\prime}^{\text{in} -} \bigr) \; . 
\end{gather}
These equations ensure that the number of anti-particles having
momentum $-\p$ equals the number of particles having momentum
$+\p$:
\begin{equation}
\langle 0_\text{in} | a_\p^{\text{out} \dagger} a_\p^\text{out} |0_\text{in} \rangle
 = \langle 0_\text{in} | b_{-\p}^{\text{out} \dagger} b_{-\p}^\text{out} |0_\text{in} \rangle \; . 
\end{equation}

By substituting eq.~\eqref{eq:Bogo0a} into eq.~\eqref{eq:spec_out0}, we
can express the momentum spectrum of produced particles in terms of
the mode functions as follows:
\begin{equation}
\frac{dN_s}{d^3 \p} = \frac{1}{(2\pi)^3 2E_\p} 
\sum_{s^\prime} \int \frac{d^3\p^\prime }{(2\pi)^3 2E_{\p^\prime}}\; 
\left| \bigl( \psi_{\p ,s}^{\text{out} +} \big| \psi_{\p^\prime ,s^\prime}^{\text{in} -} \bigr) \right|^2 \; .
\end{equation}
If we evaluate the inner product in the right hand side at $t\to
+\infty$, using the out-mode functions given by
eq.~\eqref{eq:mode_out}, the spectrum reads
\begin{equation} \label{eq:in-out_spectrum1}
\frac{dN_s}{d^3 \p} = \frac{1}{(2\pi)^3 2E_\p} 
\sum_{s^\prime} \int \frac{d^3\p^\prime }{(2\pi)^3 2E_{\p^\prime}}\; 
\lim_{x^0 \to +\infty} 
\left| u^\dagger (\p ,s) \int d^3\x \; e^{-i\p \cdot \x}\; U(x) \;\psi_{\p^\prime ,s^\prime}^{\text{in} -} (x) \right|^2 \; .
\end{equation}
This equation is the QED analogue of the scalar formula given in
eq.~(\ref{eq:N1-NLO-2}).


Although the particle number can be defined unambiguously only in the
asymptotic region where the background electromagnetic field vanishes,
it is informative to define quasi-particles at intermediate times when
there is nonzero background field.  A time-dependent spectrum can be
heuristically defined simply by removing the limit of $x^0 \to
+\infty$ from eq.~\eqref{eq:in-out_spectrum1},
\begin{equation} \label{eq:timedep_spectrum1}
\frac{dN_s}{d^3 \p} = \frac{1}{(2\pi)^3 2E_\p} 
\sum_{s^\prime} \int \frac{d^3p^\prime }{(2\pi)^3 2E_{\p^\prime}} 
\left| u^\dagger (\p ,s) \int d^3x \, e^{-i\p \cdot \x} U(t,\x ) \psi_{\p^\prime ,s^\prime}^{\text{in} -} (t,\x ) \right|^2 \; .
\end{equation}
This generalization is equivalent to computing the expectation value
of a time-dependent particle number operator:
\begin{equation}
\frac{dN_s}{d^3 \p} = \frac{2E_\p}{(2\pi)^3} \;
\langle 0_\text{in} | a_{\p ,s}^\dagger (t) a_{\p ,s} (t) |0_\text{in} \rangle \; ,
\end{equation}
where an instantaneous quasi-particle definition is introduced by the
expansion
\begin{equation} \label{eq:t-expansion}
\hat{\psi} (x) = \sum_{s=\uparrow ,\downarrow} \int \frac{d^3\p}{(2\pi)^3}\;
\left[ \psi_{\p ,s}^{(t) +} (x)\, a_{\p ,s} (t)
  +\psi_{\p ,s}^{(t) -} (x)\, b_{\p ,s}^\dagger (t) \right] 
\end{equation}
with 
\begin{equation} \label{eq:mode_t}
\begin{split}
\psi_{\p ,s}^{(t) +} (x) &= U^\dagger (t,\x )\; u(\p ,s)\; e^{-ip\cdot
  x} \; , \\ \psi_{\p ,s}^{(t) -} (x) &= U^\dagger (t,\x )\; v(\p
,s)\; e^{+ip\cdot x} \; .
\end{split}
\end{equation}
This definition of a time-dependent spectrum naturally interpolates
between the zero particle state at $t\to -\infty$ and the final state
at $t\to +\infty$.  At intermediate times when the gauge field is not
a pure gauge, the gauge link $U(x)$ can depend on the path chosen to
define the line integral.  Therefore, one must keep in mind that the
spectrum evaluated in a region where the background is not a pure
gauge suffers from this unavoidable ambiguity of the particle definition.

\subsubsection{Uniform electrical field}
Let us now restrict ourselves to a spatially homogeneous electric
field which can be given by a gauge field that depends only on time:
\begin{equation} \label{eq:uni_gauge}
A^0 = 0 \; , \ A^i = A^i (t) \; . 
\end{equation}
Since the background gauge field has no spatial dependence, the
spatial dependence of the mode functions can be trivially factorized as
\begin{equation}
\psi_{\p ,s}^{\text{in} \pm} (t,\x ) = \widetilde{\psi}_{\p
  ,s}^{\text{in} \pm} (t)\; e^{\pm i\p \cdot \x} \; .
\end{equation}
The gauge factor $U(x)$ is path-independent and it simply reads
\begin{equation}
U(x) = e^{-ie \A (t) \cdot \x} \; .  
\end{equation}
The $\x$-integration in the inner products between the free mode
functions at the time $t$ and the in-mode functions can be performed
analytically, resulting in a delta function of momentum of the form
$\delta(\p+e\A-\p')$.  The Bogoliubov transformation between the
in-particles and the quasi-particles at time $t$ reads
\begin{align}
a_{\p ,s} (t) &= 
\sqrt{\frac{E_{\p +e\A}}{E_\p}} \sum_{s^\prime} 
\left[ \alpha(t;\p ,s,s^\prime )\, a_{\p +e\A ,s^\prime}^\text{in} 
+\beta (t;\p ,s,s^\prime )\, b_{-\p-e\A ,s^\prime}^{\text{in} \dagger} \right] \; , 
\label{eq:Bogo2a} \\
b_{-\p ,s}^\dagger (t) &= 
\sqrt{\frac{E_{\p +e\A}}{E_\p}} \sum_{s^\prime} 
\left[ \alpha^* (t; \p ,s,s^\prime )\, b_{-\p -e\A ,s^\prime}^{\text{in} \dagger} 
-\beta^* (t, \p ,s,s^\prime )\, a_{\p +e\A ,s^\prime}^\text{in} \right] \; , 
\label{eq:Bogo2b}
\end{align}
with the following time-dependent Bogoliubov coefficients
\begin{gather}
  \alpha (t ;\p ,s,s^\prime) \equiv
  \frac{e^{iE_\p t}}{2\sqrt{E_\p E_{\p +e\A}}} \;
 u^\dagger (\p ,s)\; \tilde{\psi}_{\p +e\A ,s^\prime}^{\text{in} +} (t) \; , \label{eq:t-dep_Bogo1a} \\
\beta (t ;\p ,s,s^\prime) = \frac{e^{iE_\p t}}{2\sqrt{E_\p E_{\p +e\A}}} \;
 u^\dagger (\p ,s)\; \tilde{\psi}_{-\p -e\A ,s^\prime}^{\text{in} -} (t) \; . \label{eq:t-dep_Bogo1b}  
\end{gather}
Note that the momentum label $\p$ of the mode functions
$\tilde{\psi}_{\pm\p ,s^\prime}^{\text{in} \pm} (t)$ is shifted by the
gauge field $e\A$ because of the insertion of the gauge factor.  This
shift amounts to changing from the canonical momentum to the kinetic
momentum, and it is crucial in order to describe properly the
acceleration of the produced particles by the electrical field.  

From the time independence of the anti-commutation relations, it follows
that the following bilinear combination of the Bogoliubov coefficients
is also constant:
\begin{gather} \label{eq:Bogo_norm}
\sum_\sigma \left[ \alpha (t ;\p ,s,\sigma) \alpha^* (t ;\p ,s^\prime ,\sigma)
 +\beta (t ;\p ,s,\sigma) \beta^* (t ;\p ,s^\prime ,\sigma) \right] = \delta_{s,s^\prime} \; . 
\end{gather}
In terms of the Bogoliubov coefficients, the momentum spectrum of
produced particles can be expressed as
\begin{equation} 
\frac{dN_s}{d^3 \p} = \frac{V}{(2\pi)^3}\; \sum_{s^\prime} \left| \beta (t; \p ,s,s^\prime) \right|^2 \; ,
\end{equation}
where $V=(2\pi)^3 \delta ({\boldsymbol 0})$ is the volume of the
system.  Thanks to eq.~\eqref{eq:Bogo_norm}, the occupation number
$f_\p$ is always smaller than unity
\begin{equation}
f_\p\equiv \frac{(2\pi)^3}{V}\; \frac{dN_s}{d^3 \p} = \sum_{s^\prime} \left| \beta (t; \p ,s,s^\prime) \right|^2 \leq 1 \; ,
\end{equation}
as required by the Pauli exclusion principle. 

\subsubsection{Sauter potential} \label{subsubsec:Sauter}
As an explicit example to illustrate the method of Bogoliubov
transformations, we derive the spectrum of particles produced by the
Sauter-type pulsed electrical field\footnote{ Originally, Sauter
  \cite{sauter1932kleinschen} studied a space-dependent electrical
  field $E (x) = E/\cosh^2 (x/a)$ in the context of the Klein
  paradox. Both the space-dependent and the time-dependent Sauter
  electrical fields are amenable to an explicit analytic
  solutions. The comparison of these two situations is in fact very
  interesting to understand the essential differences between temporal
  inhomogeneities (that increase the particle yield) and spatial
  inhomogeneities (that reduce the particle yield). We will further
  comment on this difference in the subsection \ref{subsec:instanton},
  when we discuss the worldline instanton approximation.}:
\begin{equation} \label{eq:Sauter_E}
E_z (t) = \frac{E}{\cosh^2 (t/\tau)} \; ,
\end{equation}
where $\tau$ stands for the pulse duration.
\begin{figure}[htbp]
  \begin{center}
    \resizebox*{7cm}{!}{\includegraphics{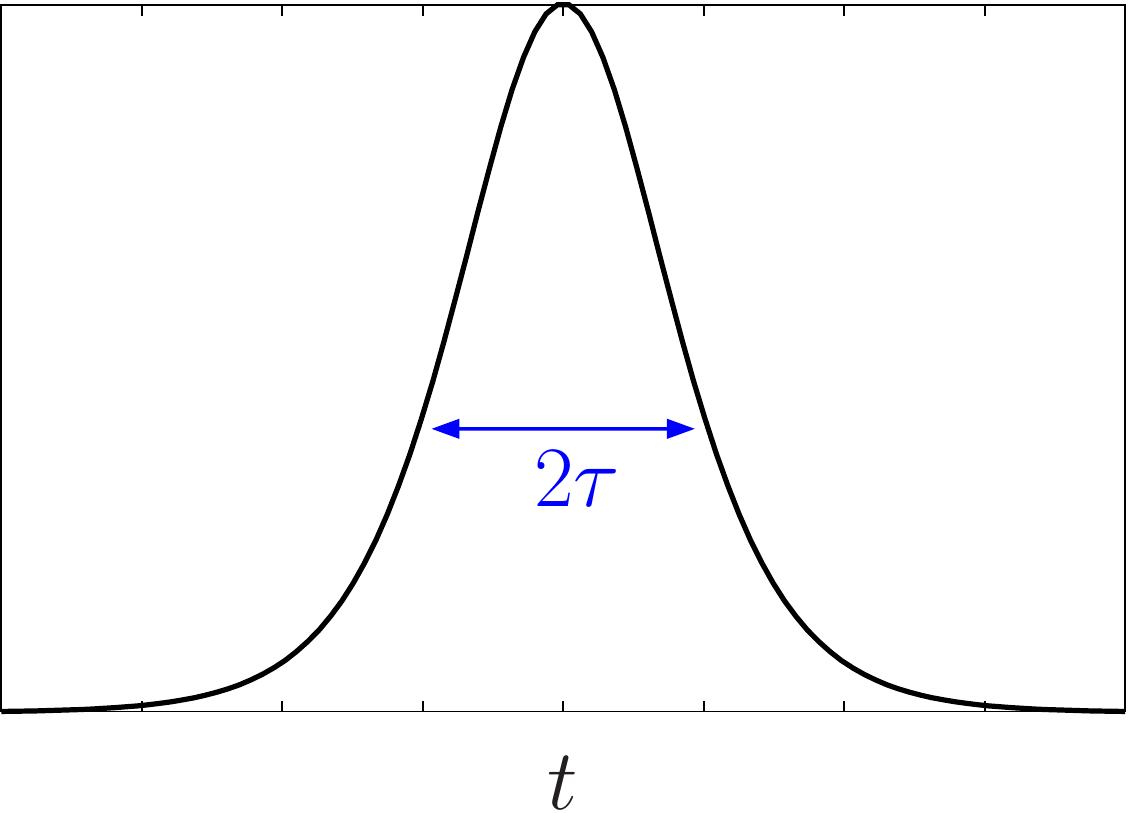}} 
  \end{center}
  \vspace{-5mm}
  \caption{\label{fig:Sauter_E}
  The temporal profile of the Sauter electrical field given in eq.~\eqref{eq:Sauter_E}. } 
\end{figure}
The temporal profile of this field is shown in the figure
\ref{fig:Sauter_E}.  Since the electrical field of
eq.~\eqref{eq:Sauter_E} vanishes exponentially at $t\to \pm \infty$, a
proper particle definition is available without any ambiguity at
asymptotic times.  Under the influence of this electrical field, the
Dirac equation is analytically solvable \cite{Narozhnyi:1970uv}, and
one can compute the spectrum in closed form:
\begin{equation} \label{eq:Sauter_spec}
\frac{(2\pi)^3}{V}\; \frac{dN}{d^3 \p} 
 =  \frac{\sinh \left[ \pi ( \lambda +\mu -\nu )\right] 
     \sinh \left[ \pi ( \lambda -\mu +\nu ) \right] }%
     {\sinh \left( 2\pi \mu \right) \sinh \left( 2\pi \nu \right)} \; ,
\end{equation}
where we have defined
\begin{gather}
\mu \equiv \frac{\tau}{2} \sqrt{m^2+p_\perp^2 +(p_z -2eE\tau )^2} \; , \\
\nu \equiv \frac{\tau}{2} \sqrt{m^2+p_\perp^2 +p_z^2} \; , \\
\lambda \equiv eE\tau^2 \; . 
\end{gather}
Note that $\p$ in these equations is the physical kinetic momentum,
while in some works the spectrum is given in terms of a
gauge-dependent canonical momentum.

\begin{figure}[htbp]
  \begin{center}
    \resizebox*{10cm}{!}{\includegraphics{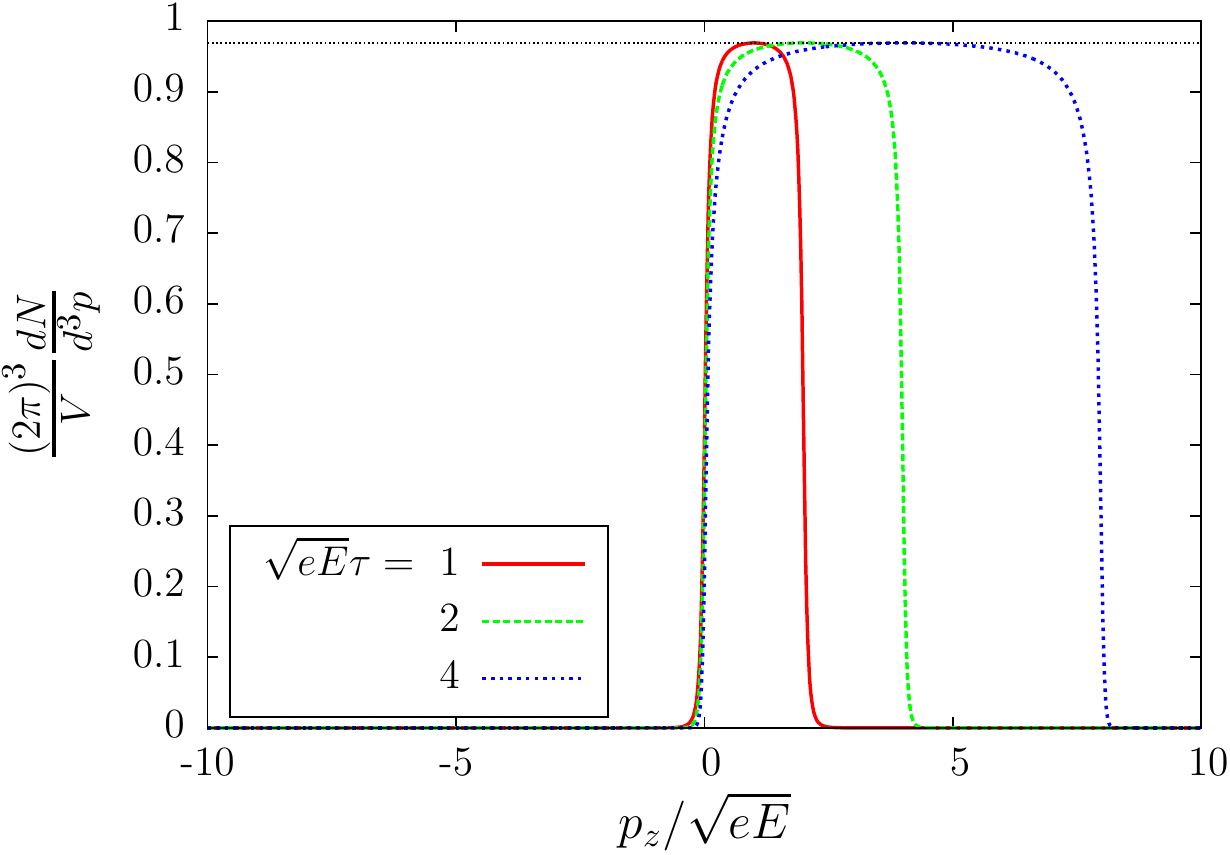}}
  \end{center}
  \caption{\label{fig:Sauter_analytic1} The spectrum
    \eqref{eq:Sauter_spec} as a function of $p_z$ for various values
    of $\tau$.  The transverse mass is fixed to $m_\perp
    /\sqrt{eE}=0.1$.  A thin black horizontal line denotes $\exp
    ( -{\pi m_\perp^2}/{(eE)})$. }
\end{figure}

In the figure \ref{fig:Sauter_analytic1}, the spectrum of
eq.~\eqref{eq:Sauter_spec} is plotted as a function of $p_z$, for a
fixed transverse mass $\sqrt{{(m^2+p_\perp^2)}/eE} =0.1$, and for
various values of $\tau$.  The spectrum has a width of order $2eE\tau$
in the $p_z$ direction.  This can be understood as follows: particles
are produced with a nearly zero momentum because the spatially
homogeneous electrical field carries no momentum.  After being
produced, they are accelerated by the electrical field following the
classical equation of motion $p_z =\int^t eE_z (t^\prime ) dt^\prime
\simeq eEt$.  Since the particle production and the acceleration
mostly happen in the time interval $t\in[-\tau,+\tau]$, most of the
particles are distributed in the range $p_z \in[0, 2eE\tau]$.

\begin{figure}[htbp]
  \begin{center}
    \resizebox*{10cm}{!}{\includegraphics{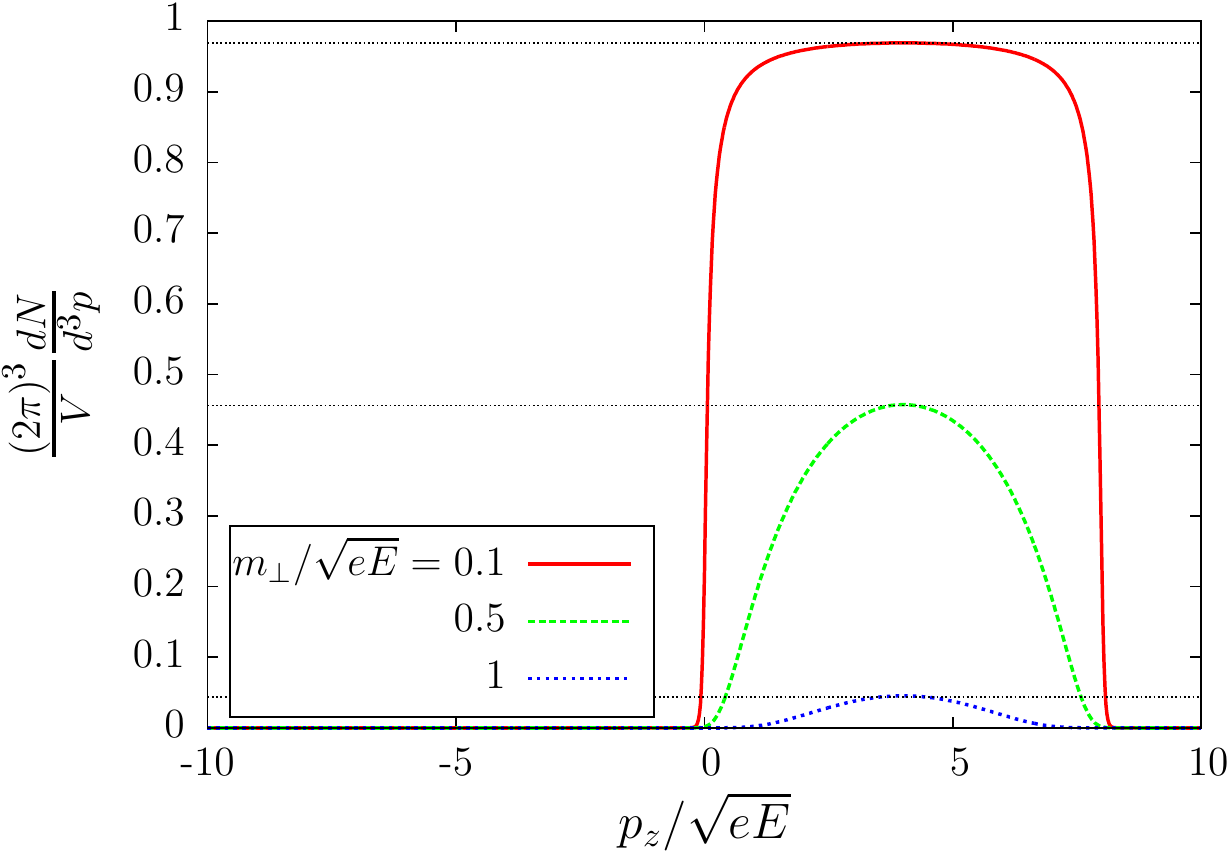}}
  \end{center}
  \caption{\label{fig:Sauter_analytic2} The spectrum
    \eqref{eq:Sauter_spec} as a function of $p_z$ for various values
    of $m_\perp$.  The pulse duration is fixed to $\sqrt{eE} \tau =4$.
    Thin black horizontal lines denote $\exp( -{\pi
      m_\perp^2}/{(eE)})$.}
\end{figure}

The figure \ref{fig:Sauter_analytic2} shows the $p_z$-spectrum for
$\sqrt{eE}\tau=4$, and for various values of the transverse mass
$m_\perp\equiv\sqrt{m^2+p_\perp^2}$.  The factor $\exp( -{\pi
  m_\perp^2}/{(eE)})$ is also indicated by thin black lines, which
shows that the peak value of the spectrum agrees well with this
exponential factor.  In this range of parameters, the exponential
$m_\perp$-dependence of the spectrum indicates that the particle
production is dominated by the non-perturbative Schwinger effect.  As
we will discuss in the section~\ref{subsec:perturbative}, there is
another range of parameters where the perturbative particle production
is the dominant one.

\subsection{Quantum kinetic equation} \label{subsec:QKE}
In the previous subsection, we have shown that the particle spectrum
can be expressed in terms of the Bogoliubov coefficients, which can be
computed by solving the Dirac equation for the mode functions.  When
the electrical field is uniform, it is possible to derive equations
that directly describe the evolution of the Bogoliubov coefficient
\cite{popov1973e+}.  Furthermore, if the direction of the electrical
field is time-independent, one can derive an equation for the
distribution function, which is called \emph{quantum kinetic equation}
\cite{Smolyansky:1997fc,Schmidt:1998vi,Kluger:1998bm}.

Let us derive first the equations for the Bogoliubov coefficients in
the presence of an uniform and time-dependent gauge field
\eqref{eq:uni_gauge}, whose direction may vary in time.  The
Bogoliubov coefficients are given by eqs.~\eqref{eq:t-dep_Bogo1a} and
\eqref{eq:t-dep_Bogo1b}.  Note that the momentum $\p$ appearing in
these equations is the kinetic momentum, which is related to the
canonical momentum $\q$ by
\begin{equation}
\p = \q -e\A \; . 
\end{equation}
Under the influence of a spatially uniform gauge field such as
\eqref{eq:uni_gauge}, the canonical momentum is a constant of motion,
while the kinetic momentum depends on time due to the acceleration by
the electrical field.  When we take the time derivative of the
Bogoliubov coefficients, the kinetic momentum $\p$ must be regarded as
a time dependent quantity such that $\frac{\partial \p}{\partial t} =
e\E$.  In order to simplify the notations, let us treat the Bogoliubov
coefficients as a $2\times 2$ matrix whose indices are the spin
indexes $(s,s^\prime )$.
By taking the time derivative of the Bogoliubov coefficients given in
eqs.~\eqref{eq:t-dep_Bogo1a} and \eqref{eq:t-dep_Bogo1b}, one can
derive
\begin{equation} \label{eq:eq_alpha}
\frac{d \alpha (t;\p )}{dt}
 = i\,\left[ \epsilon (t) +S(t) \right]\; \alpha (t;\p )
    -e^{2iE_\p t}\; P(t)\; \beta^* (t;\p ) \; ,
\end{equation}
and 
\begin{equation} \label{eq:eq_beta}
\frac{d\beta (t;\p )}{dt}
 = i\,\left[ \epsilon (t) +S(t) \right]\; \beta (t;\p )
    +e^{2iE_\p t}\; P(t) \;\alpha^* (t;\p ) \; ,
\end{equation}
where we have defined
\begin{equation}
\epsilon (t) \equiv \frac{e\E \cd \p}{E_\p}\; t \; ,
\end{equation}
\begin{equation}
P (t) \equiv \frac{eE_z E_\p^2 -(e\E \cd \p )\,p_z }{2E_\p^2 m_\perp}\; {\bs1}
          +i\,\frac{eE_x p_y -eE_y p_x}{2E_\p m_\perp}\; {\bs\sigma}^3 
          +i\,\frac{m}{2E_\p m_\perp} (eE_x \,{\bs\sigma}^2 + eE_y\, {\bs\sigma}^1 ) \; ,
\end{equation}
and
\begin{equation}
S(t) \equiv \frac{m (eE_x\,{\bs\sigma}^2 +eE_y \,{\bs\sigma}^1)}{2E_\p (E_\p -p_z )} 
         +\frac{(eE_x p_y -eE_y p_x) \,{\bs\sigma}^3 }{2E_\p (E_\p -p_z )} \; . 
\end{equation}
(The ${\bs \sigma}^i$ are the Pauli matrices.)  The matrix $S(t)$
represents the precession of the spin in the electrical field, while
the matrix $P(t)$ describes the rate of mixing between the
coefficients $\alpha$ and $\beta$, which is closely related to the
rate of pair production.  The detailed form of $P(t)$ and $S(t)$
depends on the choice of a spin basis.  Here we have used the free
spinors whose spin basis diagonalizes the interaction with a uniform
electrical field along the $z$-direction.  If instead we use spinors
whose spin basis is given by the eigenstates of ${\bs\sigma}^3$ in the
rest frame, the form of these equations becomes essentially equivalent
to those in ref.~\cite{popov1973e+}.  Of course, spin-averaged
observables do not depend on the choice of the spin basis.

Let us now restrict ourselves to the case where the direction of the
electrical field is fixed to be the $z$-direction, and derive the
kinetic equation for the distribution function.  In this case, the
matrices $P(t)$ and $S(t)$ receive important simplifications, namely
$P(t)$ is proportional to the unit matrix and $S(t)=0$.  Consequently
the Bogoliubov coefficients are also proportional to the unit matrix.
By taking the time derivative of the distribution function
(occupation number) defined by
\begin{equation}
f_\p(t) \equiv \frac{(2\pi)^3}{V}\; \frac{1}{2} \sum_s \frac{dN_s}{d^3 \p} 
 = \left| \beta (t,\p ) \right|^2 \; ,
\end{equation}
one obtains
\begin{equation} \label{eq:eq_f}
\frac{d\,f_\p(t)}{dt} 
 = \frac{eE_z m_\perp}{E_\p^2}\; {\rm Re}\,(g_\p(t)) \; ,
\end{equation}
where we have introduced the \emph{anomalous distribution}:
\begin{equation}
g_\p(t) \equiv e^{-2iE_\p t}\; \alpha (t,\p ) \beta (t,\p )\; .
\end{equation}
Eq.~\eqref{eq:eq_f} must be supplemented by an equation for the
anomalous distribution,
\begin{equation} \label{eq:eq_g}
\frac{d\,g_\p(t) }{dt}
 = -2iE_\p\; g_\p(t) 
    +\frac{eE_z m_\perp}{2E_\p }\; \left[ 1-2\,f_\p(t)\right] \; .
\end{equation}
The latter equation admits the following formal solution
\begin{equation} \label{eq:sol_g}
g_\p(t) = \int^t dt^\prime \; \frac{eE_z (t^\prime ) m_\perp}{2E_\p^2 (t^\prime )}\;
 \left[ 1-2\,f_\p(t^\prime)\right]\; e^{-i( \theta (t)-\theta (t^\prime ))} \; ,  
\end{equation}
where we denote
\begin{equation}
\theta (t) \equiv 2\int^t d\tau \; E_\p (\tau ) \; . 
\end{equation}
The factor $1-2\,f_\p(t^\prime)$ encodes the effect of Pauli
blocking\footnote{In case of scalar QED, this factor would be replaced by a
  Bose enhancement factor $1+2\,f_\p(t^\prime)$.}.  By substituting
eq.~\eqref{eq:sol_g} into eq.~\eqref{eq:eq_f}, we obtain a closed
equation for the distribution function itself,
\begin{equation} \label{eq:QKE}
\frac{d\,f_\p(t)}{dt} 
 = \frac{eE_z (t) m_\perp}{2E_\p^2 (t)} 
    \int^t dt^\prime \; \frac{eE_z (t^\prime ) m_\perp}{E_\p^2 (t^\prime )}\;
    \left[ 1-2\,f_\p(t^\prime)\right]\, \cos \left( \theta (t)-\theta (t^\prime )\right) \; . 
\end{equation}
This equation is called \emph{quantum kinetic equation}
\cite{Smolyansky:1997fc,Schmidt:1998vi,Kluger:1998bm}.  From its derivation, it
should be obvious that this formalism is equivalent to solving the
Dirac equation for the mode functions, as long as the background
electrical field is uniform and its direction is fixed.  For practical
purposes in numerical calculations, it is easier to solve
eqs.~\eqref{eq:eq_f} and \eqref{eq:eq_g} as associated equations
rather than eq.~\eqref{eq:QKE} which is non-local in time. The
non-locality in time of eq.~\eqref{eq:QKE}, obtained after eliminating
$g_\p(t)$ to get a closed equation for $f_\p(t)$, is reminiscent of
the quantum nature of the process under consideration.  Similar
equations to eqs.~\eqref{eq:eq_f} and \eqref{eq:eq_g} for the particle
production by parametric resonance in a $\phi^4$ scalar theory can be
found in ref.~\cite{Micha:2004bv}. In this seemingly unrelated
problem, the large zero mode of the field acts as a time dependent
background field for the non-zero modes. Because of this analogy, this
problem is amenable to a treatment which is very similar to that of
the Schwinger mechanism.

\subsection{Wigner formalism} \label{subsec:Wigner}
The \emph{Wigner formalism} is another approach which has been applied
to studies of particle production by the Schwinger mechanism
\cite{BialynickiBirula:1991tx,Levai:2009mn,Hebenstreit:2010vz,Hebenstreit:2010cc,Hebenstreit:2011wk,Berenyi:2013eia}.
This approach shares the same spirit as the quantum kinetic approach
in the sense that equations for one-particle distributions are
obtained instead of an equation for the elementary fields, though it
is more general than the quantum kinetic approach as is applicable to
inhomogeneous background fields.  Under the influence of inhomogeneous
backgrounds, the momentum distribution function is not a well-defined
quantity.  Instead one can consider the Wigner function defined by
\begin{equation}
 \mathcal{W} (t;\x ,\p) \equiv
  -\frac{1}{2} \!\int \!d^3 \s\, e^{-i\p \cdot \s}\; 
 \big< 0{}_{\rm in}\big| e^{-ie \int_{-1/2}^{1/2} d\lambda\,\A (\x +\lambda \s ) \cdot \s }\, 
 \left[ \hat\psi (t,\x +\s/2 ) , \overline{\hat\psi} (t,\x -\s/2 ) \right]
 \big|0{}_{\rm in}\big> \; .
 \label{eq:W-def}
\end{equation} 
A Wilson line factor is inserted between the two field operators in
order to preserve the invariance under space-dependent gauge
transformations (the temporal gauge condition $A^0=0$ is assumed).
Here, we take a straight segment to connect the two points. As already
mentioned earlier, if the background is not a pure-gauge potential
(i.e. if there are non-zero electrical or magnetic fields), the Wilson
line depends on the path one chooses. Therefore, one should keep in
mind that a certain amount of arbitrariness is present here (but
nothing worse than our earlier attempts to define a particle spectrum
in a non-pure gauge background).

Note that since the uncertainty principle does not allow the
simultaneous measurement of the position and momentum of a particle,
the Wigner distribution $\mathcal{W}(t;\x,\p)$ is not a proper probability
distribution. But its integrals over $\x$ or $\p$ are probability
distributions in $\p$ or $\x$, respectively. In fact, even the
positivity of the Wigner function is not guaranteed. However, it is
generally the case that the support in phase-space of the negative
values of $\mathcal{W}$ is of order $\hbar$. After a coarse graining of
phase-space into cells of size $\hbar$ or more, these negative regions
usually disappear and a probabilistic interpretation becomes
plausible.

From the Dirac equation for the fermion field operator, one can derive
an evolution equation for the Wigner distribution.  If the gauge field
is a quantum field operator, the equation is not closed and it depends
on higher order correlation functions.  Since here we regard the gauge
field as a classical background field, a closed equation can be
derived:
\begin{equation} \label{eq:eq_W}
D_t \mathcal{W} = -\frac{1}{2} {\boldsymbol D}_\x \cd \left[ \gamma^0 {\boldsymbol \gamma} ,\mathcal{W} \right] 
 -im \left[ \gamma^0 ,\mathcal{W} \right] 
 -i{\boldsymbol P} \cd \left\{ \gamma^0 {\boldsymbol \gamma} ,\mathcal{W} \right\} \; ,   
\end{equation}
where $D_t$, ${\boldsymbol D}_\x$, and ${\boldsymbol P}$ are non-local
operators defined by
\begin{align}
  D_t \equiv \partial_t +e \int_{-1/2}^{1/2} d\lambda \;
  \E (t, \x +i\lambda \nabla_\p ) \cd \nabla_\p \; , \\
     {\boldsymbol D}_\x \equiv \nabla_\x +e \int_{-1/2}^{1/2} d\lambda \; \B (t, \x +i\lambda \nabla_\p ) \times \nabla_\p \; , \\
     {\boldsymbol P} \equiv \p -ie \int_{-1/2}^{1/2} d\lambda \; \lambda \;
     \B (t, \x +i\lambda \nabla_\p ) \times \nabla_\p \; . 
\end{align}
Since eq.~\eqref{eq:eq_W} is derived from the Dirac equation
\eqref{eq:Dirac_eq1} without any approximation, this approach is
equivalent to directly solving the Dirac equation (e.g. via the mode
function method).  An advantage of this approach over the mode
function method is that eq.~\eqref{eq:eq_W} makes gauge invariance
more manifest because it depends on the electric and magnetic fields
but not on the gauge field (but keep in mind the caveat mentioned in
the paragraph following eq.~\eqref{eq:W-def}).  Although
eq.~\eqref{eq:eq_W} is valid for arbitrary space-dependent
electromagnetic fields, the numerical implementation of the non-local
operators is difficult to achieve.  Effects of a small spatial
inhomogeneity have been studied with a derivative expansion in
ref.~\cite{Hebenstreit:2010vz}.  When the electric field is uniform
with a fixed direction and the magnetic field is absent,
eq.~\eqref{eq:eq_W} can be reduced to the quantum kinetic equation
\eqref{eq:QKE} \cite{Hebenstreit:2010vz}.

\section{Numerical methods on the lattice}
\label{sec:latt}

\subsection{Real-time lattice numerical computations} \label{subsec:lattice}
As discussed in the section \ref{subsec:Bogoliubov}, any observable
may in principle be computed once we have obtained the mode functions
$\psi_{\p ,s}^\pm (x)$ by solving the Dirac equation.  This method is
applicable to completely general space and time dependent background
fields.  Although we have mainly discussed the momentum spectrum of
the produced particles in the previous sections, other observables
like the charged current and the energy-momentum tensor can also be
expressed in terms of the mode functions.  For example, the vacuum
expectation value of a fermion bilinear operator $\hat{\psi}^\dagger
(x) {\bs M} \hat{\psi} (y)$, where ${\bs M}$ is a matrix in the spinor
and coordinate spaces, can be represented in terms of the mode
functions as\footnote{Note that only the negative energy mode
  functions appear in this vacuum expectation value.  The positive
  energy mode functions are necessary to compute, for instance, the
  expectation value of the symmetrized charged current operator,
  $J^\mu_\text{sym} (x) \equiv \frac{1}{2} e\gamma^\mu [\bar{\psi} (x) ,\psi(x)]$.
  However, if the system is charge neutral, the expectation value of
  the symmetrized charged current is the same as the expectation value
  of the unsymmetrized operator, $J^\mu (x) \equiv e\bar{\psi} (x)
  \gamma^\mu \psi (x)$. }
\begin{equation} \label{eq:bilinear_exp}
\big< 0{}_{\rm in} \big| \hat{\psi}^\dagger (x) {\bs M} \hat{\psi} (y) \big |0{}_{\rm in} \big> 
 = \sum_s \int \frac{d^3 \p}{(2\pi)^3 2E_\p } \; \psi_{\p ,s}^{\text{in}-\, \dagger} (x)\, {\bs M}\, \psi_{\p ,s}^{\text{in}-} (y) \; . 
\end{equation} 
Also the momentum spectrum \eqref{eq:timedep_spectrum1} can be
expressed in this form.  In the following, we will omit the index
`in', since the out-mode functions do not appear.

Our problem is thus reduced to solving the Dirac equation for the mode
functions with the initial condition \eqref{eq:mode_in}.  For general
space and time dependent background fields, it is impossible to solve
the Dirac equation analytically, and we therefore need to resort to
numerical computations.  In this subsection, we briefly explain a
possible lattice setup for this numerical implementation in SU($N_c$)
gauge theory \cite{Aarts:1998td,Aarts:1999zn}.  We assume the temporal
gauge condition $A^0 =0$, and treat the time variable $t$ as a
continuum variable.  We divide the 3-dimensional volume, that we take
of size of $L_x \times L_y \times L_z$, into $N_x \times N_y \times
N_z$ lattice sites.  The space coordinates are labeled by integers
$n_i$ $(i=x,y,z)$ as follows
\begin{equation}
(x,y,z) = (n_x a_x ,n_y a_y , n_z a_z )\; ,
\end{equation}
where the numbers $a_i =L_i /N_i$ are the lattice spacings.  It is
common practice to use periodic boundary condition (in space) for the
fields.

On the lattice, it is more convenient to consider the link variables
\begin{equation}
\mathcal{U}_i (x) \equiv e^{ia_ig A_i (x)}\; ,
\end{equation}
that are Wilson lines spanning one elementary edge of the lattice, as
the fundamental variables, instead of the gauge fields $A_i (x)$.
After the gauge fixing by the temporal gauge condition, there is a
residual invariance under gauge transformations that depend only on
the spatial coordinates. It is highly desirable to preserve exactly
this residual invariance through the discretization.  Under such a
gauge transformation, the fermion fields and the link variables are
transformed as
\begin{equation}
\psi (t,\x ) \longrightarrow \Omega (\x )\; \psi (t,\x ) \; ,
\end{equation}
and
\begin{equation}
  \mathcal{U}_i (t,\x ) \longrightarrow \Omega (\x )\;
  \mathcal{U}_i (t,\x )\; \Omega^\dagger (\x +\hat{i} ) \; .  
\end{equation}
Therefore, a natural definition for the covariant derivative applied
to the fermion field reads
\begin{equation} \label{eq:center-diff}
  D_i \psi (x) \equiv \frac{1}{2a_i }
  \left[ \mathcal{U}_i (x)\; \psi (x+\hat{i} ) -\mathcal{U}_i^\dagger (x-\hat{i} )\; \psi (x-\hat{i} )\right] \; ,
\end{equation}
where we have used a centered difference in order to preserve the
unitarity of the theory. This covariant derivative transforms as
expected under the residual gauge transformations. With this
definition, the Dirac equation on the lattice reads
\begin{equation} \label{eq:latt_Dirac}
\left[ i\gamma^0 \partial_0 +i\gamma^i D_i -m \right] \psi (x) = 0 \; . 
\end{equation}

If we regard the background gauge fields as generated from some given
sources, we need to solve the Yang-Mills equations on the lattice in
addition to the Dirac equation.  The lattice Yang-Mills equation can
be derived from the lattice Hamiltonian for the SU($N_c$) gauge
fields,
\begin{equation}
\mathcal{H}_\text{gauge} = \sum_{i=1}^3 {\rm tr}\, \big(E_i E_i\big)
 +\frac{N_c}{g^2} \underset{(j\neq i)}{\sum_{j=1}^3} \frac{1}{a_i^2 a_j^2}
 \left\{ 1-\frac{1}{N_c} {\rm Re}\, \left[ {\rm tr}\,\big( \mathcal{U}_{i,j} (x)\big) \right] \right\} \; ,
\end{equation}
where the $E_i (x)$ are the electrical fields, and the variables
$\mathcal{U}_{ij} (x)$, called \emph{plaquettes}, are defined by
\begin{equation}
  \mathcal{U}_{i,j} (x) \equiv
  U_i (x)\, U_j (x+\hat{i})\, U_i^\dagger (x+\hat{j})\, U_j^\dagger (x) \; . 
\end{equation}
(Plaquettes are Wilson loops spanning an elementary square on the
lattice.)  The Hamilton equations then read
\begin{equation} \label{eq:EOM_U}
\partial_t\, \mathcal{U}_i (x) = iga_i\, E_i (x)\; \mathcal{U}_i (x) \; , 
\end{equation}
and
\begin{equation} \label{eq:EOM_E}
  \partial_t\, E_i (x) =
  -\frac{1}{g\,a_i} \underset{(j\neq i)}{\sum_{j=1}^3} \frac{1}{a_j^2}\; 
 {\rm Im}\, \left[ \, \mathcal{U}_{i,j} (x) +\mathcal{U}_{i,-j} (x) \right] -\text{(trace)} \; ,
\end{equation}
where
\begin{equation}
  \mathcal{U}_{i,-j} (x) \equiv
  \mathcal{U}_i (x)\, \mathcal{U}_j^\dagger (x+\hat{i}-\hat{j})\, \mathcal{U}_i^\dagger (x-\hat{j})\, \mathcal{U}_j (x-\hat{j} ) 
\end{equation}
and where, for an element $X$ of the fundamental representation of the
SU($N_c$) algebra, the notation \lq\lq $X-$(trace)\rq\rq\ means
\begin{equation}
X-\text{(trace)} = X-\frac{1}{N_c}\, {\rm tr}\,\big( X\big)\; .
\end{equation}
For a U(1) theory such as QED, the $-$(trace) term must of course be ignored. 

Because we employ the centered difference in
eq.~\eqref{eq:center-diff}, the lattice momentum for the fermion
fields suffers from the problem of doublers.  In $d$-dimensional
momentum space, there is a $2^d$-fold degeneracy.  If the background
field is spatially homogeneous, it does not carry any nonzero momenta.
Therefore, fermions in different momentum modes do not interact with
each other, and thus all these degenerated modes remain independent.
In this case, we can simply eliminate the effect of the doublers by
dividing expectation values by the number of degenerated modes, as
long as all the degenerated modes contribute to the expectation value
equally\footnote{This condition is satisfied for the energy-momentum
  tensor and the charge current, but the chiral charge is an
  exception.}.  However, when the background gauge field is
inhomogeneous, it carries nonzero momenta that cause interactions
between fermions in different momentum modes.  In this case, a
non-trivial mixing happens between degenerate doubler modes, and the
fermionic observables may be contaminated by the doublers.  One way to
suppress the doubler modes is to add the \emph{Wilson term} to the
Dirac equation, which now reads
\begin{equation}
\left( i\gamma^0 \partial_0 +i\gamma^i D_i -m \right) \psi (x)
 +\frac{r}{2} \sum_{i=1}^3 \frac{1}{a_i} \left[ \mathcal{U}_i (x)\, \psi (x+\hat{i} ) -2\psi(x) +\mathcal{U}_i^\dagger (x-\hat{i}) \,\psi (x-\hat{i} ) \right] 
 = 0 \; ,
\end{equation}
where $r$ is a positive parameter.  Thanks to the addition of the
Wilson term, the doubler modes acquire a heavy mass of the order of
the lattice ultraviolet cutoff $1/a$, and therefore they decouple from
the dynamics when $a\to 0$.  Note that the numerical results presented
in the subsections \ref{subsec:numerical} and \ref{subsec:das_numerical}
are computed without the Wilson term since the background gauge fields
considered there are uniform.

\subsection{Statistical sampling method} \label{subsec:statistical}
\subsubsection{Formulation} 
In the previous subsection, we have presented a lattice formulation of
the Dirac and Yang-Mills equations.  Each mode function $\psi_{\p
  ,s}^\pm (x)$ obeys the Dirac equation
\begin{equation} \label{eq:latt_Dirac2}
\left[ i\gamma^0 \partial_0 +i\gamma^i D_i -m \right] \psi_{\p ,s}^\pm (x) = 0 \; . 
\end{equation}
This equation must be solved for each momentum modes (and other
quantum numbers as well).  Therefore, the numerical cost to solve the
equation is proportional to $N_\text{latt}^2 N_t$, where
$N_\text{latt}\equiv N_xN_yN_z$ is the number of lattice sites and
$N_t$ is the number of time steps.  This cost has a very unfavorable
scaling with the size of the lattice, especially in 3 dimensional
space.  If the system has some spatial symmetry, this rather expensive
cost can be reduced.  For instance, if the system is completely
uniform, the $\x$-dependence of the mode functions is analytically
known to be $e^{i\p \cdot \x}$.  In this case, we do not need to treat
the space-dependence of the field numerically, and thus the cost is
reduced to $N_\text{latt} N_t$.

In the absence of any such symmetries, a way to reduce the numerical
cost is to replace the exhaustive listing of the momentum modes by a
Monte Carlo sampling method, which is applicable to any space and time
dependent background. However, instead of picking randomly a subset of
the set of momenta, one can exploit the linearity of the Dirac
equation and use random spinors that are linear superpositions (with
random coefficients) of \emph{all} the momentum modes
\cite{Gelis:2013oca,Gelis:2015eua}. This approach provides a better
sampling of the entire momentum space.

Let us introduce a stochastic field which is a linear superposition of
the mode functions with random number coefficients:
\begin{equation} \label{eq:stochastic}
  \psi_c^- (x) = \frac{1}{\sqrt{V}} \sum_{\k,s} \frac{1}{\sqrt{2E_\k}}\;
   c_{\k ,s}\;\psi_{\k ,s}^- (x) \; , 
\end{equation}
where the random numbers are Gaussian distributed with zero mean value
and the following variance:
\begin{equation} \label{eq:variance}
\langle c_{\k ,s} c_{\k^\prime ,s^\prime}^* \rangle_\text{ens} = \delta_{\k ,\k^\prime} \delta_{s,s^\prime} \; .
\end{equation}
$\langle \cdots \rangle_\text{ens}$ denotes the ensemble average over
these random numbers.  Using this stochastic field, we can express the
expectation value of a bilinear operator \eqref{eq:bilinear_exp} as
follows:
\begin{equation} \label{eq:ensemble1}
\big< 0{}_{\rm in} \big| \hat{\psi}^\dagger (x)\,{\bs  M}\, \hat{\psi} (y) \big|0{}_{\rm in}\big> 
 = \langle \psi_c^{-\, \dagger} (x)\, {\bs M}\, \psi_c^- (y) \rangle_\text{ens} \; . 
\end{equation} 
Because the Dirac equation is linear, the stochastic fields
\eqref{eq:stochastic} obey the same Dirac equation as the mode
functions.  Therefore, expectation values at time $t$ can be computed
by the following procedure:
\begin{itemize}
\item[\textbf{i.}] At an initial time $t=t_0$, draw random numbers
  following the variance \eqref{eq:variance}, and compute a
  stochastic field \eqref{eq:stochastic} by using the initial value
  for the mode functions $\psi_{\p ,s}^- (t_0 ,\x )$. In most
  practical cases, the mode functions at $t_0$ are known analytically.
\item[\textbf{ii.}] Solve the Dirac equation for the stochastic field
  until the time $t$, and compute the contribution of this field to
  the observable.
\item[\textbf{iii.}] Repeat the steps \textbf{i} and \textbf{ii} in to
  perform the ensemble average until satisfactory statistical errors
  are reached.
\end{itemize}
The numerical cost of the step \textbf{i} is proportional to
$N_\text{latt}^2 $ since the sum over the momentum modes must be done
at each spatial coordinate\footnote{If the background field is
  independent of one or more spatial coordinates, then the mode
  functions are plane waves in these coordinates, and the sum over the
  corresponding momenta can be done very efficiently by a Fast Fourier
  Transform, thereby trading a factor $N^2$ into $N\ln(N)$ for each
  spatial direction for which such a simplification happens.}.  This
cost scales unfavorably with $N_\text{latt}$.  However, this must be
computed only at the initial time.  The cost for the time evolution is
lower than that of the direct mode function method, since it is
proportional to $N_\text{latt} N_t$ for each configuration of the
stochastic field.  If we compute $N_\text{conf}$ configurations for
the stochastic field, the total cost is roughly proportional to
$(N_\text{latt}^2 +N_\text{latt} N_t) N_\text{conf}$.  This cost is
lower than that of the direct mode function method provided that
\begin{equation*}
N_\text{conf} \ll N_\text{latt} \quad \text{and} \quad N_\text{conf} \ll N_t \; , 
\end{equation*}
both of which are easily satisfied in practice.

\subsubsection{Statistical errors}  \label{subsubsec:errors}
Eq.~\eqref{eq:ensemble1} is exact only if the number of configurations
$N_\text{conf}$ is infinite.  In practice, we have to use a finite
$N_\text{conf}$, which implies nonzero statistical errors.  In this
subsection, we describe a method to evaluate the statistical errors \cite{Gelis:2015eua}.
We consider the following vacuum expectation:
\begin{equation} \label{eq:expectationO}
  \big< 0{}_{\rm in}\big| \hat{\mathcal{O}} \big|0{}_{\rm in}\big>
  \equiv \sum_{X,f}
  \big< 0{}_{\rm in}\big| \hat{\psi}^\dagger (x)\,{\bs M}\, \hat{\psi} (y) \big|0{}_{\rm in} \big> \; ,
\end{equation}
where the sum over $X$ denotes possible summation over $x$ and $y$
coordinates (for instance in a Fourier transform to compute a
spectrum), and the sum over $f$ is over some other quantum numbers
that the matrix ${\bs M}$ may carry.  For instance, in the case of the
momentum spectrum \eqref{eq:timedep_spectrum1}, the matrix ${\bs M}$
is
\begin{equation}
  {\bs M}_\text{spec} =\frac{1}{(2\pi)^3 2E_\p}\;
  u(\p ,s) u^\dagger (\p ,s)\; U^\dagger (t,\x )\, U(t,\y )\; e^{-i\p \cdot (\x -\y )} \; , 
\end{equation}
and $X$ stands for $\{\x,\y\}$.  If we compute the spin-averaged
spectrum, $f$ would be the final spin $s$.

With a finite number $N_\text{conf}$ of configurations, the
statistical evaluation of the expectation value
\eqref{eq:expectationO} can be expressed as
\begin{equation} \label{eq:ensemble3}
\mathcal{O}_{N_\text{conf}} 
= \frac{1}{V} \sum_{X,f} \sum_{J, J^\prime}
\frac{1}{2\sqrt{E_J E_{J^\prime}}}\;  
    C_{N_\text{conf}} (J, J^\prime )\;
\psi_{J}^{-\, \dagger} (x)\,{\bs M}\, \psi_{J^\prime}^{-} (y)\; \; ,
\end{equation}
where $J$ is a shorthand notation for $(\k ,s)$.  The coefficient
$C_{N_\text{conf}}$ contains the average over $N_\text{conf}$ samples
of the Gaussian random numbers $c_{J}^{(n)}$:
\begin{equation}
C_{N_\text{conf}} (J, J^\prime ) 
 \equiv \frac{1}{N_\text{conf}} \sum_{n=1}^{N_\text{conf}} c_J^{(n)*} c_{J^\prime}^{(n)} \; .
\end{equation}
$C_{N_\text{conf}}$ itself is a random number, whose fluctuations
determine the statistical error of the evaluation
\eqref{eq:ensemble3}.  From the variance of the random coefficients
$c^{(n)}_J$, we obtain immediately the following mean value for
$C_{N_\text{conf}}(J,J')$
\begin{equation}
\langle C_{N_\text{conf}} (J ,J^\prime ) \rangle = \delta_{J,J^\prime} \; .
\end{equation}
Furthermore, if we assume for simplicity that the $c_J^{(n)}$ are
Gaussian distributed (with no correlations if $n\not=n'$), we obtain
the following variance
\begin{equation}
\langle C_{N_\text{conf}} (J ,J^\prime ) C_{N_\text{conf}} (K ,K^\prime ) \rangle
 - \langle C_{N_\text{conf}} (J ,J^\prime ) \rangle \langle C_{N_\text{conf}} (K ,K^\prime ) \rangle
    =\frac{1}{N_\text{conf}} \delta_{J,K^\prime} \delta_{K,J^\prime} \; ,
\end{equation}
which decreases as one increases the number of configurations.  By
using this equation, we can compute the variance of the statistical
evaluation \eqref{eq:ensemble3} by
\begin{equation} \label{eq:error_estimate}
\begin{split}
\langle \mathcal{O}_{N_\text{conf}}^2 \rangle
-\langle \mathcal{O}_{N_\text{conf}} \rangle^2
 = \frac{1}{N_\text{conf}} \frac{1}{V^2} \sum_{J,J^\prime} 
    \frac{1}{4E_J E_{J^\prime}}\; 
    \Big| \sum_{X,f} \psi_J^{-\, \dagger} (x)\,{\bs M}\, \psi_{J^\prime}^{-} (y) \Big|^2 \; .
\end{split}
\end{equation}
The square root of the right hand side provides an estimate of the
statistical error.  We can compute it by the statistical method in the
following procedure:
\begin{itemize}
\item[\textbf{i.}]
 Prepare two stochastic fields
\begin{equation}
  \psi_{1,2}^- (x) = \frac{1}{\sqrt{V}} \sum_J \frac{1}{\sqrt{2E_J}}\;
   c_J^{(1,2)}\,\psi_J^- (x) \; , 
\end{equation}
with uncorrelated random numbers $c_J^{(1)}$ and $c_J^{(2)}$. 
\item[\textbf{ii.}] 
Evolve these two stochastic fields by solving their Dirac equation. 
\item[\textbf{iii.}]
Compute
\begin{equation} 
\Big| \sum_{X,f} \psi_1^{-\, \dagger} (x)\,{\bs M}\, \psi_2^{-} (y) \Big|^2 \; . 
\end{equation}
\item[\textbf{iv.}]  Repeat the steps \textbf{i}-\textbf{iii} in order
  to average over the random numbers $c_J^{(1)}$ and $c_J^{(2)}$.
  Because this is merely for an error estimate, one does not need a large
  number of samples.
\item[\textbf{v.}]  At the last step, take the square root and divide
  the result by $\sqrt{N_\text{conf}}$.
\end{itemize}
Since the summand of the sum over $X,f$ in
eq.~\eqref{eq:error_estimate} is generally a complex number, phase
cancellations may happen in the summation over $X$ and $f$, making the
statistical error smaller.  For example, the transverse spectrum,
which is obtained by summing over $f\equiv(p_z ,s)$, should contain
smaller statistical error compared with the full momentum spectrum
that depends on $p_z$ and $\p_\perp$.

\subsubsection{Relation to the low-cost fermion method} \label{subsubsec:MF}
In real-time lattice simulations for fermion fields, another method
called the \emph{low-cost fermion method} \cite{Borsanyi:2008eu} has
been used in several works
\cite{Saffin:2011kc,Saffin:2011kn,Berges:2013oba,Kasper:2014uaa}.  In
the low-cost fermion method, instead of using one stochastic field
\eqref{eq:stochastic}, one employs two kinds of stochastic fields
called ``male'' and ``female'' fields:
\begin{align}
\psi_\text{M} (x) &\equiv \frac{1}{\sqrt{2V}} \sum_{\k,s} \frac{1}{\sqrt{2E_\k}}\; 
 \left[ c_{\k ,s}\,\psi_{\k ,s}^+ (x)  +  d_{\k ,s}\,\psi_{\k ,s}^- (x) \right]\; , \\
\psi_\text{F} (x) &\equiv \frac{1}{\sqrt{2V}} \sum_{\k,s} \frac{1}{\sqrt{2E_\k}}\; 
 \left[ c_{\k ,s}\,\psi_{\k ,s}^+ (x)  - d_{\k ,s}\,\psi_{\k ,s}^- (x)  \right]\; ,
\end{align}
where $c_{\k,s}$ and $d_{\k,s}$ are independent random numbers which
have the same variance as \eqref{eq:variance}.  Combining these two
fields, one can obtain the vacuum expectation value of a symmetrized
bilinear operator by an ensemble average as follows:
\begin{equation} \label{eq:MFmethod}
\begin{split}
  \big< 0{}_{\rm in}\big|
  \frac{1}{2} \left[ \hat\psi^\dagger (x) ,{\bs M}\,\hat\psi (y)\right]
  \big|0{}_{\rm in}\big>
  &= \frac{1}{2V} \sum_{\k,s}
  \frac{1}{2E_\k }\; \left[ \psi_{\k ,s}^{-\, \dagger} (x)\,{\bs M }\,\psi_{\k ,s}^- (y) 
      -\psi_{\k ,s}^{+\, \dagger} (x)\,{\bs  M}\, \psi_{\k ,s}^+ (y) \right] \\
 &= -\big< \psi_\text{M}^\dagger (x)\,{\bs  M}\, \psi_\text{F} (y) \big>_\text{ens} \; .  
\end{split}
\end{equation}
The two kinds of fields are necessary in order to obtain the minus
sign in front of the second term in the right hand side of the first
line, which originates from the anti-commutation relation for the
fermionic operators.  For bosonic fields, one would need only one kind
of stochastic field \cite{Gelis:2013oca}.

By using the completeness relation
\begin{equation}
\frac{1}{V} \sum_{\k,s} \frac{1}{2E_\k}\; \left[ \psi_{\k ,s}^+ (t,\x
  ) \psi_{\k ,s}^{+\, \dagger} (t,\y ) +\psi_{\k ,s}^- (t,\x )
  \psi_{\k ,s}^{-\, \dagger} (t,\y ) \right] = V\,\delta_{\x ,\y}\,
\unit
\end{equation}
($\unit$ is the unit matrix in the spinor space), we can relate the
quantities evaluated in the simple statistical method
\eqref{eq:ensemble1} and that in the low-cost fermion method
\eqref{eq:MFmethod} by 
\begin{equation}
\big< \psi_c^{-\, \dagger} (x)\,{\bs M}\, \psi_c^- (y) \big>_\text{ens}
 = -\big< \psi_\text{M}^\dagger (x)\,{\bs M}\, \psi_\text{F} (y) \big>_\text{ens}
   +\frac{V}{2}\; \text{tr}\,\left({\bs M}\,\delta_{\x ,\y}\right) \; . 
\end{equation}
Therefore, the two methods provide the same result for the vacuum
expectation value up to a zero-point vacuum contribution.  However,
the method \eqref{eq:ensemble1} has an advantage over the low-cost
fermion method: it leads to smaller statistical errors for the
evaluation of the spectrum if the value of the spectrum is much
smaller than one.  In the statistical method \eqref{eq:ensemble1}, the
spectrum is directly obtained by the statistical ensemble.  On the
other hand, in the low-cost fermion method \eqref{eq:MFmethod}, one
gets instead a direct access to $\frac{1}{2} -f_\p$ ($f_\p$ being the
fermion occupation number), and the zero-point occupation number 1/2
must be subtracted.  Because this vacuum 1/2 also contains statistical
errors, the low-cost fermion method suffers from relatively larger
statistical errors when the occupation number is much smaller than
1/2.

\subsection{Numerical example} \label{subsec:numerical}
\begin{figure}[htbp]
  \begin{center}
    \resizebox*{9cm}{!}{\includegraphics{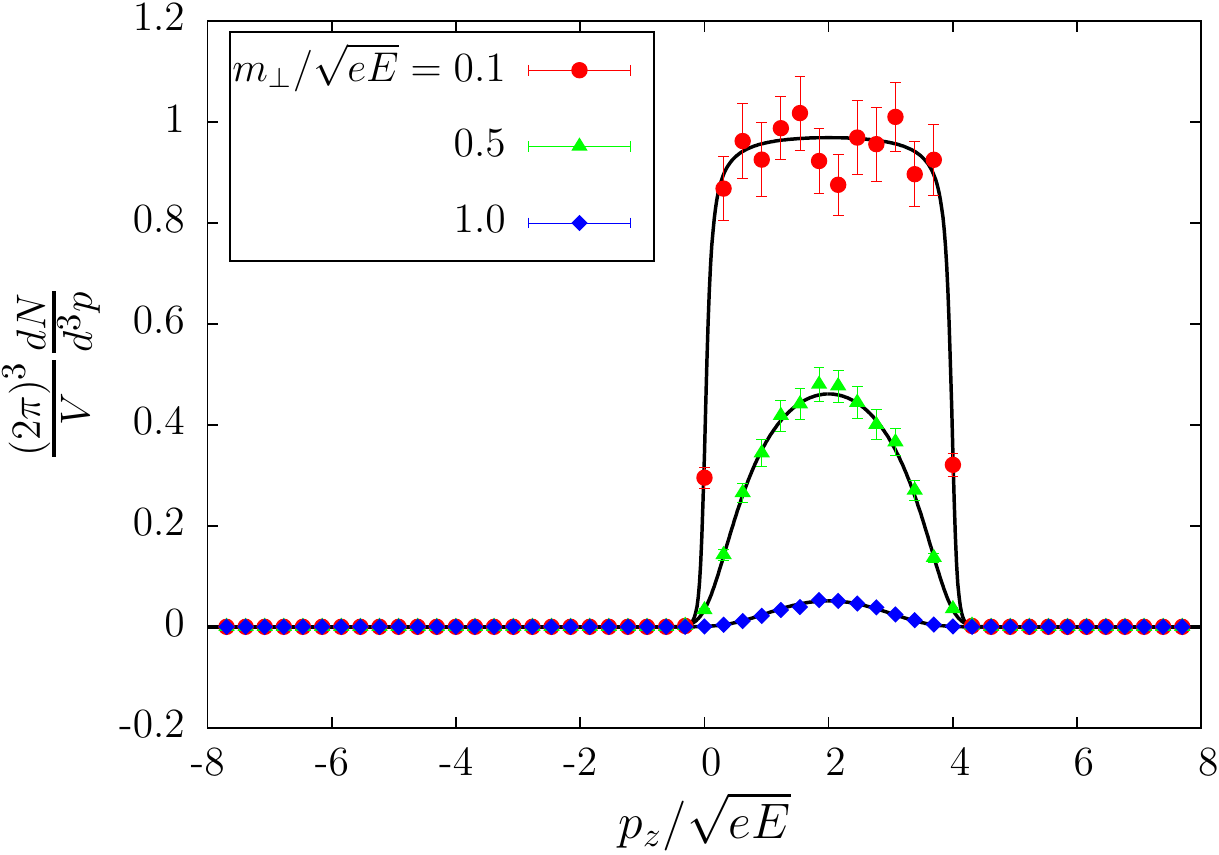}}
  \end{center}
  \caption{\label{fig:sto1} Comparison of the spectrum computed by the
    statistical method and the analytic result (black lines) in the
    Sauter electrical field with $\sqrt{eE} \tau =2$.  This plot shows
    the $p_z$-spectra for fixed values of $m_\perp$.  The lattice
    parameters are $N_z=128$, $\sqrt{eE} a_z = 0.16$, and
    $N_\text{conf}=256$. }
\end{figure}

To demonstrate the efficiency of the statistical method, we use it in
order to compute the momentum spectrum of particles produced in the
Sauter electrical field \eqref{eq:Sauter_E}. For this background
field, one can use the analytic expression \eqref{eq:Sauter_spec} to
monitor the accuracy of the numerical evaluation.  In the
figure~\ref{fig:sto1}, the $p_z$-spectrum computed by the statistical
method based on eq.~\eqref{eq:stochastic} is compared with the
analytical result.  The lattice parameters used for this computation
are\footnote{In fact, the values of the transverse ($x$ and $y$)
  lattice parameters are irrelevant for the $p_z$-spectrum with fixed
  transverse momentum $p_\perp =\sqrt{p_x^2+p_y^2}$, since there is no
  dynamics in the direction transverse to the electrical field
  \eqref{eq:Sauter_E}.}  $N_x =N_y=48$, $N_z=128$, $\sqrt{eE}\, a_x =
\sqrt{eE}\, a_y = 0.42$, $\sqrt{eE}\, a_z = 0.16$, and $N_\text{conf}
=256$.  The statistical errors evaluated by the formula
\eqref{eq:error_estimate} are indicated by error bars.  As shown in
this plot, the evaluation by the statistical method is in good
agreement with the exact results.

\begin{figure}[htbp]
  \begin{center}
    \resizebox*{9cm}{!}{\includegraphics{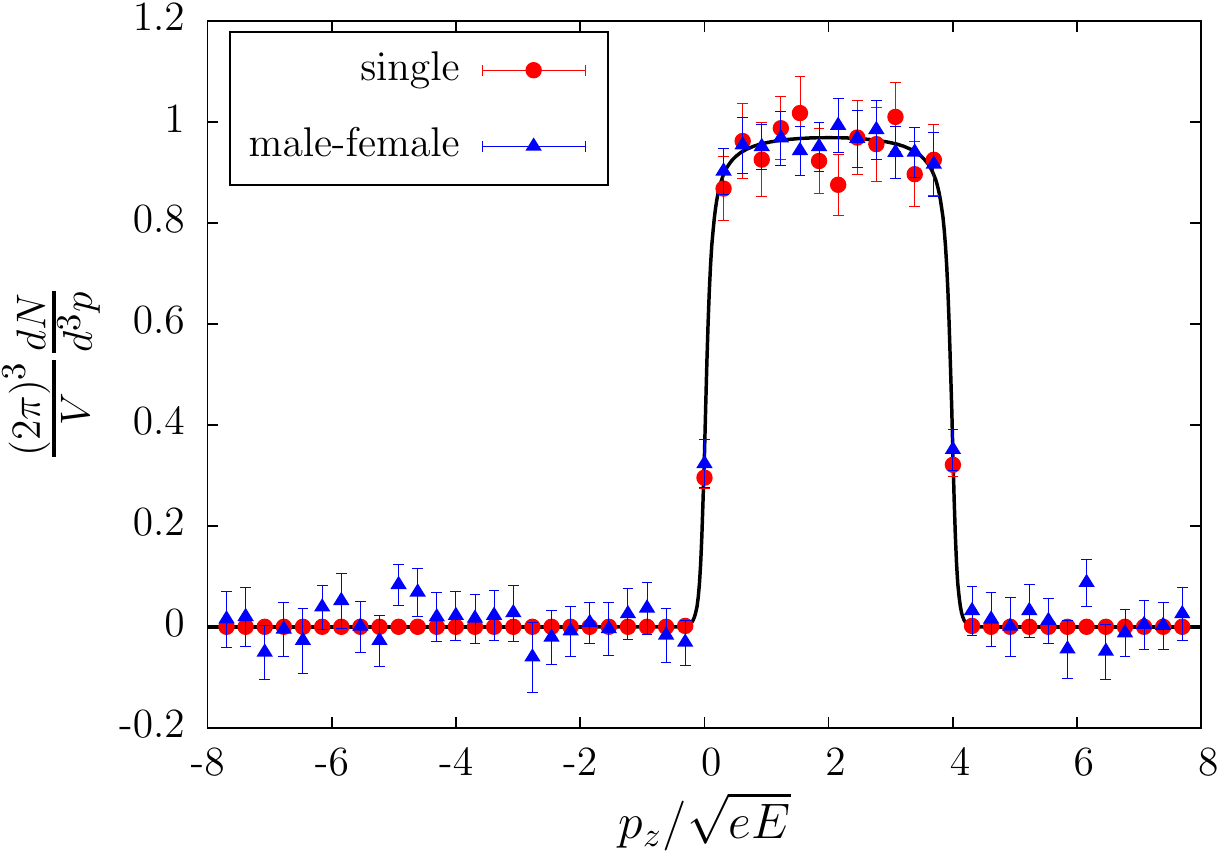}}
  \end{center}
  \caption{\label{fig:sto2} Comparison of the two statistical methods;
    one using the single kind of stochastic field
    \eqref{eq:stochastic} and the other using the male-female fields
    \eqref{eq:MFmethod}.  $\sqrt{eE}\, \tau =2$ and $m_\perp /\sqrt{eE}
    =0.1$.  The lattice parameters are the same as those in
    the figure \ref{fig:sto1}. }
\end{figure}

\begin{figure}[htbp]
  \begin{center}
    \resizebox*{9cm}{!}{\includegraphics{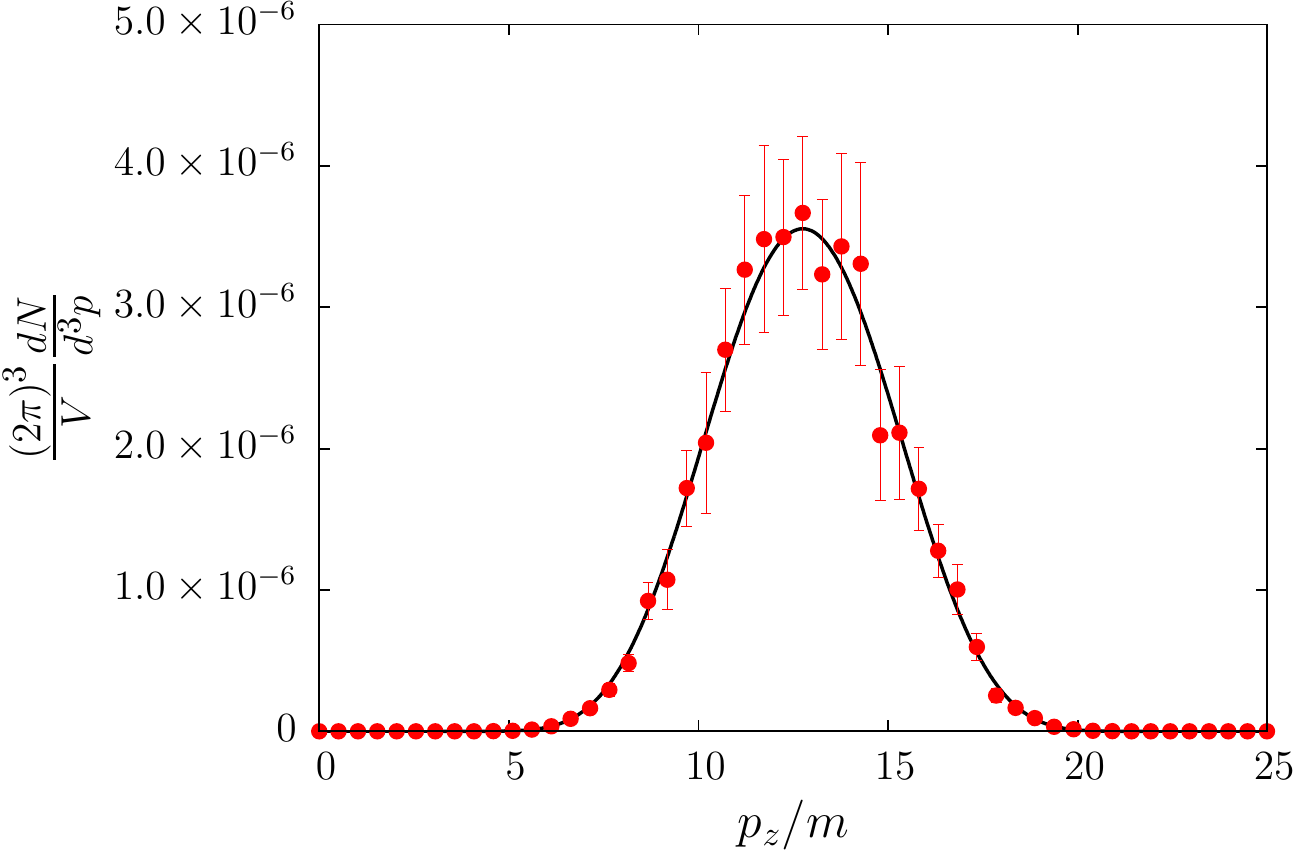}}
  \end{center}
  \caption{\label{fig:sto3} Comparison of the spectra computed by the
    statistical method and the analytic result (black line) in a weak
    field regime, $eE =0.25\,m^2$, $\sqrt{eE}\, \tau=25.5$.  The lattice
    parameters are $N_z=256$, $m a_z = 0.048$, and
    $N_\text{conf}=48$. }
\end{figure}

In the figure~\ref{fig:sto2}, the spectrum computed by the statistical
method \eqref{eq:stochastic} is compared to the result of the low-cost
fermion method \eqref{eq:MFmethod}.  Firstly, we see that both methods
agree well with the exact result.  As we have pointed out in the
section~\ref{subsubsec:MF}, the statistical errors for the
single-field stochastic method are smaller than those of the
male-female method in the region where the spectrum is smaller than
1/2.  This difference becomes crucial when we compute the particle
production in the weak field (or high momentum) regime such that $eE <
m_\perp^2$.  In such weak fields, the occupation number of produced
particle is much smaller than one.  In order to resolve these small
occupation numbers, we need to ensure that the statistical errors are
smaller than the occupation number itself.  This is easily attainable
if we use the single stochastic field method.  In the
figure~\ref{fig:sto3}, we display the spectrum computed by the single
stochastic field method in a weak field $eE =0.25\,m^2$,
$\sqrt{eE}\, \tau=25.5$.  The lattice parameters are $N_z=256$, $m a_z
= 0.048$, and $N_\text{conf}=48$.  The occupation number of the order
of $10^{-6}$ that one obtains with these parameters is accurately
reproduced within statistical errors.  If one had used the male-female
method, an extremely large number of configurations of the order of
$10^{12}$ would have been necessary in order to achieve a similar
statistical accuracy, because large statistical errors arise from the
vacuum $1/2$.

\subsection{Back reaction} \label{subsec:backreaction}
In the previous subsection, we have shown the momentum spectra of
particles produced in a gauge field whose evolution is controlled by a
given external current.  At the level of accuracy at which this
calculation was done, the total energy of the system (the
electromagnetic field and the produced particles) is not conserved.
In order to conserve the total energy, we must take into account the
back reaction of the produced particles on the gauge field.  When a
particle-antiparticle pair is produced, the gauge field is weakened
because it looses the energy that the pair takes away and eventually
decays \cite{Gavrilov:2007hq,Gavrilov:2008fv}.  In other words, when a
particle-antiparticle pair is produced, the background gauge field is
screened by the produced charges.  This screening effect is crucial
for the description of string breaking in the color flux tube model
\cite{Casher:1974vf,Glendenning:1983qq,Hebenstreit:2013baa}.  In
addition to static charges, the screening can also be caused by moving
charges (conduction current) \cite{Gatoff:1987uf}.  These effects can
be taken into account if the gauge field is coupled to a current
induced the by produced particles:
\begin{equation} \label{eq:br-maxwell}
\partial_\mu F^{\mu \nu} (x) = \langle \hat{J}^\nu (x) \rangle \; ,
\end{equation}
where $\hat{J}^\nu (x) $ is charged current operator, and $\langle
\cdots \rangle$ denotes the quantum expectation value by an initial
state.  In the case of the lattice formulation, the current must be
added to the right hand side of eq.~\eqref{eq:EOM_E}.

The description of the back reaction by eq.~\eqref{eq:br-maxwell} can
be justified in the mean field approximation based on the large-$N_f$
expansion \cite{Cooper:1989kf}, or by the classical-statistical
approximation for strong gauge fields \cite{Kasper:2014uaa}.
Numerical computations including the back reaction problem have been
conducted in several studies
\cite{Kluger:1991ib,Kluger:1992gb,Kluger:1992md,Cooper:1992hw,Kluger:1998bm,Dawson:2009cn,Asakawa:1990se,Vinnik:2001qd,Bloch:1999eu,Tanji:2008ku,Tanji:2010eu,Hebenstreit:2013qxa,Kasper:2014uaa}.

\begin{figure}[htbp]
  \begin{center}
    \resizebox*{9cm}{!}{\includegraphics{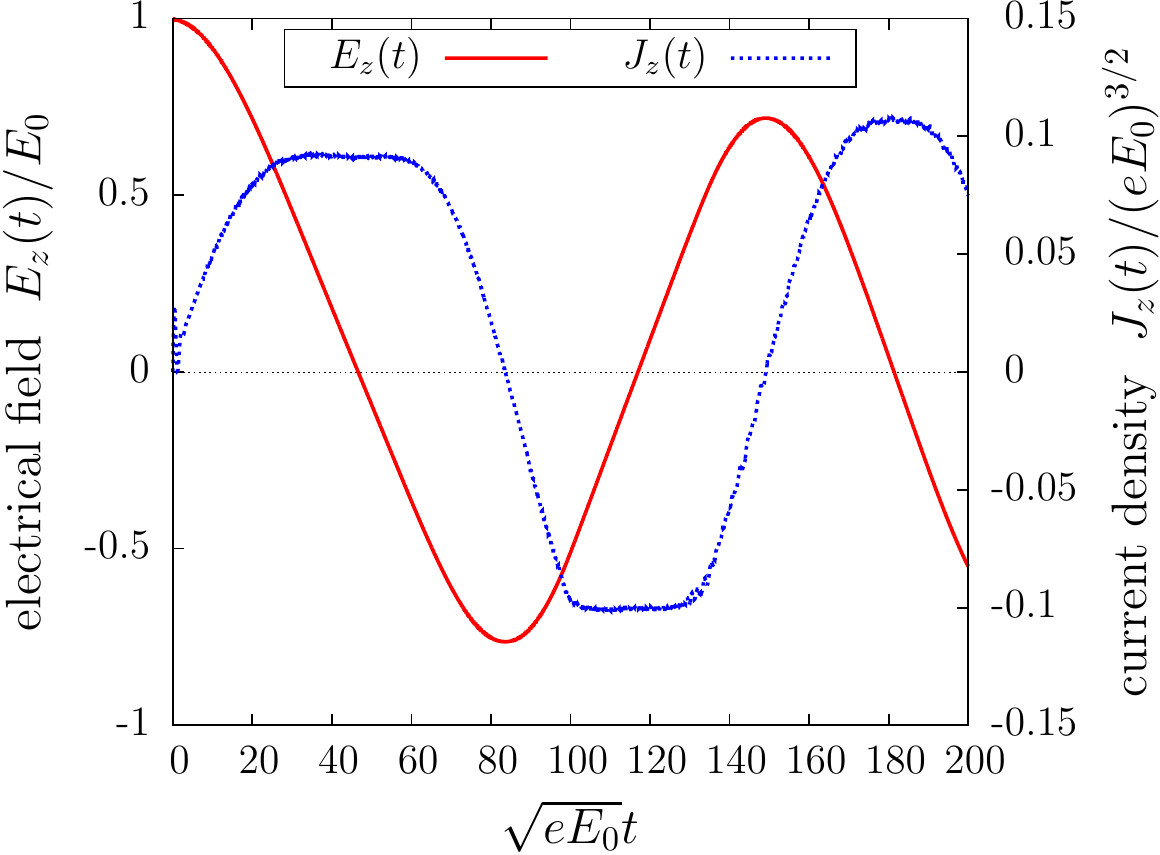}}
  \end{center}
  \caption{\label{fig:br-field-current} The time evolution of the
    electrical field $E_z (t)$ scaled by its initial value $E_0$
    (solid line, left axis), and the induced charge current $J_z (t)$
    divided by $(eE_0)^{3/2}$ (dotted line, right axis).  The
    parameters are $e=0.3$, $m/\sqrt{eE_0}=0.1$, $N_x =N_y=48$,
    $N_z=512$, $\sqrt{eE_0}\, a_x =\sqrt{eE_0}\, a_y = 0.62$,
    $\sqrt{eE_0}\, a_z = 0.029$.}
\end{figure}

\begin{figure}[htbp]
  \begin{center}
    \resizebox*{9cm}{!}{\includegraphics{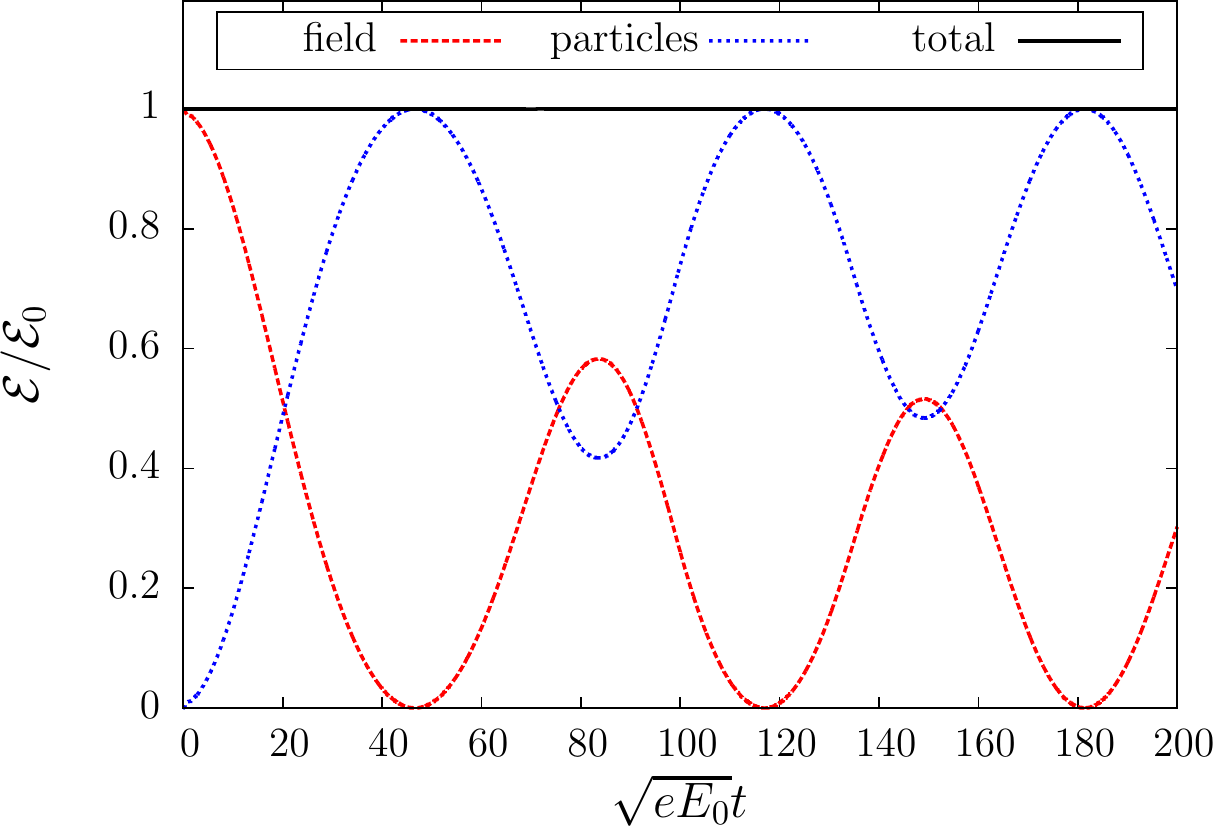}}
  \end{center}
  \caption{\label{fig:br-energy} The time dependence of the energy
    density divided by $\mathcal{E}_0 =\frac{1}{2}E_0^2$.  The energy
    density of the electrical field and that of produced particles are
    individually plotted as well as total energy density.}
\end{figure}

As an illustration of the effects of back reaction, we show numerical
results obtained in real-time lattice computations.  We have solved
(the lattice version of) the Maxwell equation \eqref{eq:br-maxwell}
with a uniform electrical field, $E_z (0)=E_0$, as initial condition.
For the fermionic sector, the mode functions are directly computed by
solving the Dirac equation, and the charge current appearing in the
right hand side of eq.~\eqref{eq:br-maxwell} is calculated at every
time step.  The parameters used in the computation are $e=0.3$,
$m/\sqrt{eE_0}=0.1$, $N_x =N_y=48$, $N_z=512$, $\sqrt{eE_0}\, a_x
=\sqrt{eE_0}\, a_y = 0.62$, $\sqrt{eE_0}\, a_z = 0.029$.  In the
figure \ref{fig:br-field-current}, the time evolution of the
electrical field $E_z (t)$ and the induced current density $J_z (t)$
are shown.  As particles are produced and accelerated by the
electrical field, a positive charge current is induced, causing the
reduction of the electrical field strength through the Maxwell
equation \eqref{eq:br-maxwell}.  At some point ($\sqrt{eE_0}\, t\simeq
45$), the field strength vanishes. However, since particles are
already accelerated to high momenta, the decrease of the electrical
field does not stop and its direction is flipped.  Then, the particles
start to be accelerated in the opposite direction.  The repeat of such
processes results in plasma oscillations.
Thanks to the back reaction, the total energy is indeed conserved
(see the figure \ref{fig:br-energy}).

\begin{figure}[htbp]
\begin{center}
  \resizebox*{8.5cm}{!}{\includegraphics{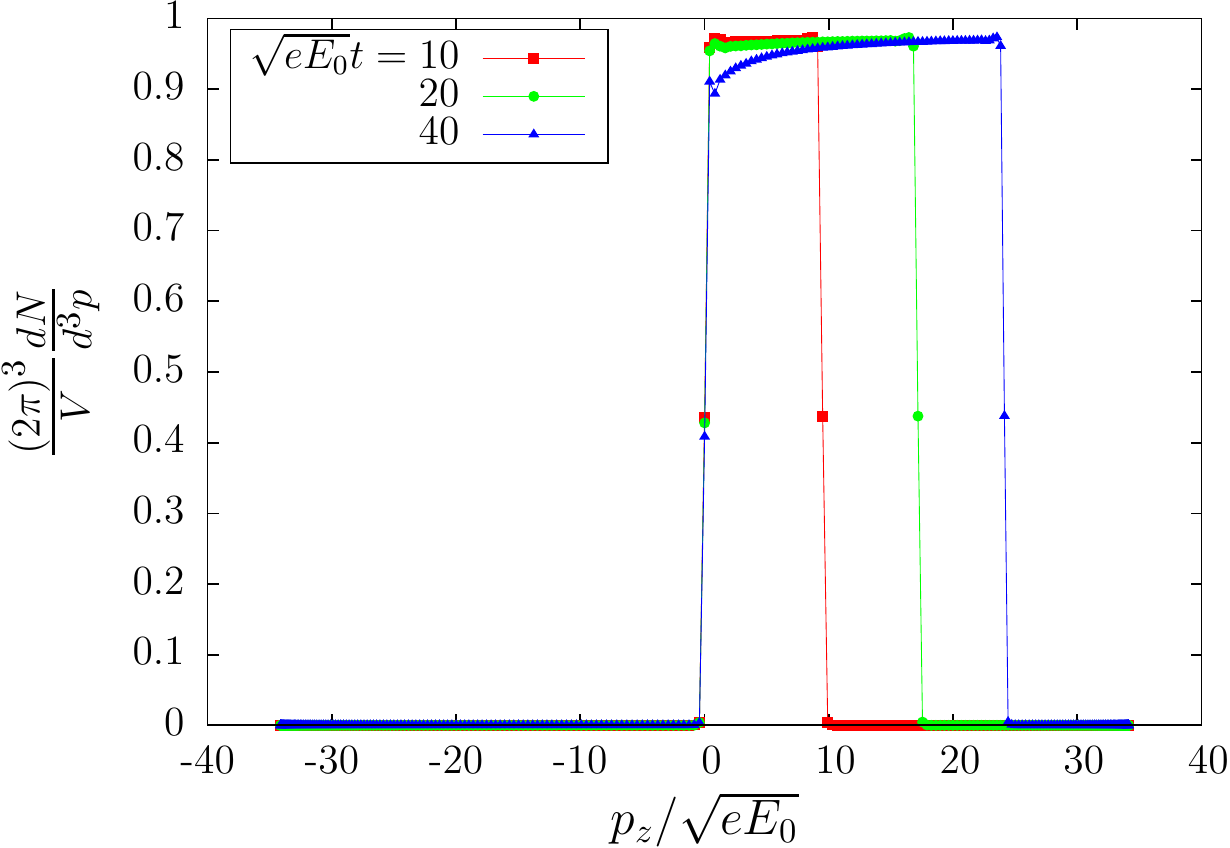}}
\end{center}
\vskip 4mm
\begin{center}
\resizebox*{8.5cm}{!}{\includegraphics{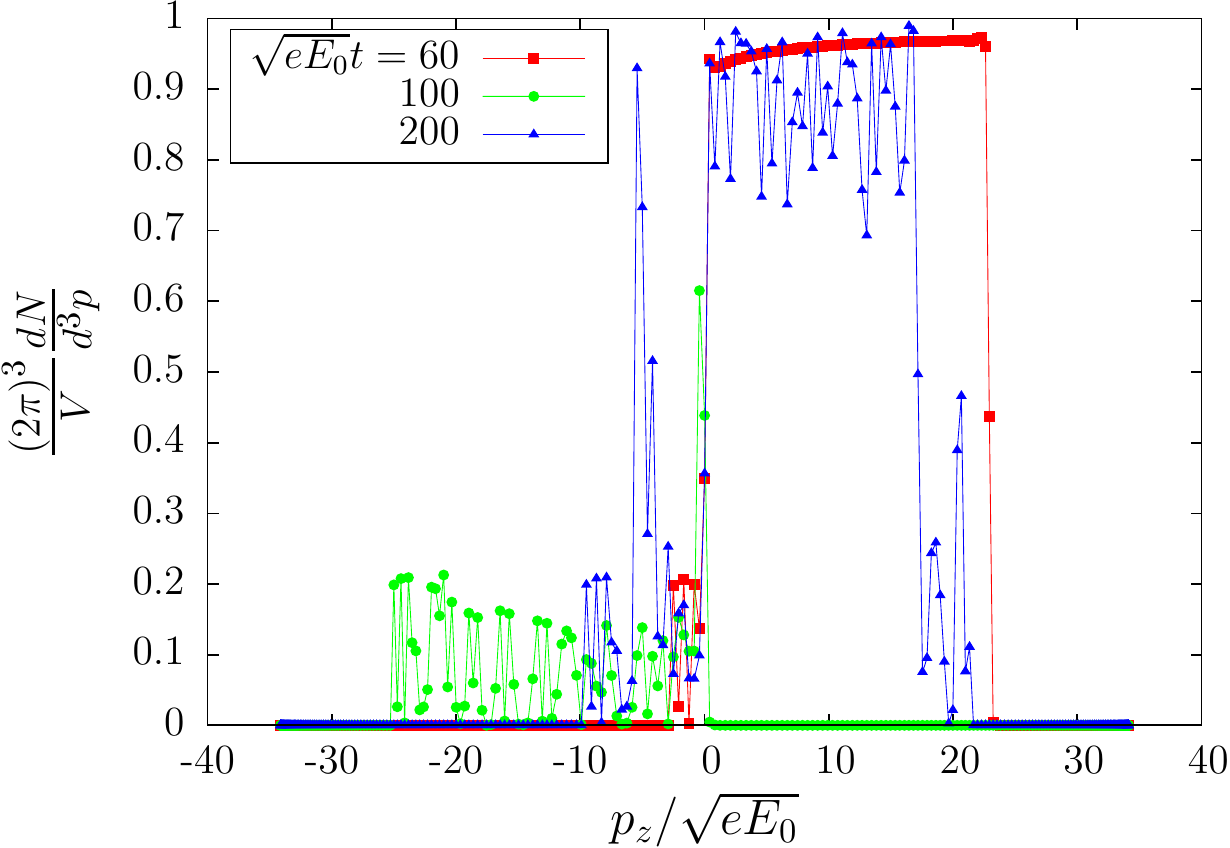}}
\end{center}
\caption{\label{fig:br-spectrum} Plots of the $p_z$-spectrum with
  fixed transverse momentum $p_\perp=0$ with back reaction taken into
  account.  The spectrum at several times is plotted (Top: early
  times, Bottom: later times).  The parameters are the same as those in
  the figure~\ref{fig:br-field-current}.}
\end{figure}

The $p_z$-spectrum of produced particles is shown in the figure
\ref{fig:br-spectrum}.  In the top figure, the spectrum at early
times before the first minimum of the electrical field happens is
plotted.  In this time range, the behavior of the spectrum is very
similar to that obtained in the Sauter electrical field, that we have
shown in the section~\ref{subsubsec:Sauter}: Particles are produced
with a nearly zero longitudinal momentum and then accelerated by the
electrical field. Their occupation number is given by $\exp (-\pi
m_\perp^2/(eE))$.

After the direction of the electrical field is flipped, non-trivial
quantum interference phenomena occur.  Because the direction of the
electrical field is negative from $\sqrt{eE_0}\, t\sim 60$ to $\sim
100$, the produced particles are accelerated in the negative $z$
direction, and thus the $p_z$-spectrum is shifted towards the negative
momentum region. While doing this, they cross the zero momentum point
$p_z =0$, where particles are still being produced by the electrical
field (for a spatially homogeneous field, the particles are created at
nearly zero momentum). Pauli blocking and interferences with the
pre-existing particles lead to a spectrum that displays distinct
features in the region $p_z<0$: (i) its magnitude is significantly
smaller than $\exp (-\pi m_\perp^2/(eE))$ and (ii) it shows rapid
oscillation in $p_z$ (see ref.~\cite{Tanji:2010eu}).  Because of this
interference phenomenon that repeats with each oscillation of the
field, the spectrum at $\sqrt{eE_0}\, t=200$ is not smooth, in contrast
to the spectrum at early times shown in the top figure.  Even though
this system is closed and the equations of motion are symmetric under
time-reversal, an apparent irreversibility of the energy flow from
coherent fields to fluctuating quantum modes emerges
\cite{Habib:1995ee}.  In the figure \ref{fig:br-energy}, we can indeed
see that the oscillation amplitude of the electrical field is slightly
decreasing in time.

\section{Worldline formalism}
\label{sec:WL}
Up to this point, our discussion of the Schwinger mechanism was
following a fairly standard quantum field theoretical approach based
on diagrammatic expansions and field equations of motion. The
non-perturbative nature of the phenomenon resulted in the necessity of
summing infinite sets of Feynman graphs, the sum of which may be
re-expressed in terms of solutions of some partial differential
equations (classical field equations of motion at Leading Order, or
the equation of motion for the mode functions over a background field
at one-loop).

In the past ten years, a radically different approach --the
\emph{worldline formalism}-- has been applied to the study of the
Schwinger mechanism. The worldline formalism first emerged from ideas
around the limit of infinite string tension in string theory
\cite{Metsaev:1987ju,Bern:1987tw,Bern:1990cu} but its potential for
performing efficiently 1-loop calculations in quantum field theories
was soon realized. This formalism was later establish of purely field
theoretical grounds in ref.~\cite{Strassler:1992zr}, in an approach
reminiscent of Schwinger's proper time method. Reviews on this
formalism can be found in
refs.~\cite{Schubert:1996jj,Schubert:2001he}.  Its main application,
of direct interest to us in this review, if the calculation of 1-loop
effective actions in a background field
\cite{Reuter:1996zm,Schmidt:1993rk,Schmidt:1994aq,Gies:2005sb}, with
special emphasis on pair creation in
refs.~\cite{Dunne:2005sx,Dunne:2006st}, but is has also been used in
studies of the Casimir effect \cite{Gies:2003cv,Gies:2006cq}.

Although the result for the Schwinger mechanism should not depend on
the method used to obtain it, the worldline formalism offers very
interesting new insights about the space-time development of the
particle production process, and provides completely novel numerical
approaches to these calculations.  In this section, we will present
several aspects of the worldline approach by returning to the simple
example of scalar QED in an external field for illustration purposes.

\subsection{Worldline formalism for pair production}
As we have discussed in the section \ref{sec:extsource}, the vacuum to
vacuum transition amplitude can be written as an exponential,
\begin{equation}
\big<0_{\rm out}\big|0{}_{\rm in}\big>=e^{i\,{\cal V}}\; ,
\end{equation}
where $i\,{\cal V}$ is the sum of all the connected vacuum
diagrams. The possibility of particle production is intimately related
to the imaginary part of ${\cal V}$, since the total probability of
producing particles reads
\begin{equation}
\sum_{n=1}^\infty P_n = 1-P_0=1-e^{-2\,{\rm Im}\,{\cal V}}\; .
\end{equation}
In scalar QED, the graphs made of one scalar loop embedded in a
background electromagnetic field lead to the following contribution to
${\cal V}$,
\begin{equation}
{\cal V}_{\rm 1\ loop}=\ln\,{\rm det}\,\left(g_{\mu\nu}D^\mu D^\nu+m^2\right)\; ,
\end{equation}
where $D^\mu$ is the covariant derivative in the background field and
$g_{\mu\nu}$ is Minkowski's metric tensor. For technical reasons
related to the definition of the worldline formalism, we should
introduce the Euclidean analogue of this quantity,
\begin{equation}
  {\cal V}_{_E,\rm 1\ loop}\equiv\ln\,{\rm det}\,\left(-D^i D^i+m^2\right)
  ={\rm Tr}\,\ln\left(-D^i D^i+m^2\right)\; ,
\end{equation}
where the index $i$ runs over the values $1,2,3,4$ (4 being the
Euclidean time). The worldline formalism can be derived from several
points of view. A simple starting point is Schwinger's proper time
representation of inverse propagators,
\begin{equation}
  \left(-D^i D^i+m^2\right)^{-1}
  =
  \int_0^\infty {d\tau}\;\exp\left(-\tau\big(-D^i D^i+m^2\big)\right)\; ,
  \label{eq:S-prop}
\end{equation}
that leads to the following formula for the logarithm
\begin{equation}
  \ln\left(-D^i D^i+m^2\right)
  =-\int_0^\infty\frac{d\tau}{\tau}\;\exp\left(-\tau\big(-D^i D^i+m^2\big)\right)\; ,
\end{equation}
up to some irrelevant integration constant. Using standard
manipulations, the trace of the exponential can then be rewritten as a
path integral over closed loops in Euclidean space-time:
\begin{equation}
  {\cal V}_{_E,\rm 1\ loop}=-\int_0^\infty\frac{d\tau}{\tau}\;
  e^{-m^2\tau}\;
  \int\limits_{x^i(0)=x^i(\tau)}\big[{\cal D}x^i(\tau')\big]\;
  \exp\Big(-\int_0^\tau d\tau'\;\big(\frac{\dot{x}^i\dot{x}^i}{4}+ie\,\dot{x}^i A^i(x)\big)\Big)\; .
  \label{eq:WL-V}
\end{equation}
This formula involves a double integration: a path integral over all
the \emph{worldlines} $x^i(\tau')$, i.e. closed paths in Euclidean
space-time parameterized by the fictitious time $\tau'\in[0,\tau]$,
and an ordinary integral over the length $\tau$ of these paths. The
dot denotes a derivative with respect to the fictitious time. The sum
over all the worldlines can be viewed as a materialization of the
quantum fluctuations in space-time, and the prefactor $\exp(-m^2\tau)$
suppresses the very long worldlines that explore regions of space-time
that are much larger than the Compton wavelength of the particles
(making obvious the role of the mass as an infrared regulator). In
contrast, the ultraviolet properties of the theory under consideration
are controlled by the short worldlines, i.e. the limit $\tau\to
0$. Note that going to Euclidean space-time was necessary in order to
obtain convergent integrals in eq.~(\ref{eq:WL-V}). In the vacuum,
i.e. when there is no background field, one can use the following
standard result in $d$ space-time dimensions,
\begin{equation}
  \int\limits_{x^i(0)=x^i(\tau)}\big[{\cal D}x^i(\tau')\big]\;
  \exp\Big(-\int_0^\tau d\tau'\;\frac{\dot{x}^i\dot{x}^i}{4}\Big)
  =\frac{1}{(4\pi \tau)^{d/2}}\empile{=}\over{d=4}\frac{1}{(4\pi\tau)^2}\; .
\end{equation}

In eq.~(\ref{eq:WL-V}), the path integral can be factored into an
integral over the barycenter $X^i$ of the worldline and the position
$r^i(\tau')$ about this barycenter,
\begin{equation}
  x^i(\tau')\equiv X^i+r^i(\tau')\quad,\qquad
  \int_0^\tau d\tau'\; r^i(\tau')=0\; .
\end{equation}
After this separation, all the information about the background field
contained in eq.~(\ref{eq:WL-V}) comes via a Wilson line,
\begin{equation}
  W_{_X}\big[r(\tau')\big]\equiv
  \exp\Big(-ie\int_0^\tau d\tau'\; \dot{r}^i(\tau')A^i(X+r(\tau'))\Big)\; ,
\end{equation}
averaged over all closed loop of length $\tau$,
\begin{equation}
  \left<W_{_X}\right>_\tau\equiv
  \frac{\int\limits_{r^i(0)=r^i(\tau) }\big[{\cal D}r^i(\tau')\big]\;
    W_{_X}\big[r(\tau')\big]\;\exp\Big(-\int_0^\tau d\tau'\; \frac{\dot{r}^i\dot{r}^i}{4}\Big)}
       {\int\limits_{x^i(0)=x^i(\tau)}\big[{\cal D}x^i(\tau')\big]\;
  \exp\Big(-\int_0^\tau d\tau'\;\frac{\dot{x}^i\dot{x}^i}{4}\Big)}\; .
  \label{eq:Wavg}
\end{equation}
In this path average, the exponential containing the integral of the
squared velocity tends to suppress the long paths. Therefore, the
average is dominated by an ensemble of loops (sometimes called a
\emph{loop cloud}) localized around the barycenter $X^i$, and
$\left<W_{_X}\right>_\tau$ encapsulates the local properties of the
quantum field theory in the vicinity of $X^i$ (roughly up to a
distance of order $\tau^{1/2}$). In terms of this averaged Wilson
loop, the 1-loop Euclidean connected vacuum amplitude reads
\begin{equation}
  {\cal V}_{_E,\rm 1\ loop}=-\frac{1}{(4\pi)^2}\int d^4X
  \int_0^\infty\frac{d\tau}{\tau^3}\;
  e^{-m^2\tau}\; \left<W_{_X}\right>_\tau\; .
\end{equation}
(In this formula, the prefactor and power of $\tau$ in the measure
assume 4 spacetime dimensions.)  Note that this expression should in
principle be supplemented by some 1-loop counterterms determined by
the renormalization conditions of the parameters of the bare
Lagrangian. The imaginary part of ${\cal V}_{_E,\rm 1\ loop}$ comes
from poles located at real values of the fictitious time $\tau$, and
can be written as
\begin{equation}
  {\rm Im}\,\left({\cal V}_{_E,\rm 1\ loop}\right)
  =
  \frac{\pi}{(4\pi)^2}\int d^4X
  \;
    {\rm Re}\sum_{{\rm poles\ }\tau_n}
    \frac{e^{-m^2\tau_n}}{\tau_n^3}\;{\rm Res}\,\left(\left<W_{_X}\right>_{\tau_n},\tau_n\right)\; .
\end{equation}

\subsection{Constant electrical field}
It is instructive to reconsider first the well known case of a
constant electrical field $E$. Since one can choose a gauge potential
which is linear in the coordinates $X^i,r^i$, the path integral
that gives the average Wilson loop is Gaussian and can therefore be
performed in closed form, leading to
\begin{equation}
  \left<W_{_X}\right>_\tau = \frac{eE\tau}{\sin(eE\tau)}\; .
  \label{eq:Wavg-constE}
\end{equation}
This expression has an infinite series of single poles along the
positive real axis, located at $\tau_n=n\pi/(eE)$ ($n=1,2,3,\cdots$),
that give the following expression for the imaginary part:
\begin{equation}
{\rm Im}\,\left({\cal V}_{_E,\rm 1\ loop}\right)
=
\frac{V_4}{16\pi^3}(eE)^2\sum_{n=1}^\infty \frac{(-1)^{n-1}}{n^2}\,e^{-n\pi m^2/(eE)}
\; .
\end{equation}
In this formula, $V_4$ is the volume in space-time over which the
integration over the barycenter $X^i$ is carried out. This formula is
identical to the standard result for the vacuum survival probability
$P_0=\exp(-2\,{\rm Im}\,{\cal V})$, that we have already recalled in
eq.~(\ref{eq:P0-constE}). By comparing the origin of the term of index
$n$ in the subsection \ref{sec:constE} and in the present derivation
in the worldline formalism, one sees that the Bose-Einstein
correlations (i.e. higher orders in the occupation number) are encoded
in the poles $\tau_n$ that are more distant from $\tau=0$, while the
first pole $\tau_1$ only contains information about the uncorrelated
part of the spectrum. This observation gives some substance to the
intuitive image that quantum fluctuations and correlations are encoded
in the fact that the worldlines explore an extended region around the
base point $X^i$. From this correspondence, we see that increasingly
intricate (the index $n$ is the number of correlated particles)
quantum correlations come from worldlines that explore larger and
larger portions of space-time.

\subsection{Numerical approaches}
The worldline formalism can be implemented numerically
\cite{Gies:2001zp,Gies:2001tj} in order to study situations, such as
inhomogeneous background fields, for which no analytical solution is
readily available.

\begin{figure}[htbp]
  \begin{center}
    \resizebox*{9.5cm}{!}{\includegraphics{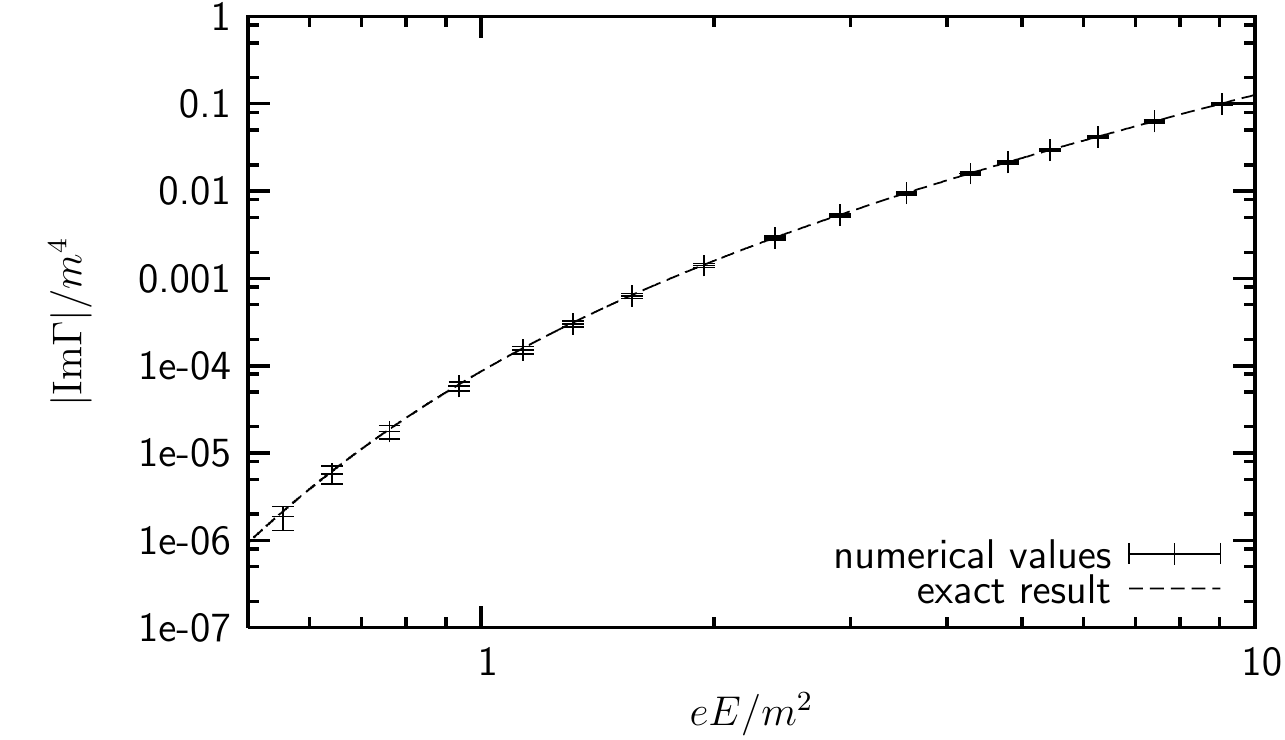}}
  \end{center}
  \caption{\label{fig:imgamma}Comparison of the exact 1-loop result
    and the result of a numerical worldline computation based on the
    Fourier transform method. From ref.~\cite{Gies:2005bz}.}
\end{figure}

Let us first present a simple algorithm specific to the case of a
constant and uniform background field. Although this algorithm cannot
be readily generalized to inhomogeneous background fields, its
discussion is useful in order to illustrate the difficulties awaiting
us in the general case. For an electrical field $E$ in the $z$
direction, we can choose a gauge in which the Euclidean gauge
potential reads
\begin{equation}
A^i=(0,0,0,-iEx^3)\; ,
\end{equation}
and the Wilson loop reads
\begin{equation}
  W_{_X}[r(\tau')]=e^{-eE {\cal A}}\; ,\quad\mbox{where we denote }
  {\cal A}\equiv \int_0^\tau d\tau'\; \dot{r}_4(\tau')r_3(\tau')\; .
  \label{eq:W-constE}
\end{equation}
(Since the loop is closed, there is no term in $X_3$ left in the
exponential.)  The quantity ${\cal A}$ is the projected area of the
loop on the plane $34$ in Euclidean space-time. Indeed, by Stokes'
theorem, the abelian Wilson loop can be rewritten as the integral of
${\cal F}_{\mu\nu}d\sigma^{\mu\nu}$ over a surface whose boundary is the loop
under consideration, where ${\cal F}_{\mu\nu}$ is the strength of the
background field and $d\sigma^{\mu\nu}$ is the measure on this
surface. Note that ${\cal A}$ is an algebraic area, i.e. it is
weighted by the winding number of the loop (and consequently it can
also be negative). Since the probability distribution of the
worldlines is Gaussian in $\dot{r}$ (see eq.~(\ref{eq:Wavg})), it is
easy to perform the path integral to obtain the probability
distribution for the area ${\cal A}$,
\begin{equation}
  {\cal P}_\tau({\cal A})=\frac{\pi}{4\tau}\,\frac{1}{\cosh^2\left(\frac{\pi{\cal A}}{2\tau}\right)}\; .
  \label{eq:probA}
\end{equation}
Thus, for a given $\tau$, the typical worldlines have an area that
reaches values up to ${\cal A}\sim\tau$. It is therefore natural to
rescale the area ${\cal A}$ by a factor $\tau$ by defining ${\cal
  I}\equiv \tau{\cal A}$ (this can be done by rescaling the
coordinates about the barycenter, $r^i$ by
$r^i\equiv\sqrt{\tau}\,y^i$). Introducing also a rescaled
\emph{imaginary} fictitious time $s\equiv -i\tau/eE$, we have
\begin{equation}
  {\cal V}_{_E,\rm 1\ loop}=\left(\frac{eE}{4\pi}\right)^2\int d^4X
  \int_0^\infty\frac{d s}{s^3}\;
  e^{-i(m^2/(eE)) s}\; \left<e^{-is{\cal I}}\right>\; .
\end{equation}
After this rearrangement of the formula, all the dependence on the
external field is contained in the quantity $m^2/(eE)$, while the
factor $\big<e^{-is{\cal I}}\big>$ is an (background
field-independent) average over the rescaled worldlines that can be
calculated once for all as a function of $s$. A Fourier transform of
this quantity then gives the imaginary part of the effective action
for all values of $m^2/(eE)$, at the expense of a very small
computational cost. The main difficulty of this algorithm is at small
field strength (i.e. $m^2/(eE)\gg 1$), for which the effective action
probes the $s$ dependence of $\big<e^{-is{\cal I}}\big>$ at small
values of $s$. This region is dominated by large worldlines, with
$|{\cal I}|\gtrsim m^2/(eE)$.  If the worldline average is obtained by
an unbiased Monte-Carlo sampling over the ensemble of loops, very few
large loops will be probed, thereby severely limiting the accuracy for
weak fields $eE\ll m^2$. In the figure \ref{fig:imgamma}, we show a
comparison of the exact 1-loop result and the result of such a
numerical calculation. One can clearly see the increase of the
Monte-Carlo statistical errors at small values of $eE/m^2$.

In order to try to evade this limitation, let us return to the real
valued fictitious time $\tau$. Of course, using the probability
distribution (\ref{eq:probA}) and eq.~(\ref{eq:W-constE}) leads easily
to the analytical result of eq.~(\ref{eq:Wavg-constE}). However, in
view of applications to more general situations, let us assume that we
want to perform this integral numerically. Thus, we need to worry
about its domain of convergence. The probability distribution of the
area ${\cal A}$ decreases as $\exp(-\pi|{\cal A}|/\tau)$ for large
areas, and is weighted by the Wilson loop given in
eq.~(\ref{eq:W-constE}). Convergence is therefore guaranteed only for
$|\tau|<\pi/(eE)$. This result is of course not surprising: the domain
of convergence cannot extend beyond the first pole $\tau_1$ of the
integral. For larger values of $\tau$, the integral over ${\cal A}$
needs to be obtained by an analytic continuation, but this is possible
only if the probability ${\cal P}_\tau({\cal A})$ is known
analytically. Unfortunately, in the case of an inhomogeneous
background field, such an analytical knowledge is not
available. It is still possible to parameterize the Wilson
loop as
\begin{equation}
W_{_X}[r(\tau')]=e^{-eE(X)\tau {\cal I}}\; ,
\end{equation}
provided that we generalize the definition of  ${\cal I}$ as follows
\begin{equation}
{\cal I}\equiv \frac{i}{\tau E(X)}\int_0^\tau d\tau'\; \dot{r}^i(\tau')\,A^i(X+r(\tau'))\; .
\end{equation}
Note that ${\cal A}\equiv \tau{\cal I}$ is no longer simply related to
the geometrical area of the worldline when the background field is not
homogeneous.  The probability distribution of ${\cal I}$ now depends
on the base point $X$ and the details of the background field around
this point. In order to obtain an approximate analytical expression
for the distribution of ${\cal I}$, it is possible to make an ansatz
that generalizes eq.~(\ref{eq:probA}),
\begin{equation}
  {\cal P}_{_X}({\cal I})=N\,\frac{1}{\cosh^{2\nu}\left(\frac{\pi\,\alpha{\cal I}}{2}\right)}\; ,
  \label{eq:PI-ansatz}
\end{equation}
where $\alpha$ and $\nu$ are $X$-dependent modifications of the width
and shape of the distribution (the prefactor $N$ is not independent,
since it is fixed by normalization).  These parameters can be fitted
from a large enough ensemble of loops generated by Monte-Carlo. Once
these parameters have been determined, the integral over $I$ can be
done analytically,
\begin{equation}
  \int_{-\infty}^{+\infty} d{\cal I}\;{\cal P}_{_X}({\cal I})\;e^{-eE(X)\tau {\cal I}}
  =N\,\frac{4^\nu}{\pi\alpha}\,\frac{\Gamma(\nu+\tfrac{eE(X)\tau}{\pi\alpha})\Gamma(\nu-\tfrac{eE(X)\tau}{\pi\alpha})}{\Gamma(2\nu)}\; .
\end{equation}
This expression provides the answer for all values of the background
field and all real fictitious times, thus solving the analytical
continuation problem that one would face when doing the integral over
${\cal I}$ by Monte-Carlo. The poles that are responsible for the
imaginary part of the effective action are given by the second gamma
function in the numerator,
\begin{equation}
  \tau_n=\frac{\pi\alpha(n+\nu)}{eE(X)}\; ,\quad\mbox{with residue\ \ }
      {\rm Res}\,\left(\big<W_{_X}\big>,\tau_n\right)
      =N\frac{4^\nu}{\pi\alpha}\frac{\Gamma(2\nu+n)}{\Gamma(2\nu)}
      \left.\frac{(-1)^n}{l!\frac{d}{d\tau}(\nu-\frac{eE(X)\tau}{\pi\alpha})}\right|_{\tau_n} .
\end{equation}

This algorithm has been applied in ref.~\cite{Gies:2005bz} to the
following $x^1$ dependent Sauter potential,
\begin{equation}
  A^0=-a\,\tanh(k x^1)\; ,\qquad E^1=\frac{ak}{\cosh^2(k x^1)}\; ,
  \label{eq:sauter-wl}
\end{equation}
whose known 1-loop analytical results serve as a reference for
checking the accuracy of the numerical approach. The parameter $k$ can
be varied in order to test the algorithm at various spatial scales.
\begin{figure}[htbp]
  \begin{center}
    \resizebox*{9.5cm}{!}{\includegraphics{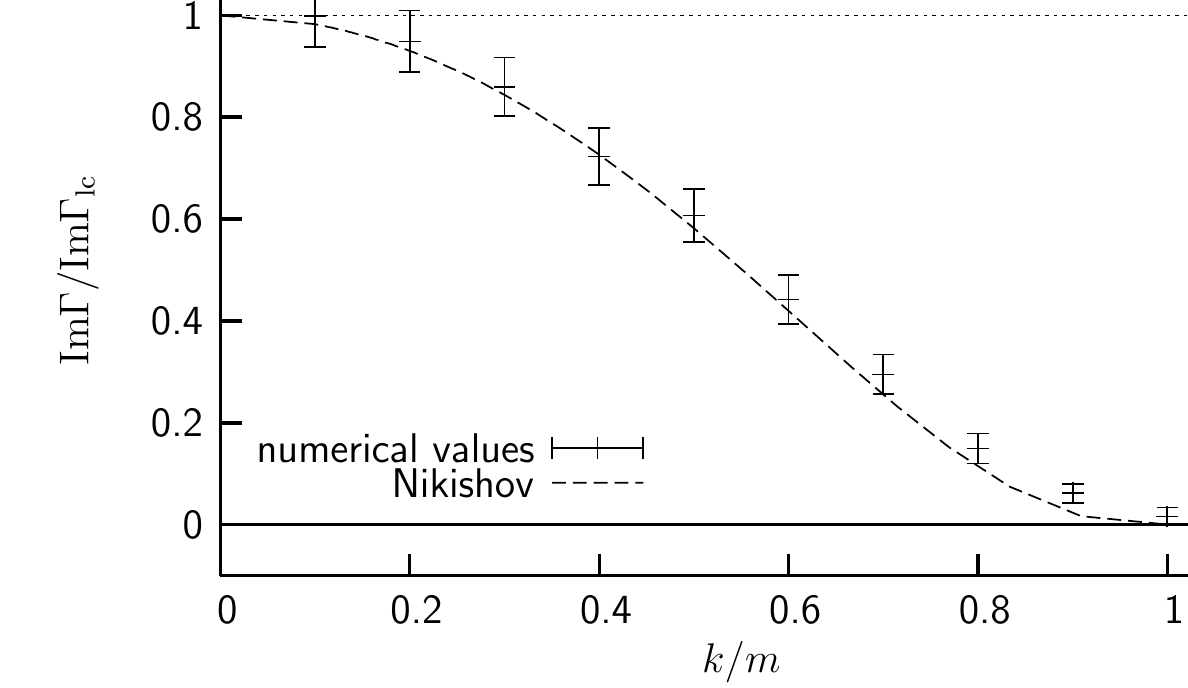}}
  \end{center}
  \caption{\label{fig:wl-sauter}Comparison of the exact 1-loop result
    for a Sauter potential of eq.~(\ref{eq:sauter-wl}) and the result
    of a numerical worldline computation based on the Ansatz of
    eq.~(\ref{eq:PI-ansatz}). From ref.~\cite{Gies:2005bz}.}
\end{figure}
The results of this test are shown in the figure \ref{fig:wl-sauter},
in which one displays the imaginary part of the effective action
normalized to the approximation of a locally constant field
(i.e. where one uses the result for a constant field with the local
value $E(X)$), as a function of the wavenumber $k$. The dashed curve
is the analytical result by Nikishov. One sees that, while the locally
constant field approximation quickly breaks down (the ratio starts
deviating from one at very small values of $k/m$), the worldline
numerical approach remains in good quantitative agreement with the
exact result up to rather large values of $k/m$. Note that the limit
$k/m\to 1$, where the wavelength of the external field equals the
Compton wavelength of the charged particles, is hard to cope with
because it requires the $e^+e^-$ pair to become extremely delocalized
in order to reach an energy sufficient to become on-shell (when $k=m$,
the particle yield is known to vanish with the Sauter background
field). Consequently, this limit is dominated by large worldlines,
whose sampling is difficult. Although imperfect, the ansatz of
eq.~(\ref{eq:PI-ansatz}) manages to capture the main behavior in this
limit.

\subsection{Lattice worldline formalism}
The algorithms described in the previous section do not require any
discretization of space-time, which is arguably the most natural setup
if the background field is known analytically.

In this subsection, we present an alternate formulation of the
worldline formalism where the (Euclidean) space-time is first
discretized on a cubic lattice \cite{Schmidt:2002yd,Schmidt:2002mt,Schmidt:2003bf,Epelbaum:2015vaa}.  Although it has not yet been used to
evaluate tunneling phenomena such as the Schwinger mechanism, such a
lattice version of the worldline formalism seems promising when
the background itself is obtained as the result of a lattice
computation. Let us call $a$ the lattice spacing (for simplicity, we
assume that the lattice has identical lattice spacings in all
directions, but it is easy to depart from this restriction).

On the lattice, it is more convenient to replace the representation of
eq.~(\ref{eq:S-prop}) for the inverse propagator by the following
discrete formula,
\begin{equation}
  \frac{2\tilde{d}}{a^2}\left(-D^i D^i+m^2\right)^{-1}
  =
  \sum_{n=0}^\infty\left(1-\frac{a^2\big(-D^i D^i+m^2\big)}{2\tilde{d}}\right)^n\; ,
\end{equation}
where we have defined $\tilde{d}\equiv d+\tfrac{1}{2}m^2 a^a$. Note
that this is an exact formula. Likewise, the logarithm can be written
as
\begin{equation}
  \ln\left(\frac{a^2}{2\tilde{d}}\,\big(-D^i D^i+m^2\big)\right)
  =
  -\sum_{n=1}^\infty\frac{1}{n}\,\left(1-\frac{a^2\big(-D^i D^i+m^2\big)}{2\tilde{d}}\right)^n\; .
\end{equation}

The trace that must be applied to the logarithm of the propagator is
best calculated in position space, as a sum over all the lattice sites,
\begin{equation}
{\rm Tr}\,{\cal O}=\sum_{{\rm sites\ }\x} \big<\x\big|{\cal O}\big|\x\big>\; .
\end{equation}
Let us define a probability distribution $P_0$ on the lattice,
localized at $\y=\x$, i.e. $P_0(\y,\x)\equiv \delta_{\x,\y}$. From
$P_0$, we can define a sequence of distributions $P_n$ defined
iteratively by
\begin{equation}
  P_{n+1}(\y,\x)
  =
  \sum_{{\rm sites\ }\z}
  \left(1-\frac{a^2\big(-D^i D^i+m^2\big)}{2\tilde{d}}\right)_{\y,\z}\,P_n(\z,\x)\; .
\end{equation}
In terms of these $P_n$, the trace of the logarithm of the inverse
propagator reads
\begin{equation}
  {\rm Tr}\,\ln\left(\frac{a^2}{2\tilde{d}}\,\big(-D^i D^i+m^2\big)\right)
  =
  -\sum_{n=1}^\infty \frac{1}{n}\sum_{{\rm sites\ }\x}P_n(\x,\x)\; .
\end{equation}

\begin{figure}[htbp]
  \begin{center}
    \resizebox*{7.5cm}{!}{\includegraphics{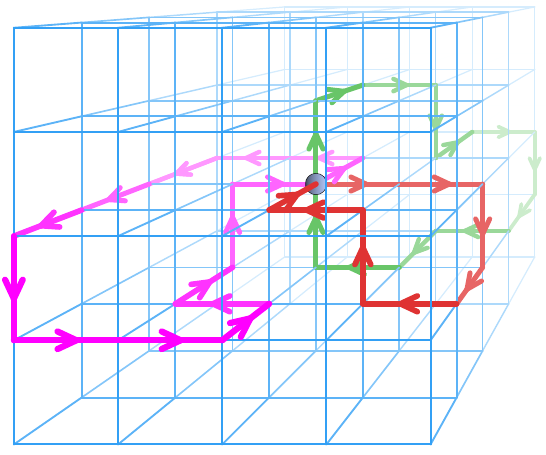}}
  \end{center}
  \caption{\label{fig:wl-lattice}Illustration of the lattice worldline
    formalism. The gray dot indicates the base point $\x$.}
\end{figure}

On the lattice, the covariant derivatives have a simple expression as
finite differences weighted by link variables that encode the
background gauge potential. For instance, for the direction $x^1$:
\begin{equation}
  [-D_1^2 f]_{i}
  =\frac{2f_{i}-U_{1,i}f_{i+1}-U_{1,i-1}^{-1}f_{i-1}}{a^2}\; ,
\end{equation}
where $U_{1,i}$ is the link variable in the direction $x^1$ from the
point $i$ to the point $i+1$ (therefore, $U_{1,i-1}^{-1}$ can be
viewed as a Wilson line oriented in the opposite direction, from the
point $i$ to the point $i-1$).  Using this definition, one obtains the
following explicit formula for $P_{n+1}$ in terms of $P_n$:
\begin{equation}
  P_{n+1}(i\cdots,\x)
  =
  \frac{1}{2\tilde{d}}\Big[
    U_{1,i\cdots} P_n(i+1\cdots,\x)
    +
    U^{-1}_{1,i-1\cdots} P_n(i-1\cdots,\x)
    +\cdots\Big]\; ,
\end{equation}
where only the first coordinate has been written explicitly. Thus,
$P_{n+1}(\y,\x)$ is obtained from the values of $P_n(\y',\x)$ on the
nearest neighbors $\y'$ of the point $\y$, with weights determined by
the link variables that start at the point $\y$ (divided by
$2\tilde{d}$). From this formula, one obtains a simple geometrical
interpretation of $P_{n}(\y,\x)$:
\begin{equation}
  P_n(\y,\x)
  =\frac{1}{(2\tilde{d})^n}\sum_{\gamma\in\Gamma_n(\y,\x)}\prod_{\ell\in\gamma}U_\ell\; ,
\end{equation}
where $\Gamma_n(\y,\x)$ is the set of all the paths of length $n$ on
the lattice that start at the point $\x$ and end at the point $\y$,
and $\prod_{\ell\in\gamma}U_\ell$ is the product of all the link
variables encountered along the path $\gamma$. Therefore, one arrives
at the following expression for the trace of the logarithm of the
inverse lattice propagator:
\begin{equation}
  {\rm Tr}\,\ln\big(-D^i D^i+m^2\big)=\mbox{const}
  -\sum_{n=0}^\infty \frac{1}{n}\,\frac{1}{(2\tilde{d})^n}\;
  \sum_{{\rm sites\ }\x}\;\sum_{\gamma\in\Gamma_n(\x,\x)} \prod_{\ell\in\gamma}U_\ell
  \; .
  \label{eq:wl-latt}
\end{equation}
The sum is over all the \emph{closed} paths on the lattice, and the
products of the link variables along such a path forms a Wilson
loop\footnote{The formula written here is for scalar QED. In the case
  of an SU($N_c$) gauge theory with scalars in the adjoint
  representations, the Wilson loop is an element of the adjoint
  representation of the SU($N_c$) algebra, that must be traced in
  order to obtain the effective action.}, which is the discrete
analogue of the Wilson loop encountered in the continuous version of
the worldline formalism exposed earlier in this section. A few of
these closed paths are shown in the illustration of the figure
\ref{fig:wl-lattice}.

Note that the number of hops $n$ must be even for closed paths on a
cubic lattice, $n=2m$. The total number of paths of length $n$ on a
$d$-dimensional lattice is $(2d)^{n}$. The mass introduces an
exponential suppression of the weight of long paths in the sum over
the length $n$ in eq.~(\ref{eq:wl-latt}), since we can bound the
summand by $(d/\tilde{d})^n=(1+\tfrac{m^2a^2}{2d})^{-n}$. The number
of \emph{closed} paths of length $n=2m$ is given by the following formula
\begin{equation}
\sum_{\gamma\in\Gamma_{2m}(\x,\x)}1=\sum_{n_1+\cdots+n_d=m}\frac{(2m)!}{n_1!^2\cdots n_d!^2}\; ,
\end{equation}
and although it is considerably smaller than the total number of
paths, it grows too quickly for an exhaustive enumeration\footnote{In
  the simple case of a constant and uniform background electrical
  field in the direction $3$, the Wilson loop depends only on the area
  of the loop projected in the plane $34$.  One can then use some
  known results on the moments of the distribution of these areas
  \cite{Mingo199855,Epelbaum:2015vaa} over the set $\Gamma_n(\x,\x)$
  in order to recover the standard result on the Schwinger mechanism
  in a constant electrical field.} of all the closed paths to reach
the cutoff imposed by the mass. In
refs.~\cite{Schmidt:2002yd,Schmidt:2002mt,Schmidt:2003bf}, it was
proposed to depart from an exhaustive listing of the loops, and to use
instead a statistical sampling that spans a much larger range of
lengths. The ensemble of loops needs to be generated only once for a
given set of lattice parameters, and can be reused for any background
field on this lattice. Note however that, it may be advantageous to
choose the ensemble of loops according to the physical length scale
relevant for the problem under consideration (which in turn depends on
the strength of the background field).

\subsection{Worldline instanton approximation}
\label{subsec:instanton}
Let us finally discuss a semi-classical approximation\footnote{Despite
  technical differences, this approach bears a lot of resemblance with
  the WKB semi-classical approximation, see e.g.
  \cite{Brezin:1970xf,popov1972pair,Kleinert:2008sj,Strobel:2013vza,Strobel:2014tha}.}
of the worldline expression of the 1-loop effective action
\cite{Dunne:2005sx,Dunne:2006st,Dunne:2006ff,Dunne:2006ur,Dumlu:2011cc,Ilderton:2015lsa},
which bears some resemblance with instantons in the fictitious
time. For this reason, these semi-classical solutions are called
\emph{worldline instantons}.  The starting point is the worldline
representation of eq.~(\ref{eq:WL-V}) for the 1-loop effective
action. By defining $\tau'\equiv \tau u$ and $m^2\tau=s$, we can
firstly rewrite it as
\begin{equation}
  {\cal V}_{_E,\rm 1\ loop}=-\int_0^\infty\frac{ds}{s}\;
  e^{-s}\;
  \int\limits_{x^i(0)=x^i(1)}\big[{\cal D}x^i(u)\big]\;
  \exp\Big(-\Big(\frac{m^2}{4s}\int_0^1 du\;\frac{\dot{x}^2}{4}+ie\int_0^1du\;\dot{x}^i A^i(x)\Big)\Big)\; ,
  \label{eq:WL-V1}
\end{equation}
where the path integration is now over closed loops of period 1. Since
the rescaled fictitious time does not appear any longer as the
integration bound inside the exponential, the integral over $s$ can be
performed in closed form, yielding a Bessel function:
\begin{equation}
  {\cal V}_{_E,\rm 1\ loop}=-2\int\limits_{x^i(0)=x^i(1)}\big[{\cal D}x^i(u)\big]\;
  K_0\left(\sqrt{m\int_0^1du \;\dot{x}^2}\right)\;
  \exp\Big(-ie\int_0^1 du\;\dot{x}^i A^i(x)\Big)\; .
\end{equation}

In the regime where
\begin{equation}
  m^2\int_0^1du\,\dot{x}^2\gg 1\; ,
  \label{eq:wl-inst-cond}
\end{equation}
whose physical significance will be clarified later, it is possible to
replace the Bessel function by the first term of its asymptotic
expansion,
\begin{equation}
K_0(z)\empile{\approx}\over{z\gg 1}\sqrt{\frac{\pi}{2}}\;\frac{e^{-z}}{\sqrt{z}}\; .
\end{equation}
This approximation leads to
\begin{eqnarray}
  {\cal V}_{_E,\rm 1\ loop}&\approx&-\sqrt{\frac{2\pi}{m}}
  \int\limits_{x^i(0)=x^i(1)}\big[{\cal D}x^i(u)\big]\;
  \left({\int_0^1 du\; \dot{x}^2}\right)^{-1/4}\nonumber\\
  &&\qquad\qquad\times\,
  \exp\Big(-\Big(m\sqrt{\int_0^1 du\;\dot{x}^2}+ie\int_0^1 du\;\dot{x}^i A^i(x)\Big)\Big)\; .
\end{eqnarray}
The remaining path integral can be approximated by a stationary phase
approximation, if we define the action
\begin{equation}
{\cal S}\equiv m\sqrt{\int_0^1 du\;\dot{x}^2}+ie\int_0^1 du\;\dot{x}^i A^i(x)\; .
\end{equation}
The stationarity condition is fulfilled for a closed path $x^i(u)$
that satisfies
\begin{equation}
  m\frac{\ddot{x}^i}{\sqrt{\int_0^1du\;\dot{x}^2}}=ieF^{ij}\dot{x}^j\; ,
  \label{eq:wl-eom}
\end{equation}
where $F^{ij}$ is the strength of the background field. Since $F^{ij}$
must be evaluated at the point $x(u)$, this equation is a very
non-trivial set of ordinary differential equations. By contracting
eq.~(\ref{eq:wl-eom}) with $\dot{x}^i$, the right hand side vanishes
thanks to the antisymmetry of $F^{ij}$ and we therefore learn that
\begin{equation}
\dot{x}^i\dot{x}^i=\mbox{const}\equiv v^2\; .
\end{equation}
The condition of eq.~(\ref{eq:wl-inst-cond}) is thus equivalent to
$mv\gg 1$.  For each solution of the stationarity condition
(\ref{eq:wl-eom}), the corresponding extremal value of the action
gives a contribution to the imaginary part of the effective action
\begin{equation}
{\rm Im}\,{\cal V}_{_E,\rm 1\ loop}\sim e^{-{\cal S}_{\rm extremum}}\; .
\end{equation}
Note that a more complicated calculation, involving the calculation of
the determinant of the Gaussian fluctuations around these stationary
solutions, is required in order to determine the prefactor in front of
the exponential.

In order to illustrate this method, let us consider the case of a
spatially dependent Sauter field
\begin{equation}
E(x^3)\equiv \frac{E}{\cosh^2(k x^3)}\; ,
\end{equation}
that can be derived from the following gauge potential
\begin{equation}
A^4=-i\,\frac{E}{k}\,\tanh(k x^3)\; .
\end{equation}
The equations of motion for the stationary solutions are
\begin{equation}
  \dot{x}^3=v\sqrt{1-\gamma^{-2}\tanh^2(kx^3)}\;,\qquad
  \dot{x}^4=-\gamma^{-1}v\,\tanh(kx^3)\; ,
\end{equation}
where $\gamma\equiv mk/(eE)$. There is a countable infinity of closed
paths that satisfy these equations can be parameterized by
\begin{eqnarray}
  x^3(u)&=&\frac{m}{eE}\,\frac{1}{\gamma}\,{\rm arcsinh}\,\left(\frac{\gamma}{\sqrt{1-\gamma^2}}\,\sin(2\pi n\,u)\right)
  \nonumber\\
  x^4(u)&=& \frac{m}{eE}\,\frac{1}{\gamma\sqrt{1-\gamma^2}}\,{\rm arcsin}\,(\gamma\,\cos(2\pi n\, u))\; ,
  \label{eq:instanton-sauter}
\end{eqnarray}
where $n$ is a strictly positive integer and $u\in[0,1]$. The index
$n$ corresponds to how many times the closed path is traveled by the
instanton solution.

\begin{figure}[htbp]
  \begin{center}
\resizebox*{9cm}{!}{\includegraphics{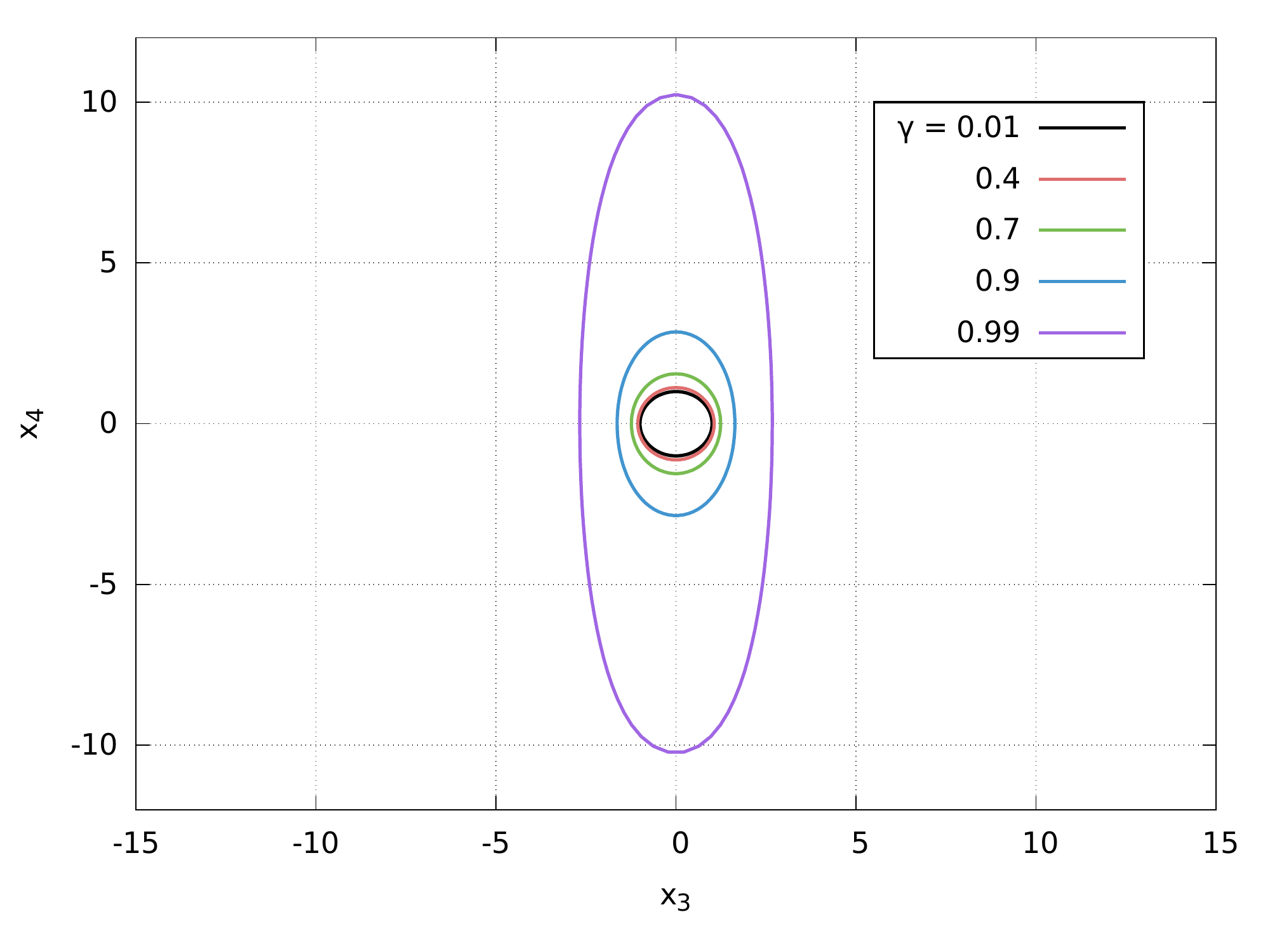}}
  \end{center}
  \caption{\label{fig:instanton}Closed paths in the $(x_3,x_4)$ plane
    of the instanton solutions in the case of an $x_3$-dependent
    Sauter electrical field, for various values of the parameter
    $\gamma\equiv mk/(eE)$. $\gamma=0$ corresponds to a circle of
    radius 1, and the closed paths become more and more elongated in
    the $x_4$ direction as $\gamma$ increases.}
\end{figure}

In the limit of an extended field, $\gamma\to 0$, these closed paths
become circular (one could have checked directly that the closed paths
that extremalize the action in the case of a constant and homogeneous
electrical field are indeed circles). In contrast, when $\gamma$
increases, these orbits become more and more elongated in the $x^4$
direction (see the figure \ref{fig:instanton}) and they become
infinitely large when $\gamma$ reaches unity (because of the prefactor
$1/\sqrt{1-\gamma^2}$ in the second of
eqs.~(\ref{eq:instanton-sauter})).
In this limit, the value of the
stationary action becomes infinite, and the rate of particle
production goes to zero. This result was of course known from the
exact solution to particle production problem in the Sauter
field. Physically, it reflects the fact that a field must exist over a
spatial domain as large as one Compton wavelength in order to be able
to produce particles. This simple academic example is an illustration
of a more general phenomenon, namely that spatial inhomogeneities tend
to decrease particle production (this reduction can become total when
the field becomes incoherent over scales comparable to the Compton
wavelength).

\section{Dynamically assisted Schwinger mechanism} \label{sec:dynamically}
\subsection{Comparison of perturbative and non-perturbative particle production} \label{subsec:perturbative}
As indicated by the form of the factor $\exp( -{\pi m^2}/{(eE)})$, the
Schwinger mechanism in a constant field is completely nonperturbative,
even in a weak field regime $eE \ll m^2$.  This is because the field
is time-independent, and therefore no process initiated by a finite
number of photons can produce the energy of a pair.  If the electrical
field has a time dependence, a perturbative component can also
contribute particle production \cite{Brezin:1970xf}.  The transition between the
perturbative regime and the non-perturbative regime can be
characterized by the following dimensionless parameter,
\begin{equation}
\gamma \equiv \frac{m}{eE \tau} \; ,
\end{equation} 
where $\tau$ is a characteristic time-scale of the time-dependence of
the electrical field.  This parameter has first been introduced by
Keldysh in the context of the ionization of atoms
\cite{keldysh1965ionization}, and is therefore called \emph{Keldysh
  parameter}\footnote{In refs.~\cite{Brezin:1970xf,Taya:2014taa}, the
  inverse of this parameter is defined as $\gamma$.  We follow the
  original definition by Keldysh.}.  In this subsection, by using an
analytically solvable time-dependent field, we explicitly show how the
transition between the perturbative and the nonperturbative regimes
happens \cite{Narozhnyi:1970uv,Taya:2014taa}.

Firstly, let us consider the distribution function obtained in a
perturbative expansion up to the first order in a general background
field.  Since the distribution function is expressed by the mode
function $\psi_{\p ,s}^{-} (x)$, we need to find the perturbative
solution of the equation for the mode function,
\begin{equation}
\left[ i\gamma^\mu D_\mu -m \right] \psi_{\p ,s}^{-} (x) = 0 \; . 
\end{equation}
It can be easily obtained from the Green's formula
\begin{equation} \label{eq:psi_Green0}
\psi (x) = \int_{y^0 =t_0} d^3 \y \; D_R^0 (x,y)\; \gamma^0\, \psi (y) 
 -ie \int_{y^0 >t_0} d^4 y \; D_R^0 (x,y)\; \slA (y) \,\psi (y) \; , 
\end{equation}
where $D_R^0 (x,y)$ is the free retarded fermion propagator:
\begin{equation}
D_R^0 (x,y) = \int \frac{d^4 q}{(2\pi)^4} \frac{i (\slq +m)}{q^2 -m^2 +iq^0 \epsilon} e^{-iq\cdot (x-y)} \; . 
\end{equation}
By taking the limit of $t_0 \to -\infty$ and using the boundary
condition \eqref{eq:mode_in}, one can see that the first term in the
right hand side of eq.~\eqref{eq:psi_Green0} is simply the free
spinor, propagated from $t_0$ to $t$.  Therefore, the Green's formula
also reads
\begin{equation} \label{eq:psi_Green}
\psi_{\p ,s}^{-} (x) = \psi_{\p ,s}^{\text{free} -} (x)
 -ie \int d^4 y \; D_R^0 (x,y)\; \slA (y)\, \psi_{\p ,s}^{-} (y) \; .
\end{equation}
This equation has an iterative solution
\begin{equation} \label{eq:pert_mode}
\psi_{\p ,s}^{-} (x)
 = \sum_{n=0}^\infty \psi_{\p ,s}^{(n) -} (x) \; , 
\end{equation}
where
\begin{equation}
\psi_{\p ,s}^{(n) -} (x) 
= (-ie)^n \int d^4 y_1 \cdots \int d^4 y_n \;
D_R^0 (x,y_1)\; \slA (y_1 ) \cdots
    D_R^0 (y_{n-1} ,y_n) \;\slA (y_n )\,  \psi_{\p ,s}^{\text{free} -} (y_n ) \; . 
\end{equation}
This formula describes the correction to the spinor due to $n$
scatterings off the external field.

From the iterative solution~\eqref{eq:pert_mode}, we can easily obtain
the spectrum \eqref{eq:in-out_spectrum1} at the first non-zero
order in $e$.  At this order, the gauge rotation factor $U(x)$ has no
effect (it alters the spectrum only in higher orders).  Therefore the
spin-averaged spectrum at the lowest-order reads
\begin{eqnarray}
\frac{dN^{(1)}}{d^3 \p} 
&=& \frac{e^2}{(2\pi)^3 2E_\p} 
 \frac{1}{2} \sum_{s,s^\prime} \int \frac{d^3\p^\prime }{(2\pi)^3 2E_{\p^\prime}} 
 \nonumber\\
 &&\times\,
 \lim_{x^0 \to +\infty}
 \Big| \bar{u} (\p ,s) \int \frac{dq^0}{2\pi}\; \frac{e^{-iq^0 x^0} }{q^0 -E_\p +i\epsilon} \, 
 \tilde{\slA} (q^0 +E_{\p^\prime} , \p +\p^\prime ) 
 v(\p^\prime ,s^\prime ) \Big|^2 \; , 
\end{eqnarray}
where we have introduced
\begin{equation}
\tilde{A}^\mu (p) \equiv \int d^4 x \; A^\mu (x)\; e^{ip \cdot x} \; . 
\end{equation}
This equation corresponds to computing the following 1-loop diagram:
\begin{center}
 \resizebox*{4cm}{!}{\includegraphics{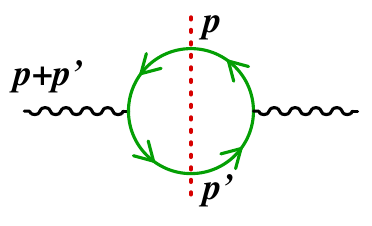}}
\end{center}
If the Fourier-transformed gauge field $\tilde{A} (q^0 +E_{\p^\prime}
,\p +\p^\prime )$ contains no pole in the $q^0$-plane and is bounded
when ${\rm Im}\,(q^0) \to -\infty$, the $q^0$-integration can be
performed by the theorem of residues by picking the pole of the
prefactor $(q_0-E_p+i\epsilon)^{-1}$ (note that the exponential factor
forces us to close the contour in lower half of the $q_0$ plane).
After summing over the initial and final spin states, we have
\begin{equation} \label{eq:LOspec}
\begin{split}
\frac{dN^{(1)}}{d^3 \p} 
&= \frac{2e^2}{(2\pi)^3 2E_\p} 
      \int \frac{d^3\p^\prime }{(2\pi)^3 2E_{\p^\prime}}\; 
      \left[ p^\mu p^{\prime \, \nu} +p^\nu p^{\prime \, \mu} -(p\cd p^\prime +m^2) g^{\mu \nu}\right] \\
&\hspace{10pt} \times 
      \tilde{A}_\mu (E_\p +E_{\p^\prime} , \p +\p^\prime )
      \tilde{A}_\nu^* (E_\p +E_{\p^\prime} , \p +\p^\prime ) \; ,
\end{split}
\end{equation}
which is gauge invariant.

As an example, let us consider a sinusoidal time-dependent electrical
field, $E_z (t) = E\,\cos (\omega t)$, for which the occupation number
of the produced electrons is
\begin{equation}
\frac{(2\pi)^3}{V} \frac{dN^{(1)}}{d^3 \p} = \frac{e^2 E^2}{16
  E_\p^2}\; \left[ 1-\left( \frac{p_z}{E_\p} \right)^2 \right]\; 2\pi
\delta (\omega -2E_\p )\, T \; ,
\end{equation}
where $T\equiv2\pi \delta (0)$ stands for the time duration of the
electrical field. The proportionality of the result to this time $T$
is nothing but a manifestation of \emph{Fermi's golden rule}.  This
lowest-order spectrum corresponds to the process $\gamma\to e^+e^-$,
where a single photon decays into an electron-positron pair.  From the
delta function, we see that this production mechanism is possible only
if the frequency $\omega$ is above the threshold $2m$, as expected
from energy momentum conservation.

To compare the lowest-order perturbative result with the
non-perturbative all-order result, we use the Sauter-type pulsed field
of eq.~\eqref{eq:Sauter_E}, which can be obtained from the gauge
potential
\begin{equation} \label{eq:Sauter_gauge_das}
A^3 (t) = E\,\tau \tanh\left(\frac{ t}{\tau}\right) \; . 
\end{equation}
The photons that make up this electrical field typically carry an
energy $\omega \sim \tau^{-1}$.  In this gauge field background, the
lowest-order perturbative spectrum reads
\begin{equation} \label{eq:Sauter_spec_pert}
\frac{(2\pi)^3}{V} \frac{dN^{(1)}}{d^3 \p} 
 = e^2 E^2\; \left[ 1-\left( \frac{p_z}{E_\p} \right)^2 \right]\;
    \frac{\pi^2 \tau^4}{\sinh^2 (\pi E_\p \tau )} \; .
\end{equation}
Because the gauge field \eqref{eq:Sauter_gauge_das} has a continuous
spectrum in Fourier space, the perturbative production does not
display any threshold behavior.  Unlike with the sinusoidal electrical
field, the particle production happens for any value of the pulse
characteristic timescale $\tau$.  However, the particle production is
most efficient if the photon energy is near the energy of the
produced pair, $2E_\p$.  Indeed, as a function of $\tau$, the spectrum
has a peak around $E_\p \tau \simeq 0.61$.

The exact result for the spectrum in the same background, that
includes the interaction with the background gauge field to all
orders, is given by eq.~\eqref{eq:Sauter_spec}.  When changing the
pulse duration $\tau$, this spectrum has a clear transition
between the perturbative regime and the non-perturbative regime
\cite{Taya:2014taa}.  Since we have three
dimensionful quantities, $m$, $eE$, and $\tau$, the system is governed
by two dimensionless parameters,
\begin{equation}
\gamma \equiv \frac{m}{eE \tau} \; , \qquad \lambda \equiv eE\tau^2 \; . 
\end{equation} 
Taking the limit $\tau \to +\infty$ (limit of constant electrical
field) while keeping $m$ and $eE$ constant corresponds to $\gamma \ll
1$ and $\lambda \gg 1 $.  One can confirm that the exact spectrum
\eqref{eq:Sauter_spec} goes to the spectrum due to a constant
electrical field in this limit\footnote{When taking the limit, one
  needs to keep the canonical momentum at the final time $p_z -eE\tau
  $ constant, in order to obtain the result \eqref{eq:long_limit},
  which is independent of the longitudinal momentum.}:
\begin{equation} \label{eq:long_limit}
  \frac{(2\pi)^3}{V} \frac{dN}{d^3 \p}\qquad
  \xrightarrow[\gamma \ll 1,\ \lambda \gg 1]{}
  \qquad\exp \left( -\frac{\pi m_\perp^2}{eE} \right) \; . 
\end{equation}

\begin{figure}[htbp]
  \begin{center}
    \resizebox*{8.5cm}{!}{\includegraphics{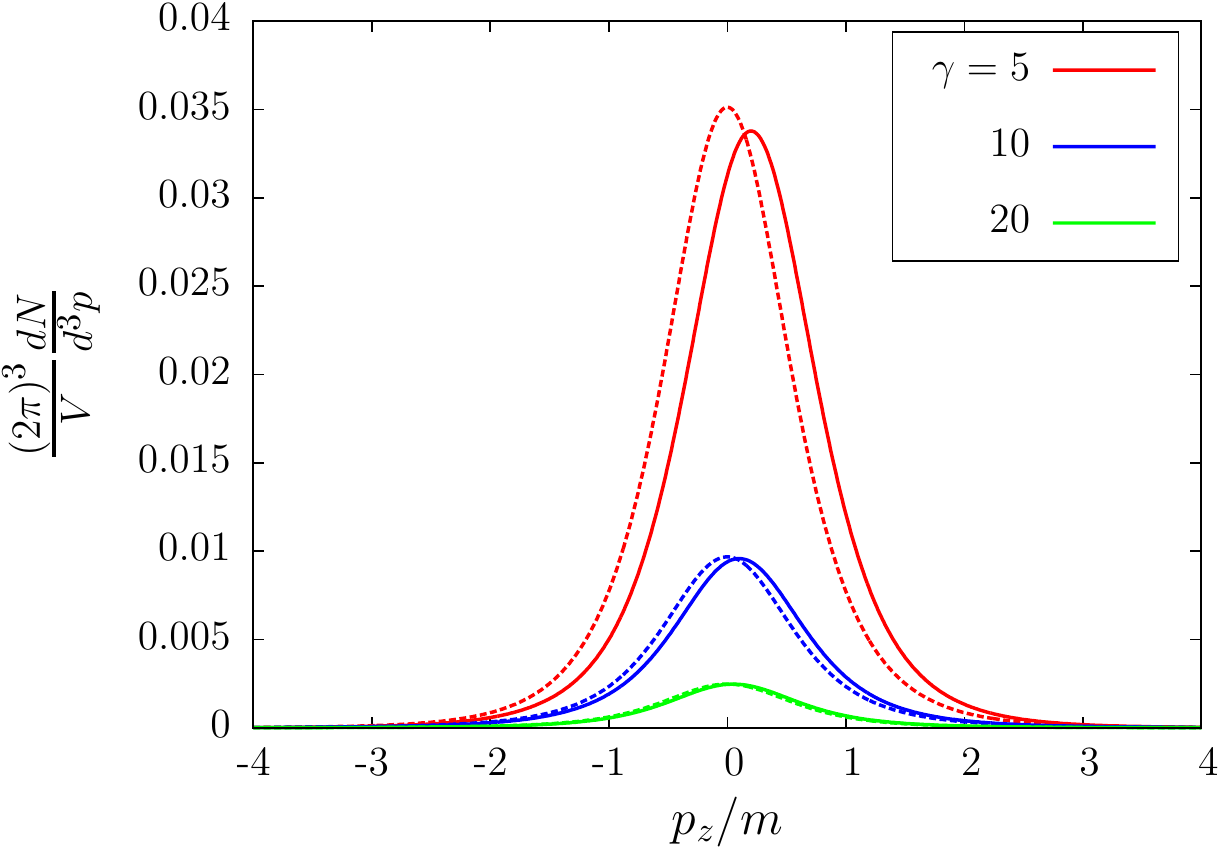}}
  \end{center}
  \caption{\label{fig:pert_comp1} $p_z$-dependence of the spectrum at
    $p_\perp=0$.  The all-order exact result \eqref{eq:Sauter_spec}
    (solid lines) and the lowest-order perturbative result
    \eqref{eq:Sauter_spec_pert} (dashed lines) are compared for
    several values of the Keldysh parameter $\gamma$.  The ratio
    $eE/m^2 =1$ is fixed.  For $\gamma=20$, the two lines are almost
    undistinguishable.}
\end{figure}

At the other extreme, let us consider the short pulse limit, $\tau \to
+\infty$, while keeping $m$ and $eE$ fixed.  In terms of the
dimensionless parameters $\gamma$ and $\lambda$, this limit
corresponds to $\gamma \gg 1 $ and $\lambda \ll 1 $.  Note that taking
the limit $eE\to 0$ while keeping $m$ and $\tau$ fixed would also
provide the same behavior of $\gamma$ and $\lambda$.  Therefore, we
expect that the lowest-order perturbative result
\eqref{eq:Sauter_spec_pert} is recovered by this limit.  Indeed, one
can confirm that
\begin{equation} \label{eq:short_limit}
  \frac{(2\pi)^3}{V} \frac{dN}{d^3 \p} \qquad
  \xrightarrow[\gamma \gg 1,\ \lambda \ll 1]{} \qquad
 e^2 E^2 \left[ 1-\left( \frac{p_z}{E_\p} \right)^2 \right]
    \frac{\pi^2 \tau^4}{\sinh^2 (\pi E_\p \tau )} \; . 
\end{equation}
In order to illustrate this behavior, we plot the exact result
\eqref{eq:Sauter_spec} and the lowest-order result
\eqref{eq:Sauter_spec_pert} for several values of the Keldysh
parameter in the figure~\ref{fig:pert_comp1}.  The $p_z$-distributions
at $p_\perp=0$ are shown in this plot.  For large enough Keldysh
parameter ($\gamma\gtrsim 20$), the two results show a good agreement.
Conversely, as $\gamma$ gets smaller, the discrepancy becomes larger.

\begin{figure}[htbp]
\begin{center}
  \resizebox*{8.5cm}{!}{\includegraphics{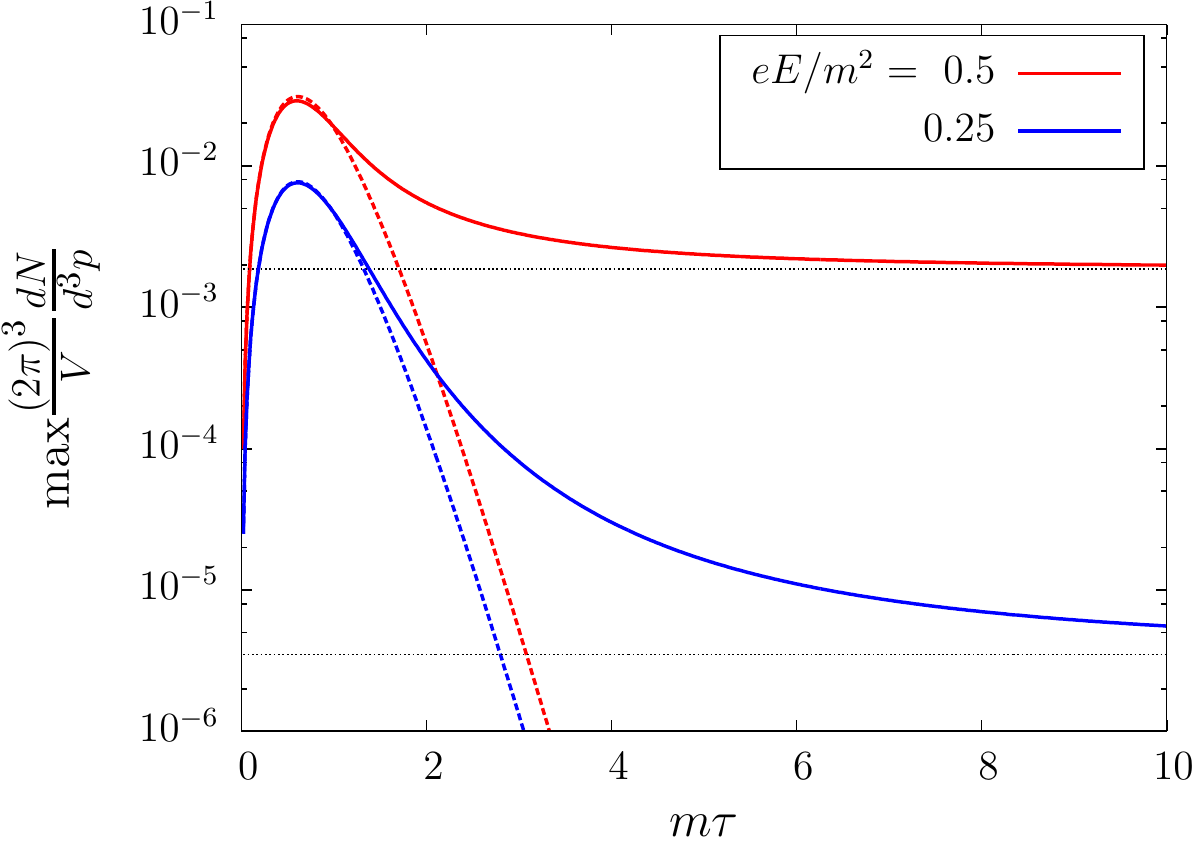}}
\end{center}
\vskip 4mm
\begin{center}
\resizebox*{8.5cm}{!}{\includegraphics{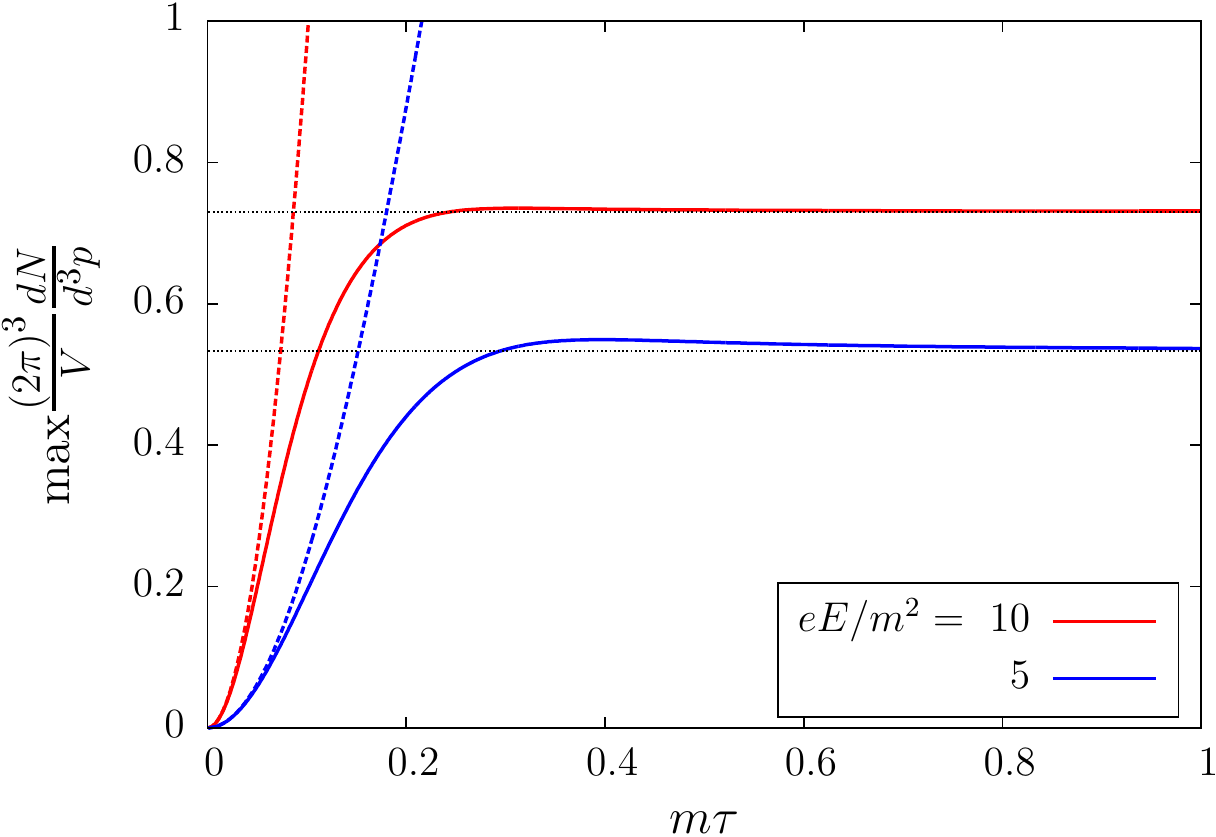}}
\end{center}
\caption{\label{fig:pert_comp2} Plots of the maximum value of the
  momentum spectrum as a function of $m\tau$ for several values of
  $eE/m^2$.  Top: subcritical field strength $eE/m^2 <1$. Bottom:
  supercritical field strength $eE/m^2 > 1$.  The all-order exact
  result \eqref{eq:Sauter_spec} (solid lines) and the lowest-order
  perturbative result \eqref{eq:Sauter_spec_pert} (dashed lines) are
  compared.  The constant field Schwinger result $\exp(
  -{\pi m^2}/(eE))$ is plotted as thin black lines. }
\end{figure}

The transition between the perturbative regime and the
non-perturbative regime is further illustrated in the
figure~\ref{fig:pert_comp2}.  The maximum values of the momentum
spectra (located at $\p =0$ for the perturbative spectrum, and at $\p
=(0,0,eE\tau)$ for the all-order one) are compared as a function of
$m\tau$, for several values of $eE/m^2$.  The constant field Schwinger
result $\exp ( -{\pi m^2}/{(eE)})$ is also indicated by thin black
lines for reference.  The top figure shows the subcritical case
$eE/m^2 <1$, and the bottom figure shows the supercritical case $eE/m^2
>1$.  For small $\tau$ (large $\gamma$), the all-order result and the
lowest-order one agree well.  For large $\tau$ (small $\gamma$), the
all-order result converges to the constant field Schwinger result,
while the perturbative computation breaks down.
What is remarkable in the left plot (subcritical case) is that the
spectrum can be several orders of magnitude larger than the constant
field Schwinger result in the regime $m\tau \sim 1$.  This means that,
in subcritical electrical fields, $eE/m^2 <1$, a time-dependence of
the electrical field dramatically amplifies the particle production,
especially if the typical energy carried by the time-dependent
electrical field is near the perturbative threshold $2m$.

\subsection{Lattice numerical results} \label{subsec:das_numerical}
In the previous subsection, we have observed that the particle
production in a time-de\-pen\-dent electrical field can be described by
the lowest order perturbative calculation if the time-dependence is
fast enough, such that $\gamma \gg 1$ and $\lambda \ll 1$.  
An important observation was
that the particle production by the perturbative process $\gamma\to
e^+e^-$ is much stronger than the non-perturbative Schwinger result if
the field is subcritical $eE < m^2$ and the typical frequency of the
time-dependent electrical field is near the threshold energy $2m$.

However, if the particle production is a purely perturbative process,
it is not much interesting since perturbative processes are well
understood theoretically and have been verified experimentally in many
circumstances.  Our main interest in this review is the particle
production in the non-perturbative regime.  As we shall see, a similar
enhancement mechanism is realized also in the non-perturbative regime
if the electrical field is a superposition of a slowly varying and
strong (but still subcritical) field and a fast and weak field
\cite{Schutzhold:2008pz}.  This phenomenon, called \emph{dynamically
  assisted Schwinger mechanism}, has been the subject of many recent
studies \cite{DiPiazza:2009py,Dunne:2009gi,Monin:2010qj,Orthaber:2011cm,%
Fey:2011if,Jansen:2013dea,Augustin:2014xga,Otto:2014ssa}, 
since it opens up the possibility that the non-perturbative 
electron-positron pair production may be observed more easily than 
expected in experiments \cite{DiPiazza:2011tq,Otto:2015gla}.

\begin{figure}[htbp]
\begin{center}
\resizebox*{8cm}{!}{\includegraphics{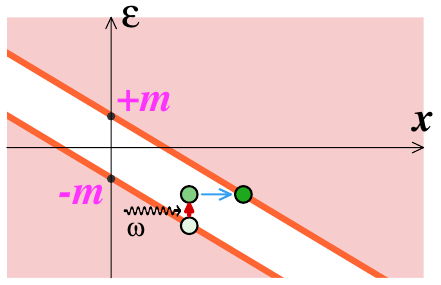}}
\end{center}
\caption{\label{fig:tunneling}
 Schematic illustration of the dynamically assisted Schwinger mechanism. }
\end{figure}

To be specific, we consider a superposition of two Sauter
time-dependent electrical fields
\begin{equation} \label{eq:double_pulse}
  E_z (t) = \frac{E_1}{\cosh^2\big(\frac{t}{\tau_1}\big)} +
  \frac{E_2}{\cosh^2 \big(\frac{t}{\tau_2}\big)} \; ,
\end{equation}
where $E_1 \gg E_2 $ and $\tau_1 \gg \tau_2$ so that the first term
represents a strong and slow field, while the second is a fast and
weak field.  In terms of the Keldysh parameter, these parameters are
chosen so that $\gamma_1 =\frac{m}{eE_1 \tau_1} \ll 1$ and $\gamma_2
=\frac{m}{eE_2 \tau_2} \gg 1$.  Therefore, the first pulse causes only
non-perturbative particle production, while the second field can
produce particles perturbatively.  By superposing these two fields
that have very different scales, an interplay between the
non-perturbative and the perturbative physics occurs, leading to an
important enhancement of the particle production.

Before presenting numerical evidence of this effect, let us 
qualitatively explain the mechanism at play. This is illustrated in
the figure \ref{fig:tunneling} (to be compared with the figure
\ref{fig:S-tunnel}). In the standard Schwinger effect, a hole
excitation must tunnel through the gap that separates the Dirac sea
from the band of positive energy states, and there is therefore an
exponential suppression in the length of this gap. As cartooned in the
figure \ref{fig:tunneling}, a photon from the short weak field $E_2$
can slightly raise the energy of a hole excitation. But this effect
alone is too weak to reach the positive energy band and produce an
on-shell electron-positron pair. However, once it has been brought to
this slightly higher energy, the hole excitation also has a shorter
length to tunnel through in order to reach the positive energy band by
quantum tunneling. Since the tunneling probability is exponentially
sensitive to this length, even a moderate change in the length can
produce an important increase in the yield. This is the essence of the
dynamically assisted Schwinger mechanism.

\begin{figure}[htbp]
  \begin{center}
    \resizebox*{10cm}{!}{\includegraphics{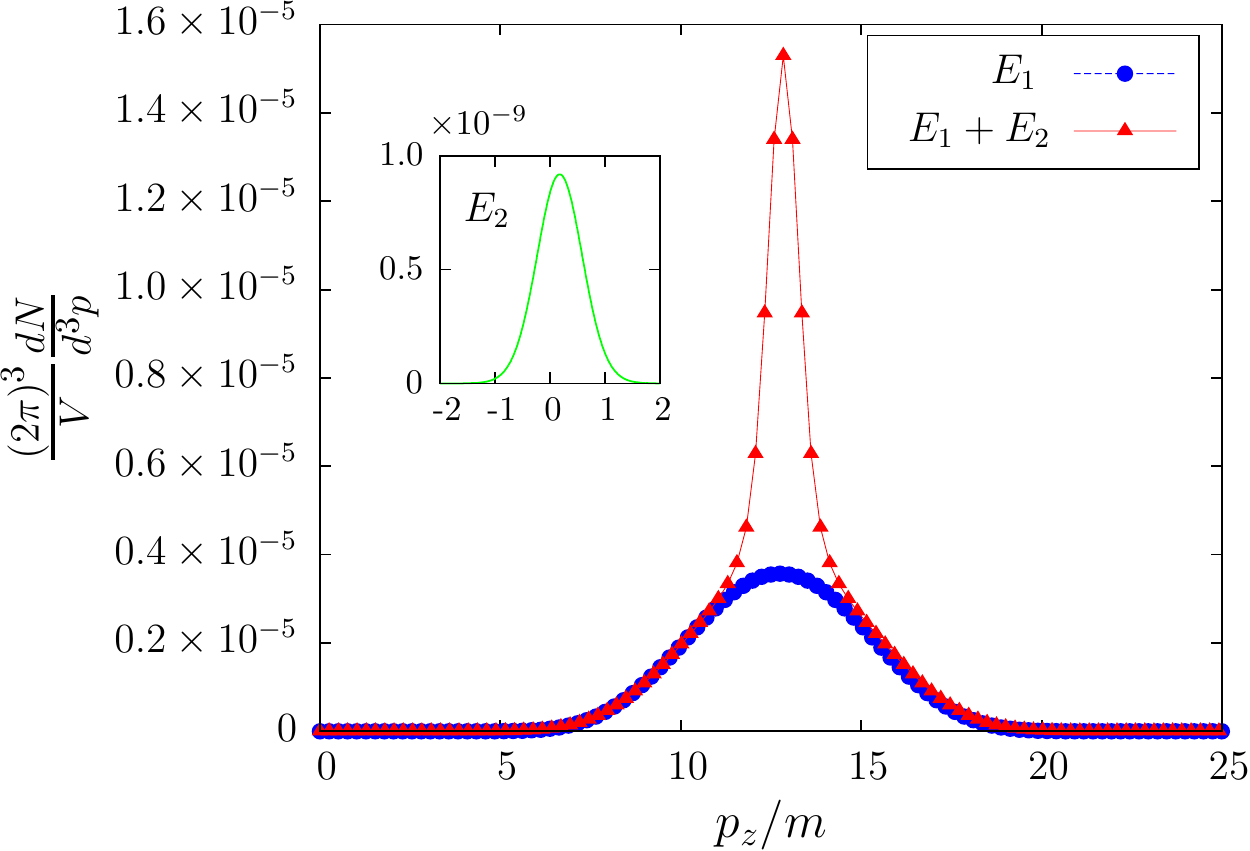}}
  \end{center}
  \caption{\label{fig:das1} The momentum spectrum of electrons
    produced by the field \eqref{eq:double_pulse} with the parameters
    $eE_1 = 0.25\,m_e^2$, $\tau_1 =10^{-4} \ \text{eV}^{-1}$ and $eE_2
    = 0.025\,m_e^2$, $\tau_2 =7\times 10^{-6} \ \text{eV}^{-1}$.
    $p_z$-dependence at $p_\perp =0$ is shown. The curve labeled
    $E_1+E_2$ shows the particle spectrum resulting from the sum of
    the two fields, while the curves labeled $E_1$ and $E_2$ show the
    particle spectrum that results from the two fields considered
    separately.}
\end{figure}

Following ref.~\cite{Orthaber:2011cm}, we show the momentum spectra of
particles produced in the electrical field \eqref{eq:double_pulse}.
While the quantum kinetic approach was used in
ref.~\cite{Orthaber:2011cm}, we present the same results computed by
real-time lattice calculations in the mode function method.  In the
figure~\ref{fig:das1}, the momentum spectrum with the parameters $eE_1
= 0.25\,m_e^2$, $\tau_1 =10^{-4} \ \text{eV}^{-1} =510/m_e$ and $eE_2
= 0.025\,m_e^2$, $\tau_2 =7\times 10^{-6} \ \text{eV}^{-1} =3.57/m_e$
is plotted.  The Keldysh parameters for each pulse are $\gamma_1
\simeq 0.078$, and $\gamma_2 \simeq 11.2$, respectively.  Therefore,
the first pulse is in the non-perturbative regime, while the second is
in the perturbative regime.  For comparison, the spectrum produced by
the strong and slow pulse (curve labeled $E_1$) alone and that by the
single weak and fast pulse (curve labeled $E_2$) alone are also
shown.  Although the single pulse $E_2$ is extremely weak and the
spectrum it produces by itself is several orders of magnitude smaller
than the spectrum produced by the pulse $E_1$, superposing this weak
and fast fields causes a dramatic enhancement of the production
yield, by a factor around $4$.

\begin{figure}[htbp]
  \begin{center}
    \resizebox*{10cm}{!}{\includegraphics{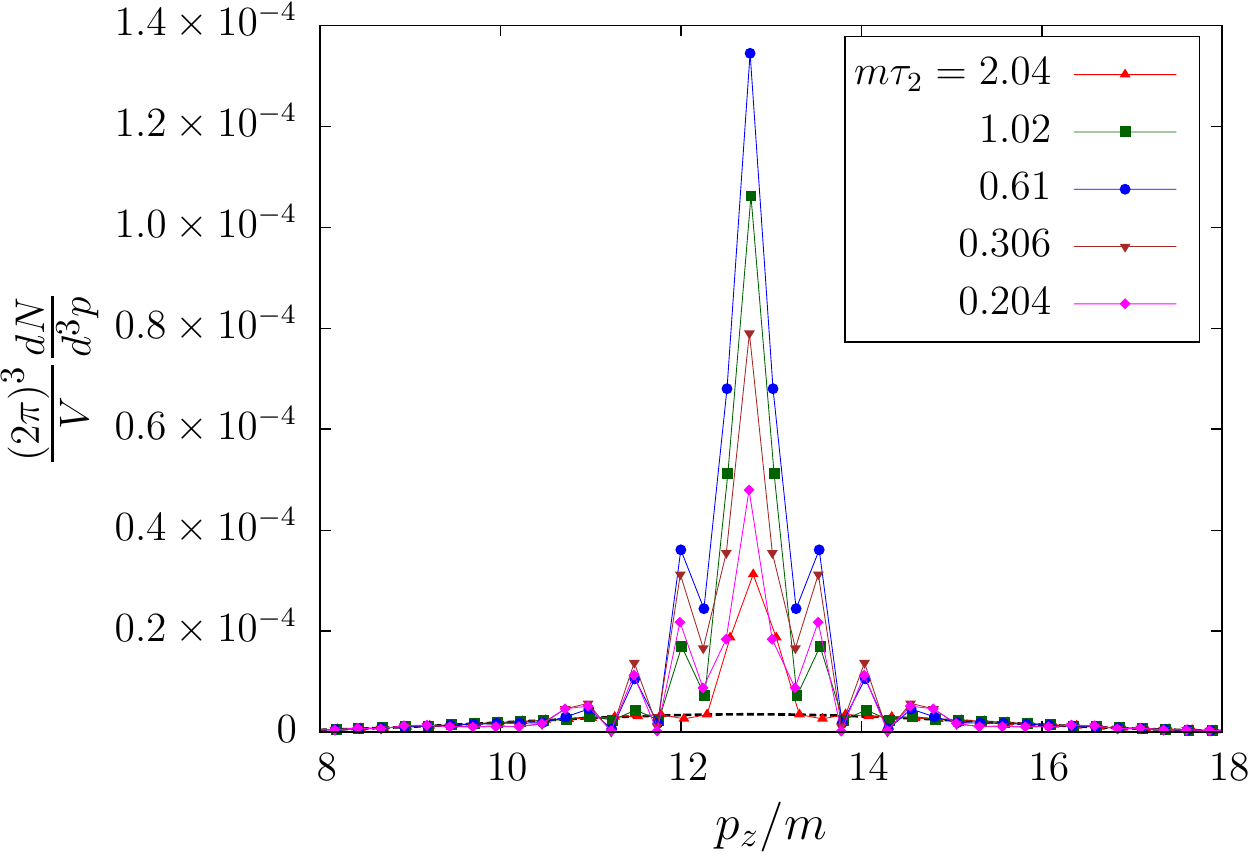}}
  \end{center}
  \caption{\label{fig:das2} The momentum spectrum of electrons
    produced by the field \eqref{eq:double_pulse} for several values
    of $m_e \tau_2$.  The other parameters are fixed to $eE_1 =
    0.25\,m_e^2$, $m_e \tau_1 =510$ and $eE_2 = 0.025\,m_e^2$.  The result
    with the single pulse $E_1$ is plotted as a black dashed line for
    comparison.  The lattice parameters are the same as those in
    the figure \ref{fig:das1}. }
\end{figure}

In the lowest-order perturbative production studied in the previous
subsection, the particle production is the most intense when the
typical energy carried by photons constituting the electrical field is
near to the threshold value $2m$.  In the case of the single
Sauter-type pulse, the particle yield by the lowest-order production
takes maximum value when $m\tau \simeq 0.61$.  Also in the case of the
dynamically assisted Schwinger mechanism in the double pulse
\eqref{eq:double_pulse}, the maximum enhancement occurs when the
typical energy carried by the weak pulse is near to $2m$.  In
Fig.~\ref{fig:das2}, the $p_z$-spectra are plotted for several values
of $m_e \tau_2$.  Other parameters are fixed to $eE_1 = 0.25\,m_e^2$,
$m_e \tau_1 =510$, and $eE_2 = 0.025\,m_e^2$.  Within this parameter
set, the enhancement is the most prominent at $m_e \tau_2 \simeq
0.61$.  Of course, the degree of the enhancement largely depends not
only on $\tau_2$ but also on other parameters and the profile of the
electrical field as well.  Optimization of the field profile to gain
the maximum particle yield has been investigated in 
refs.~\cite{Kohlfurst:2012rb,Li:2014psw,Hebenstreit:2014lra}.
 
\section{Summary and conclusions}
\label{sec:sum}
In this review, we have covered some of the recent works related to
the study of the Schwinger mechanism. Our main motivation in this
review has been to present methods that have been designed in order to
cope with more general backgrounds than those for which exact
analytical solutions exist\footnote{Besides the well known analytical
  solutions with constant and Sauter-type fields, solutions have been
  obtained for more complicated setups in $1+1$-dim massless QED,
  where the fermion fields can be elimnated by a bosonization
  procedure \cite{Chu:2010xc}.}, allowing (numerical) studies of pair production
in backgrounds with arbitrary space-time dependences (for smooth
enough dependences, one may instead use a derivative expansion of the
effective action \cite{Gusynin:1998bt,Gusynin:1995bc}).

Two main classes of methods have been discussed in this review. The
first one, that follows more conventional techniques of quantum field
theory, is based on the possibility of expressing any one-loop
amplitude in a background field in terms of \emph{mode
  functions}. Loosely speaking, such one-loop amplitudes can be
generated by small Gaussian fluctuations around a given background,
and are therefore fully determined by the knowledge of a basis of
small field perturbations about this background. This approach has a
number of variants, that differ in the formulation but share common
origins and are rigorously equivalent. The method of Bogoliubov
transformations is directly connected to the mode functions by the
fact that the Bogoliubov coefficients are obtained from the
decomposition of the mode functions in Fourier modes. For spatially
homogeneous electrical fields, it is possible to express the time
evolution directly at the level of the Bogoliubov coefficients
themselves, via quantum kinetic equations. These can be formulated
either as two local in time equations for a pair of coupled
distributions, one of which is the usual occupation number and the
other is an anomalous distribution, or as a single equation, non-local
in time, for the occupation number. When the background field is
inhomogeneous, quantum kinetic equations can be generalized into an
equation for the Wigner distribution of the system (i.e. the Wigner
transform of the density operator).

The second family of methods that we have discussed in this review
revolves around the \emph{worldline formalism}, a formulation in which
one-loop amplitudes in a background field are expressed as a path
integration over all the closed paths in Euclidean space-time,
parameterized by a extra fictitious time. Although it was originally
derived from ideas in string theory, this representation has also some
connections with Schwinger's proper time formulation for propagators.
For a constant and homogeneous background, this approach leads easily
to the well known analytical one-loop answer. In other cases, such as
the Sauter background electrical field, one can obtain approximations
valid in the weak field regime by finding extrema of a 5-dimensional
action, called worldline instantons due to their resemblance with
ordinary instanton solutions. For even more general backgrounds,
numerical methods have been developed, in which one samples
statistically the ensemble of worldlines. Although completely
equivalent to more conventional methods, the worldline formalism
emphasizes the space-time development of the production process, and
brings some useful intuition over the relevant phenomena (e.g. the
fact that in weak fields large worldlines play a very important
role).

Besides these developments in the technical tools available for
theoretical studies of the Schwinger mechanism, a number of works have
focused on finding the type of external fields that may maximize the
yield of produced particles, following the simple observation that the
superposition of two fields results in a particle spectrum which is
not simply the sum of the spectra that these two fields would produce
individually (because the Schwinger mechanism is highly non-linear in
the field). A simple yet spectacular such effect is the
\emph{dynamically assisted Schwinger mechanism}, obtained by
superimposing a slow but intense field with a fast and much weaker
field, both of which are much lower than the critical
field. Intuitively, the short pulse raises the energy of a hole
excitation, thereby shortening the tunneling length it has to overcome
in order to be produced on-shell and considerably enhancing the
yield. These phenomena could be used in order to greatly facilitate
the experimental production of pairs by the Schwinger mechanism.

Let us finish this section by listing several topics related to the
Schwinger mechanism that were not discussed in any detail in this
review. Firstly, all the discussion of pair production by external
fields in electrodynamics can be carried through to strong
interactions in quantum chromodynamics. The most direct extension
obviously concerns quark production in a chromo-electrical field
\cite{Cox:1985bu,Nayak:2005pf,Tanji:2010eu,Iwazaki:2011is,Gyulassy:1986jq},
but the Schwinger mechanism can also be relevant for gluon production
\cite{Yildiz:1979vv,Ambjorn:1982bp,Nayak:2005yv,Tanji:2011di,Iwazaki:2012xi}.
In heavy-ion collisions at high energies, the system is approximately
boost-invariant in the longitudinal beam direction. The
Schwinger mechanism in such a boost invariant expanding geometry has
been studied in refs.~\cite{Cooper:1992hw,Tanji:2010ui}. In this
review, we have only considered the case where the system is initially
in the vacuum state. This can be generalized to situations where the
system is initially non-empty, such as a thermal system
\cite{Gies:1999vb,Hallin:1994ad,Gavrilov:2006jb,Kim:2007ra,Kim:2008em,Kleinert:2012bu},
where quantum statistical effects will alter the production rate of
particles by the external field. In QCD, it has also been argued that
the Schwinger mechanism may lead to the dynamical generation of a
gluon mass, that could explain some features of the Landau gauge
propagators and verticein the soft sector \cite{Aguilar:2011xe}.

Besides the production of elementary particles by external fields in
vacuum, the Schwinger mechanism can also lead to the production of
quasi-particle excitations in more exotic materials such as graphene
\cite{Allor:2007ei,Zubkov:2012ht,Gavrilov:2012jk,Klimchitskaya:2013fpa,Fillion-Gourdeau:2015dga},
where it may be easier to achieve experimentally. The possibility of
using ultracold atoms in an optical lattice as a simulator for the
Schwinger mechanism was considered in ref.~\cite{Kasper:2015cca}.

Among the theoretical tools used to study the Schwinger mechanism, let
us mention a proposal to use stochastic quantization in
ref.~\cite{Fukushima:2014iqa}, and some of the many works where
holographic (e.g. based on the AdS/CFT correspondence) setups were
considered in order to investigate particle production by external
fields in the strong coupling regime
\cite{Semenoff:2011ng,Ambjorn:2011wz,Bolognesi:2012gr,Sato:2013pxa,Sato:2013iua,Sonner:2013mba,Hashimoto:2013mua,Dietrich:2014ala,Hashimoto:2014yya,Kawai:2015mha}.
Let us also finally mention another recent review on the Schwinger
mechanism more focused on the backreaction and applications in
astrophysics \cite{Ruffini:2009hg}, as well as some works on the
closely related question of pair creation in a curved space-time
\cite{Kim:2008xv,Garriga:2012qp,Frob:2014zka,Kim:2014iba,Cai:2014qba,Fischler:2014ama}.

\section*{Acknowledgements}
FG was supported by the Agence
Nationale de la Recherche project 11-BS04-015-01.


\providecommand{\href}[2]{#2}\begingroup\raggedright\endgroup

\end{document}